\documentclass[twocolumn]{aastex61}

\usepackage{amsmath}
\newcommand{\Mg}[5]{\mbox{$#1\,^#2{\rm #3}^{{\rm #4}}_{\rm #5}$}}

\newcommand{\Elow}{E_{\rm low}}
\newcommand{\Eup}{E_{\rm up}}
\newcommand{\Vmic}{\xi_{\rm t}}

\def\teff{T_{\rm eff}}

\newcommand{\mgi}{Mg\,{\textsc i}}
\newcommand{\mgii}{Mg\,{\textsc {ii}}}
\newcommand{\md}{\langle {\rm 3D} \rangle}
\def\mgfe{[\hbox{Mg}/\hbox{Fe}]}
\newcommand{\fei}{Fe\,{\textsc i}}
\newcommand{\feii}{Fe\,{\textsc {ii}}}
\def\feh{\hbox{[Fe/H]}}
\def\hi{H\,{\sc i}}

\accepted{5 July, 2017}
\submitjournal{ApJ}

\shorttitle{Mg abundances in cool stars}
\shortauthors{Bergemann et al.}

\begin{document}


\title{Non-local thermodynamic equilibrium stellar spectroscopy with 1D and $\md$ models - I. Methods and application to magnesium abundances in standard stars}


\correspondingauthor{Maria Bergemann}
\email{bergemann@mpia-hd.mpg.de}

\author{Maria Bergemann}
\affil{Max-Planck Institute for Astronomy, 69117, Heidelberg, Germany}

\author{Remo Collet}
\affil{Stellar Astrophysics Centre, Ny Munkegade 120, Aarhus University, DK--8000 Aarhus, Denmark}
\affil{Research School of Astronomy and Astrophysics, Australian National University, Canberra ACT 2601, Australia}

\author{Anish M. Amarsi}
\affil{Research School of Astronomy and Astrophysics, Australian National University, Canberra ACT 2601, Australia}
\affil{Max-Planck Institute for Astronomy, 69117, Heidelberg, Germany}

\author{Mikhail Kovalev}
\affil{Max-Planck Institute for Astronomy, 69117, Heidelberg, Germany}

\author{Greg Ruchti}
\affil{Lund Observatory, Box 43, SE-221 00 Lund, Sweden}

\author{Zazralt Magic}
\affil{Niels Bohr Institute, University of Copenhagen, Juliane Maries Vej 30, DK--2100 Copenhagen, Denmark}
\affil{Centre for Star and Planet Formation, Natural History Museum of Denmark, {\O}ster Voldgade 5-7, DK--1350 Copenhagen, Denmark}

\begin{abstract}
We determine Mg abundances in 6 Gaia benchmark stars using theoretical one-dimensional (1D) hydrostatic model atmospheres, as well as temporally- and spatially-averaged 3D model atmospheres ($\md$). The stars cover a range of $T_{\text{eff}}$ from $4700$~to $6500\,\mathrm{K}$, $\log{g}$~from $1.6$~to $4.4\,\mathrm{dex}$, and $\mathrm{\left[Fe/H\right]}$~from $-3.0\,\mathrm{dex}$~to solar. Spectrum synthesis calculations are performed in local thermodynamic equilibrium (LTE) and in non-LTE (NLTE) using the oscillator strengths recently published by Pehlivan Rhodin et al. We find that: a) Mg abundances determined from the infrared spectra are as accurate as the optical diagnostics, b) the NLTE effects on \mgi\ line strengths and abundances in this sample of stars are minor (although for a few \mgi\ lines the NLTE effects on abundance exceed $0.6\,\mathrm{dex}$ in $\md$ and $0.1\,\mathrm{dex}$ in 1D, c) the solar Mg abundance is $7.56 \pm 0.05\,\mathrm{dex}$~(total error), in the excellent agreement with the Mg abundance measured in CI chondritic meteorites, d) the 1D NLTE and $\md$~NLTE approach can be used with confidence to analyse optical \mgi\ lines in spectra of dwarfs and sub-giants, but for red giants the \mgi\ 5711 \AA\ line should be preferred, e) low-excitation \mgi\ lines are sensitive to the atmospheric structure; for these lines, LTE calculations with $\md$ models lead to significant systematic abundance errors. The methods developed in this work will be used to study Mg abundances of a large sample of stars in the next paper in the series.

\end{abstract}




\section{Introduction}
Chemical abundance ratios inferred from spectra of cool stars are key to understand stellar physics, to study planetary systems, and to unravel the formation and evolution of galaxies. Progress in these fields depends critically on our understanding of limitations and uncertainties of spectroscopic stellar abundance analyses. So far, several critical assumptions have underpinned the abundance diagnostics of cool stars. These approximations are (a) local thermodynamic equilibrium (LTE), (b) stationary one-dimensional (1D) geometry for radiative transfer and spectral line formation, and (c) hydrostatic equilibrium with convective energy transfer treated according to the mixing-length theory (MLT) \citep{prandtl1925,boehm1958} or equivalent formulations. The validity of the models built using these assumptions is questionable. Late-type stars possess an outer convection zone that affects the structure of the stellar atmosphere layers and the emergent stellar fluxes. Convective flows are reflected in the shapes and strengths of spectral lines \citep[e.g.][]{dravins1987a,dravins1987b,asplund2000,allende2001}. Furthermore, radiation transfer in the atmospheres of late-type stars generally takes place under non-local thermodynamic equilibrium (NLTE) conditions, rather than the idealised LTE \citep[e.g.][]{carlsson1992,asplund2005,bergemann2014}. 

A few recent studies of spectral line formation in late-type stars have used NLTE radiative transfer with 1D hydrostatic \citep[e.g.][]{short2006,mashonkina2013a}, $\md$\footnote{$\md$ models are 1D models that were derived by averaging the 3D radiative-hydrodynamical simulations of stellar convection over regions of equal optical depth and over the time series.} \citep[e.g.][]{bergemann2012b,mashonkina2013b,osorio2015}, and 3D model atmospheres \citep[e.g.][]{asplund2004,caffau2009,lind2013,nordlander2017}. 3D models are taken from the ab-initio time-dependent 3D radiation-hydrodynamics simulations of stellar surface convection \citep[e.g.][]{vogler2005,nordlund2009,freytag2012}. Convective flows develop naturally in this description without having to depend on approximate recipes.

More physically realistic modelling is more successful than the standard 1D LTE approach. Full 3D modelling is needed to describe the observed asymmetries in the line shapes, thereby improving the agreement of spectral line shapes with observations, including the flux spectra and the spatially-resolved spectra across the solar surface \citep[e.g.][]{allende2002,steffen2015,lind2017}. NLTE calculations reduce systematic abundance errors and ensure a better consistency between different spectroscopic diagnostics \citep[e.g.][]{mashonkina2007,bergemann2011}. Full 3D NLTE radiative transfer calculations are very computationally expensive, which has so far, prevented routine applications of this technique in spectroscopy of cool stars. In this respect, the $\md$ NLTE approach offers the best middle-ground between full 3D NLTE and 1D NLTE, by accounting for NLTE in model atoms of arbitrary size, and through the use of time-independent 1D structures derived from the full 3D hydrodynamic simulations, for the adiabatic cooling associated with surface convective overshooting. However, information about horizontal inhomogeneities is lost with the averaging, and therefore cannot be accounted for directly.
 
In \citet{bergemann2012b}, we began to systematically explore the effects of departures from 1D LTE on stellar parameters and abundances. In that paper, we focused on iron and studied the effect of NLTE spectral line formation with $\md$ model atmospheres on effective temperature $\teff$, surface gravity $\log g$, micro-turbulence $\Vmic$, and iron abundance (metallicity) [Fe$/$H].

In this work, we extend the methods developed in \citet{bergemann2012b} to magnesium, the element most commonly used in combination with iron to trace the star formation history of stellar populations \citep{fuhrmann1995,tolstoy2009}. As in \citet{bergemann2012b}, our main motivation is to explore the limitations of different physical models (LTE, NLTE, hydrostatic equilibrium) and to find the most robust, within the current computational capacities, diagnostics that can be used in quantitative spectroscopy and abundance determinations for cool stars. We carry out a detailed NLTE abundance analysis of Mg in the spectra of six benchmark stars using two different classes of model atmospheres, 1D hydrostatic and $\md$ models. In the follow-up paper (hereafter, Paper 2), we shall use the methods developed in this work to study the $\mgfe$ abundance ratios in a large sample of stars in the Galactic disk.

The paper is structured as follows. In Section \ref{sec:obsdata}, we present the observed dataset and the adopted stellar parameters of the program stars. Section \ref{sec:analysis} describes the 1D and $\md$ model atmospheres, atomic data, and NLTE line formation calculations. Section \ref{sec:results} compares LTE and NLTE, 1D and $\md$, Mg abundances determined for the program stars. We close the paper with conclusions in Section \ref{sec:conclusions}.
\section{Observations and basic stellar parameters}\label{sec:obsdata}
The sample of stars was selected from \citet{bergemann2012b} and \citet{hansen2013}. Some of them are Gaia-ESO and Gaia benchmark stars. For all stars in the sample, optical high-resolution spectra are available from observations with UVES spectrograph at the VLT \citep{bagnulo2003} or from the FOCES spectrograph at the 2.2m Calar-Alto telescope \citep{axer1994}. The UVES spectra have a slit-determined resolving power of $R = \lambda/\delta \lambda \sim$ 80\,000 and a signal-to-noise ratio S$/$N $\sim$ 300 near 5000 \AA. The FOCES spectra have $R \sim$ 60\,000 and a comparably high S$/$N. The solar spectrum was taken from the Kitt Peak Solar Flux Atlas \citep{kurucz1984}.

The infra-red spectra were kindly provided by Y. Takeda \citep{takeda2011,takeda2012}. The stars were observed with the IRCS spectrograph at the SUBARU telescope. The IRCS spectra cover the wavelength range 13500 $<$ $\lambda$ (\AA) $<$10900 \AA\ and have resolution $R \sim 30,000$. The H-band spectra are available from the APOGEE observations \citep{majewski2015}.

Stellar parameters for the program stars were adopted from our earlier studies \citep{bergemann2012b,hansen2013}. In brief, {\sc Hipparcos} parallaxes were used to fix the surface gravity, and stellar angular diameters to estimate $\teff$. For the metal-poor dwarf G 64-37, the effective temperature comes from photometric estimates, surface gravity, metallicity, and microturbulence from the NLTE ionization equilibrium of Fe lines \citep{hansen2013}. Based on our analysis of a metal-poor dwarf G 64-12 in \citet{bergemann2012b}, the difference between $\md$ and 1D NLTE metallicity for G 64-37 is expected to be of the order $0.03\,\mathrm{dex}$ and is neglected here. For the other stars, metallicities and microturbulence values were determined from $\md$ NLTE analysis of \fei\ and \feii\ lines (Table \ref{tab:1}).
%
%
% ------------- stellar parameter table ----------------- 
% -------------------------------- FINAL PARAM ---------------------------------
%
%
\begin{table*}
\caption{Input stellar parameters for the reference stars. The references to stellar parameter estimates and their errors are given in the source
  column. The solar metallicity is given in terms of $\rm A(\rm Fe)$.}
\label{tab:1}
\renewcommand{\tabcolsep}{4pt}
\begin{center}
\begin{tabular}{ll ccc ccc llll}
\hline\noalign{\smallskip}
Star &  HD & $T_{\rm eff}$ & $\sigma$ & $\log g$ & $\sigma$ & [Fe$/$H] & $\Vmic$    & Source   & \multicolumn{3}{c}{Observations} \\
          &             & K      &     &   dex    &      &  dex     &  kms$^{-1}$ &  & optical & Y,J-band & H-band \\
\noalign{\smallskip}\hline\noalign{\smallskip}
Sun       &             & $5777$ &   1 &  $4.44$  & 0.01 &   ~~7.44 &  1.00  & \citet{bergemann2012b} & FTS KPNO  &  FTS IR  & APOGEE$^d$  \\
Procyon   &  HD 61421   & $6543$ &  84 &  $3.98$  & 0.02 &  $-$0.03 &  2.05  & \citet{bergemann2012b} & UVES-POP$^a$  &  -     &  -      \\ 
          &  HD 84937   & $6408$ &  66 &  $4.13$  & 0.09 &  $-$2.03 &  1.38  & \citet{bergemann2012b} & UVES-POP$^a$  &  -     & APOGEE$^d$  \\
          &  HD 140283  & $5777$ &  55 &  $3.70$  & 0.08 &  $-$2.40 &  1.18  & \citet{bergemann2012b} & UVES-POP$^a$  &  IRCS$^c$    &  -      \\
          &  HD 122563  & $4665$ &  80 &  $1.64$  & 0.16 &  $-$2.57 &  1.66  & \citet{bergemann2012b} & UVES-POP$^a$  &  IRCS$^c$  & APOGEE$^d$  \\
          &  G 64-37    & $6494$ &  100  &  $4.23$  & 0.10 &  $-$3.00 &  1.40  & \citet{hansen2013}     & FOCES$^b$      &  IRCS$^c$  & -       \\
\noalign{\smallskip}\hline\noalign{\smallskip}
\end{tabular}
Note:  $^a$ \citet{bagnulo2003}, $^b$ \citet{axer1994}, $^c$ \citet{takeda2011,takeda2012}, $^d$ \citet{majewski2015}    
\end{center}
\end{table*}
%

%
% ---- ANALYSIS -----
%
\section{Analysis}{\label{sec:analysis}}
Line formation calculations and abundance determinations require an underlying model of the temperature and density stratification in the stellar atmosphere, together with the number densities of free electrons and of the most important atomic and molecular species contributing to the continuous and line background opacities. Hereafter, we describe the model atmospheres, atomic data, and the codes used in this paper for the calculations of NLTE statistical equilibria, Mg spectrum synthesis and abundance determinations.
\begin{figure}[ht!]
\includegraphics[width=0.7\columnwidth,angle=-90]{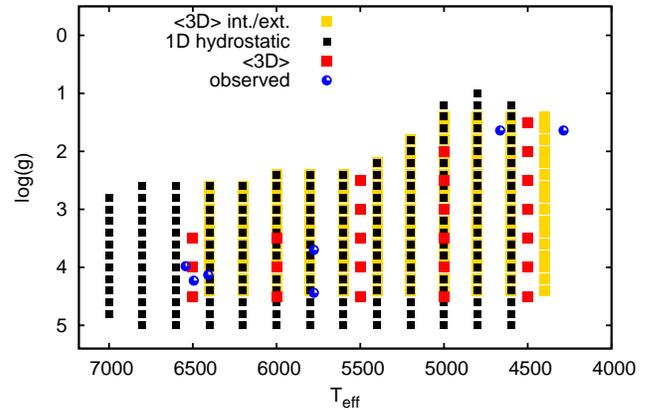}
\caption{The grid of interpolated and extrapolated $\md$ and 1D hydrostatic model atmospheres. The nodes of the original $\md$ model atmosphere grid used for the interpolation and extrapolation are also shown. The stellar parameters of the observed stars are indicated (see Section  \ref{sec:obsdata}).}
\label{grid1}
\end{figure}
\subsection{Model atmospheres}{\label{sec:atmos}}
We have used three different types of model atmospheres. The first choice are the classical 1D hydrostatic plane-parallel MAFAGS-OS model atmospheres \citep{grupp2004a,grupp2004b} employing the opacity-sampling scheme. Secondly, we use $\md$ models\footnote{We adopt the unaltered models, to which no correction in the effective temperature was applied.} that were constructed by averaging the physical structure from 3D {\sc Stagger} stellar surface convection simulations of dwarfs and giants \citep{collet2011, magic2013a} as described in more details below. Since neither MAFAGS-OS nor $\md$ models include chromosphere, which is clearly a poor approximation for the Sun, we also explore the semi-empirical model from \citet[][Table 11, hereafter MACKKL]{maltby1986}, which is a model for the quiet Sun derived using model parameters for the chromosphere from the studies by \citet{avrett1984} and \citet{avrett1985} and constrained by the solar observations in the EUV and microwave regime. This model was interpolated to a finer depth scale as described in \citet{bergemann2011}.

A description of the averaging procedure of time-dependent 3D model atmospheres is given in \citet{magic2013b}. In short, each 3D simulation consists of several (${\sim}80$-$100$) snapshots taken at regular intervals in time. For each 3D snapshot, we computed the optical depth at $5000$~{\AA}\footnote{Here and throughout the text, the parameter $\log \tau_\mathrm{5000}$ refers to the optical depth in the continuum at $5000$~{\AA}.} column by column. We then performed a cubic interpolation of the relevant physical and thermodynamic variables to a reference optical depth scale and averaged them on surfaces of constant optical depth. For gas density, gas pressure and electron number density we interpolated in the logarithm. Finally, we averaged all mean stratifications from individual snapshots to get a combined spatial and temporal averaged 3D structure, which we can use as alternative to classical 1D model stellar atmospheres.

We extended the {\sc Stagger} grid of $\md$~model atmospheres by interpolation and extrapolation. First, we interpolated the grid of the $\md$ models to overlap with the grid of MAFAGS-OS models shown in Figure \ref{grid1}. The atmospheric interpolation was done in two stages. At first, a simple and stable 2-point 1D linear interpolation was used to fill in the gaps in the original model data cube, which has 3 dimensions according to the stellar parameters $\teff$, $\log g$, and $\feh$. In other words, we search for 8 nearest models surrounding the desired $\md$ node. The algorithm runs over all grid points in this cube, and checks whether a point exists. If not, it locates two existing grid points near the missing data point, requiring that these two grid points have the same values of at least two parameters and differs only in the last parameter. We always try to interpolate first, and extrapolate only if interpolation is impossible; when extrapolation is necessary, we always consider several neighbour points that are close to the grid limit. Next, interpolation coefficients were calculated, and all values of the atmospheric thermodynamic variables, temperature, electron number density, gas pressure, opacity, and density were evaluated, for all depth points in the model atmosphere. This procedure was repeated until a complete cube in the full parameter space was obtained. This provided a more finely-sampled grid of $2112$ $\md$ models that can be used with the same algorithm as 1D static model in the spectrum synthesis and abundance analysis codes. The grid details are: $\teff$ step of 200 K, $\log g$ step of 0.2 dex and $\feh$ step of 0.3 dex. The grid covers the following range of stellar parameters: $4400 \leq \teff \leq 6400$, $1.4 \leq \log g \leq 4.6$, and $-3.1 \leq \feh$ $\leq 0.3$. In the second stage, an 8-point 3D linear interpolation scheme identical to the one described above was used to interpolate within the complete grid cube in the spectrum synthesis code. This is done to provide the model atmospheres for the stellar parameters of the target stars, which are used to perform detailed radiative transfer in the Mg lines.
\begin{figure*}
\begin{minipage}[h]{1.0\linewidth}
	\centering
	\hbox{
	\includegraphics[width=0.5\textwidth, angle=0]{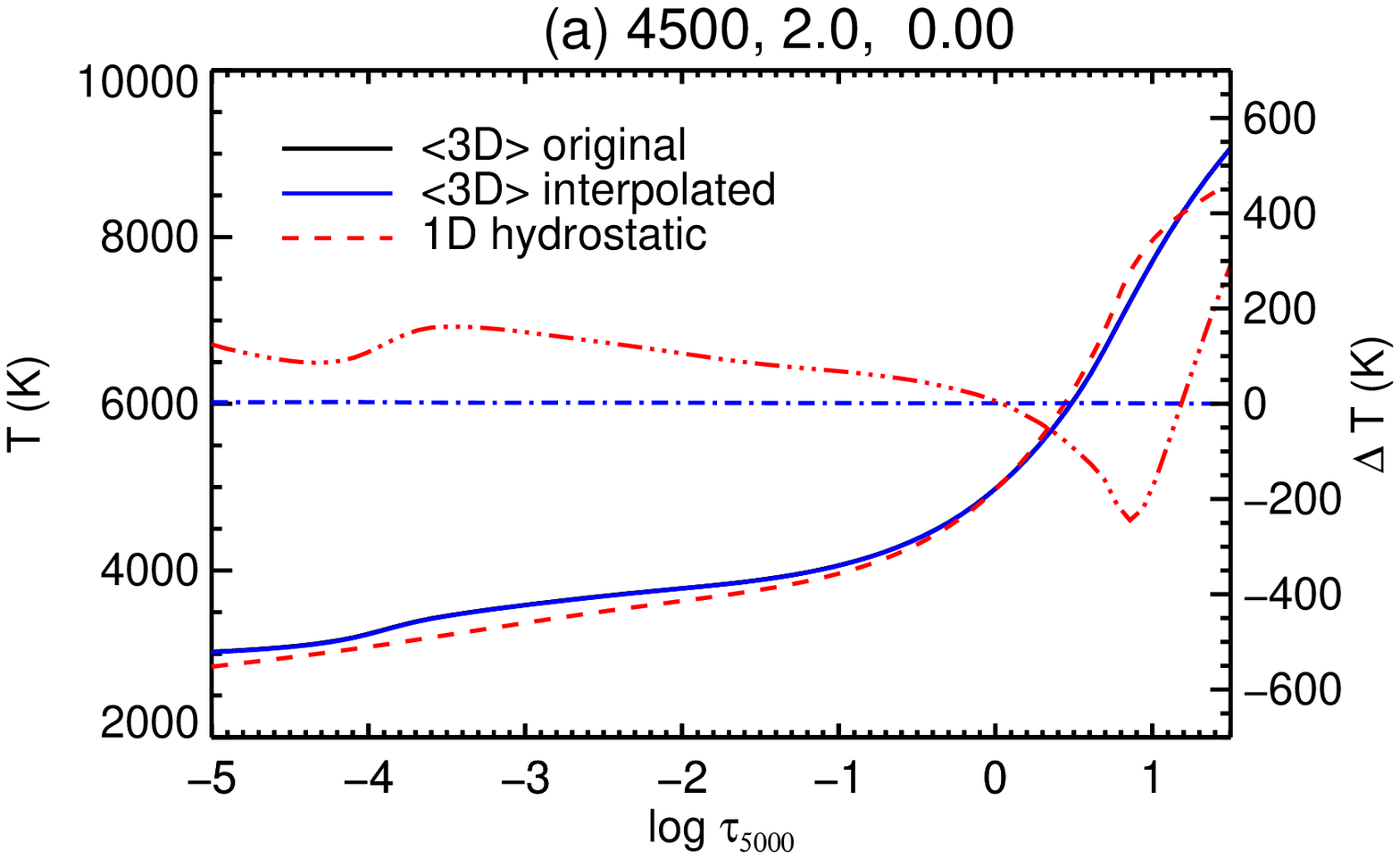}
	\includegraphics[width=0.5\textwidth, angle=0]{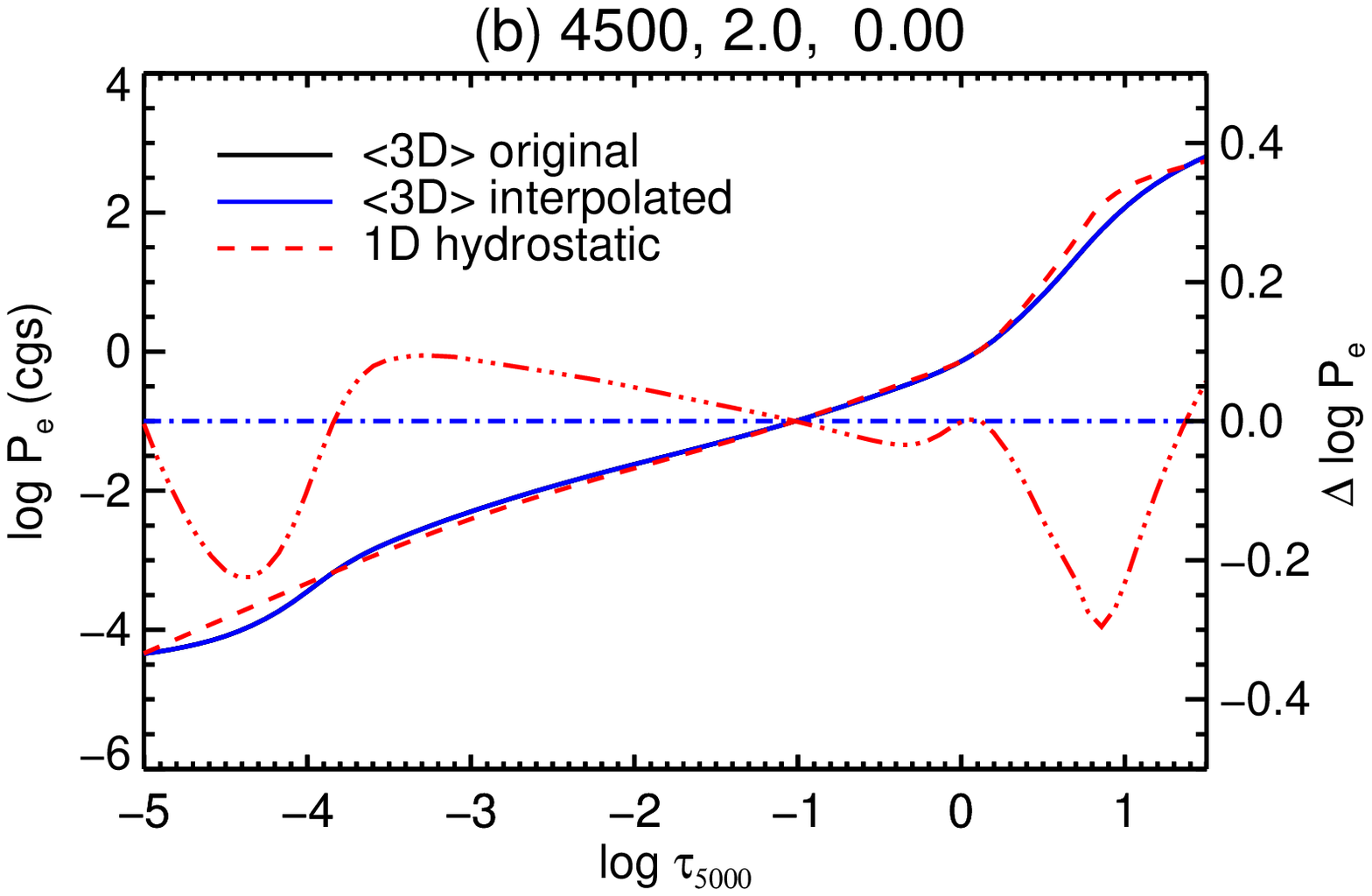}
	}
	\vspace{0.1cm}
\end{minipage}
\begin{minipage}[h]{1.0\linewidth}
	\centering
	\hbox{
	\includegraphics[width=0.5\textwidth, angle=0]{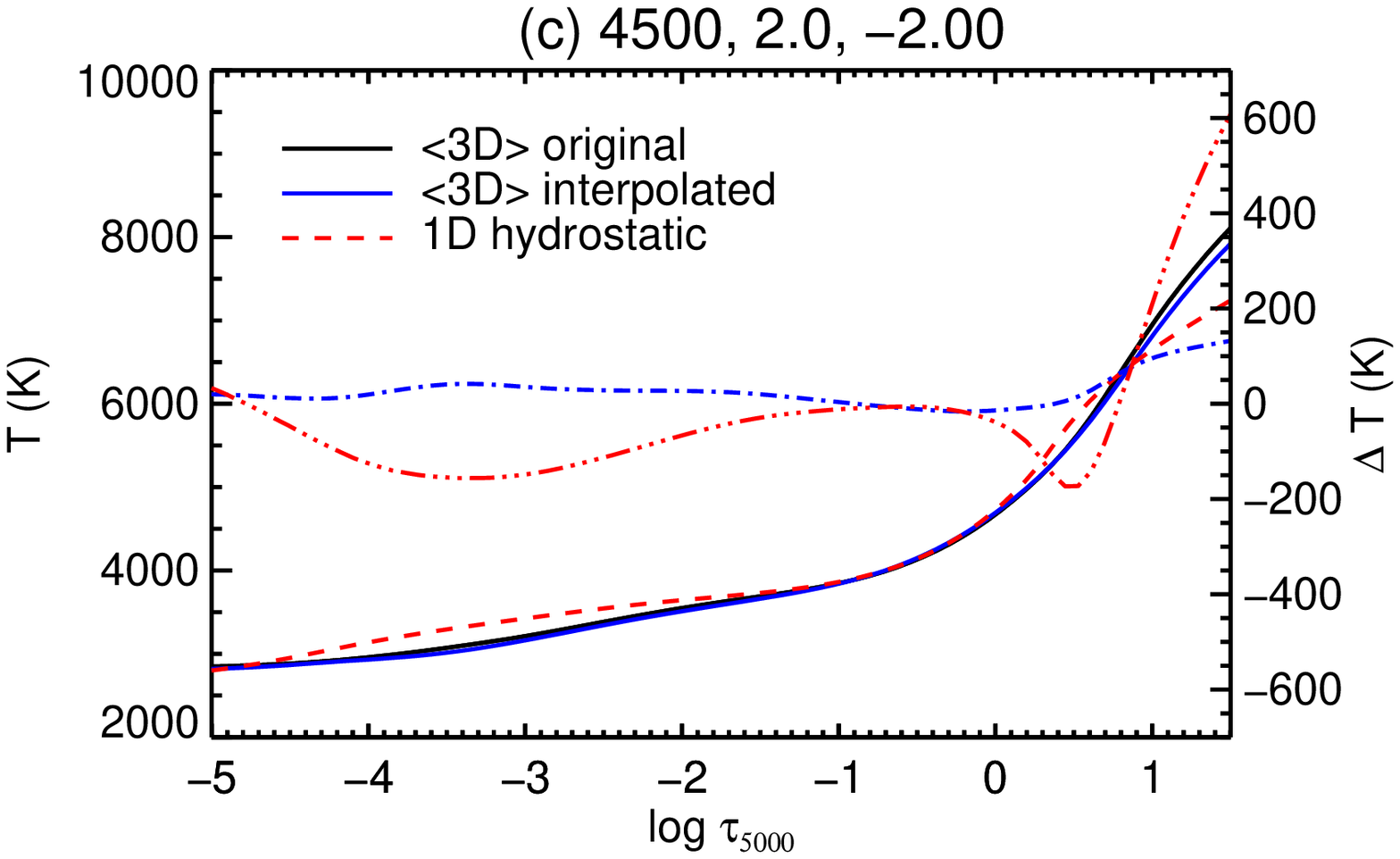}
	\includegraphics[width=0.5\textwidth, angle=0]{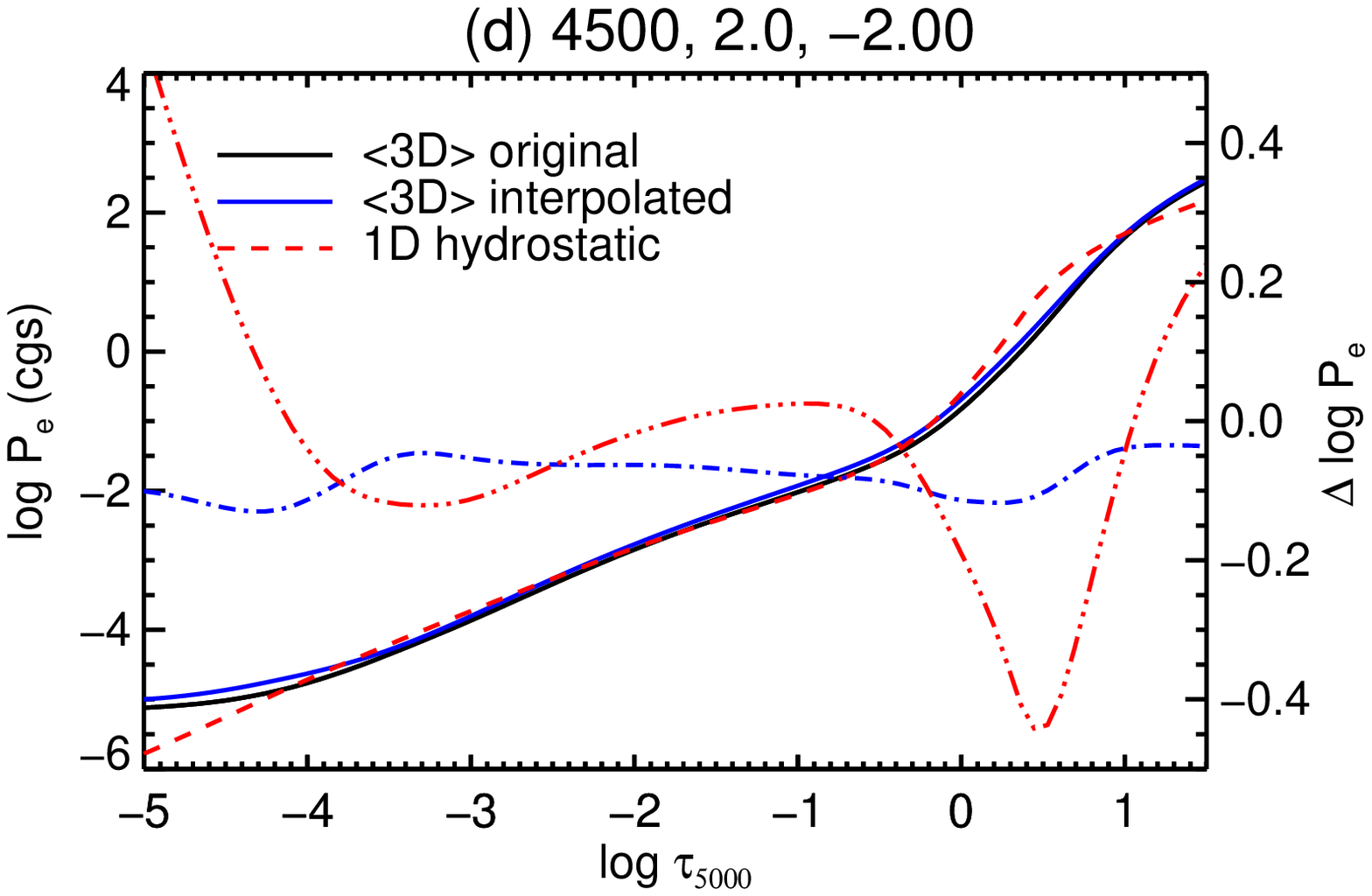}
	}
	\vspace{0.1cm}
	\end{minipage}
\begin{minipage}[h]{1.0\linewidth}
	\centering
	\hbox{
	\includegraphics[width=0.5\textwidth, angle=0]{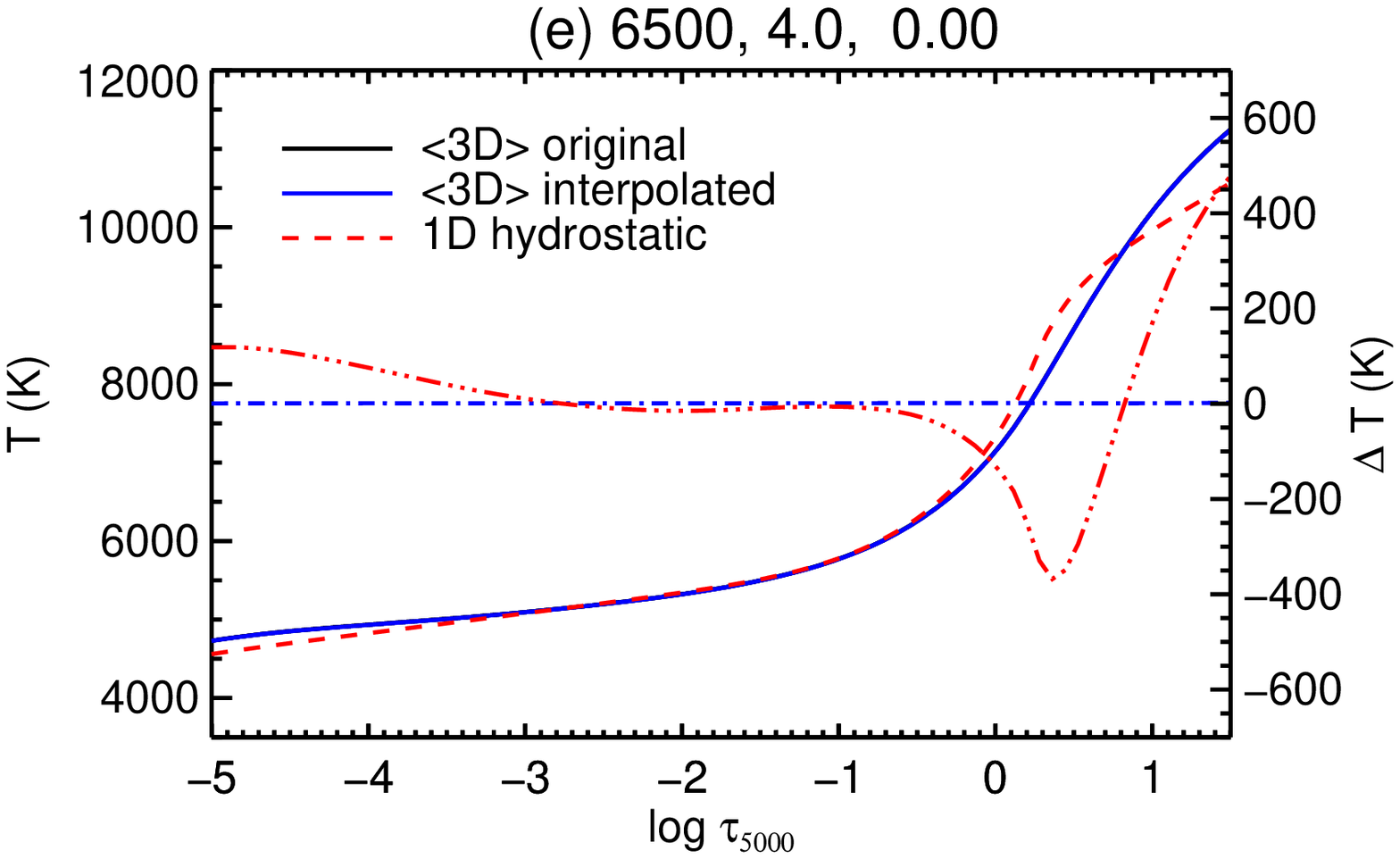}
	\includegraphics[width=0.5\textwidth, angle=0]{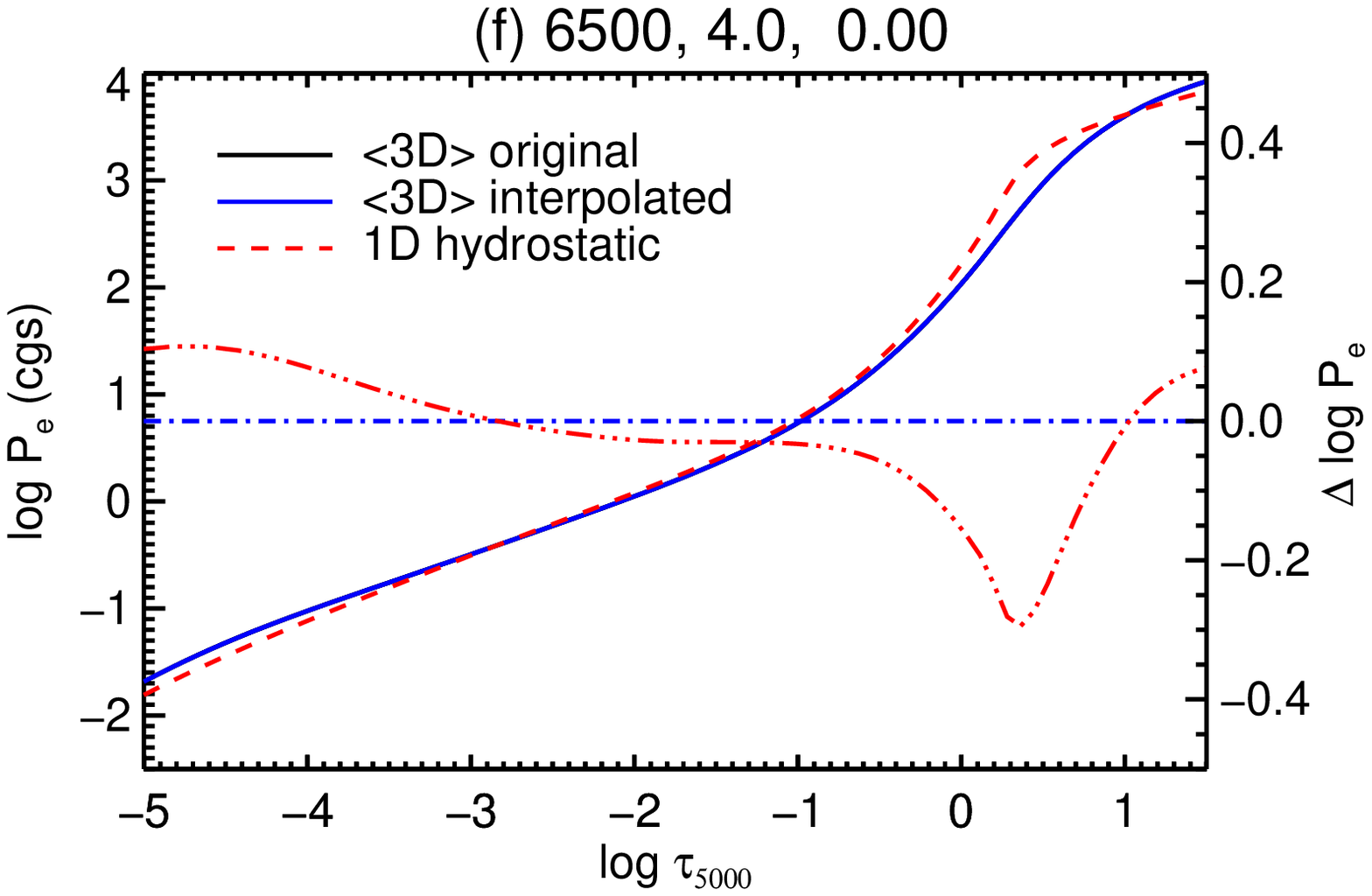}
	}
	\vspace{0.1cm}
	\end{minipage}
\begin{minipage}[h]{1.0\linewidth}
	\centering
	\hbox{
	\includegraphics[width=0.5\textwidth, angle=0]{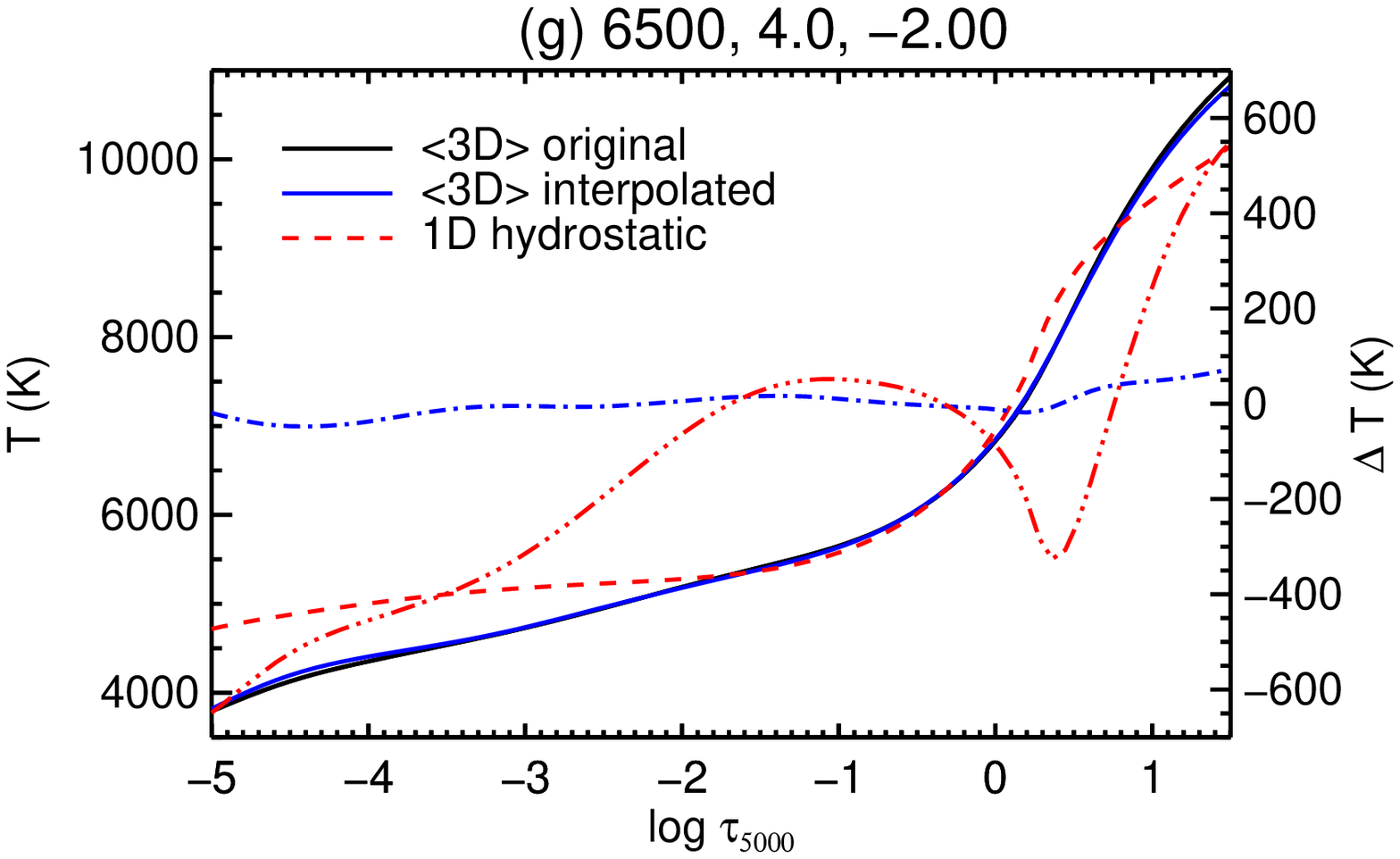}
	\includegraphics[width=0.5\textwidth, angle=0]{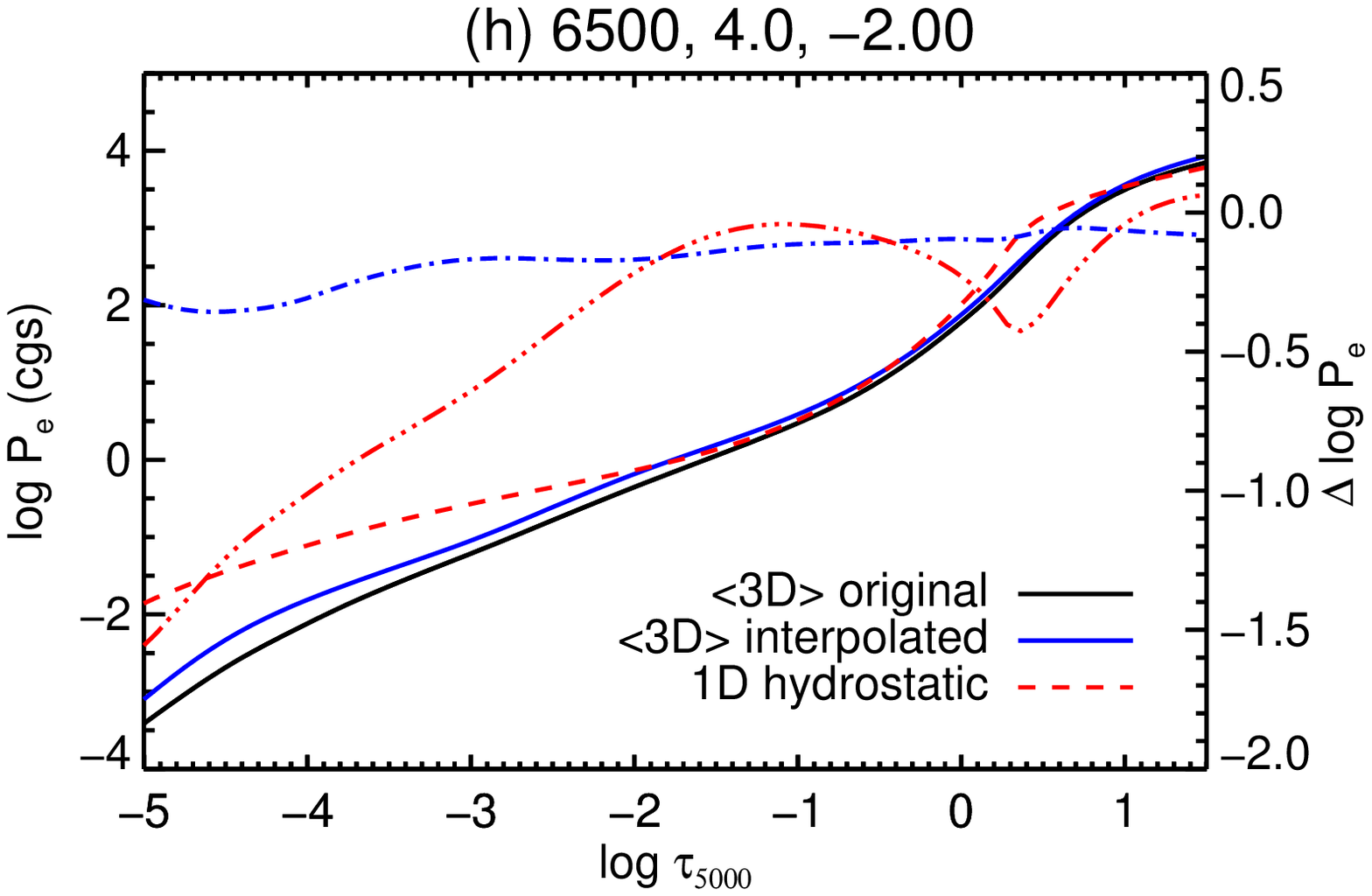}
	}
	\vspace{0.1cm}
	\end{minipage}
\caption{Temperature and electron pressure structures in the 1D hydrostatic (dashed lines) and directly averaged $\md$ model atmospheres (solid red lines) plotted as a function of optical depth $\log \tau_{5000}$. We also show $\md$ models derived through interpolation in the resampled grid (solid blue lines). Stellar parameters of the models, $\teff$, $\log g$, and $\feh$ are given in the plot titles. Also shown are the differences between the original $\md$ and 1D hydrostatic (red dash dot dot dot line), as well as the original $\md$ and interpolated $\md$ (blue dashed-dotted) models. See section \ref{sec:atmos}.} 
\label{grid2}
\end{figure*}
\subsection{Comparison of the $\md$ and 1D hydrostatic model structures}{\label{sec:compmodels}}
In Figure \ref{grid2} we show the thermodynamic structures of the $\md$ model atmospheres derived from the re-sampled grid, the directly averaged 3D  models, and the 1D hydrostatic models.  The models are chosen at $\feh = -2, 0\,\mathrm{dex}$, $\teff = 4500, 6500$ K, and $\log g = 2, 4\,\mathrm{dex}$, because the stellar parameters are representative of the sample studied in Paper 2, and directly averaged 3D models are available that allows us to check the interpolation scheme described in Section \ref{sec:atmos}.

Comparison of the hydrostatic and $\md$ models shows that including realistic convection has a different effect on the atmospheric structure of metal-rich and metal-poor models. Whereas in the metal-rich hydrostatic models, $\feh = 0.0$, the outer regions are cooler than the $\md$ structures, the behavior is reversed in the metal-poor regime. This effect of surface 'warming' in the solar-metallicity $\md$ models is not an artefact of the simulations and was also found in other studies \citep[e.g. the {\sc CO5BOLD} simulations, ][]{ludwig2012}. The metal-poor $\md$ models of dwarfs are significantly, by up to $600\,\mathrm{K}$(Figure \ref{grid2}, panel (g)), cooler than their hydrostatic counterparts. The effect of adiabatic cooling on the average structure of metal-poor red giant models (Fig. \ref{grid2}, panel (c)) is not so extreme: the outer layers of the $\md$ models are $\sim 200\,\mathrm{K}$cooler than the 1D hydrostatic models. Differences in the electron pressure $P_{\rm e}$ are of the order $50 \%$ in the inner layers, $0 \lesssim \log \tau_\mathrm{5000} \lesssim 1$, but amount to more than an order of magnitude in the dwarf models at low metallicity.

\begin{figure}
	\includegraphics[width=0.85\columnwidth, angle=-90]{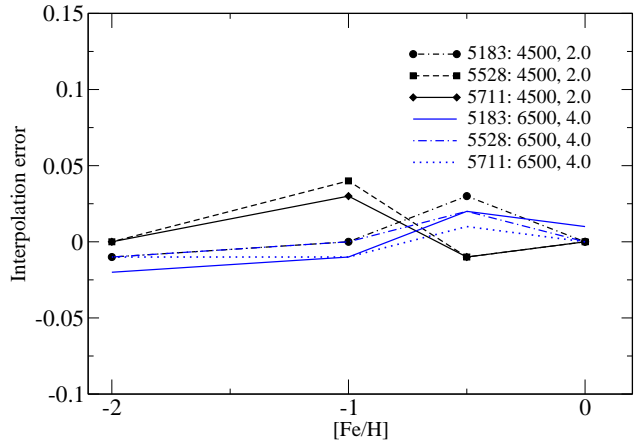}
\caption{Abundance errors for the \mgi\ lines at 5183, 5528, and 5711 \AA\ caused by the interpolation in the resampled $\md$ model atmosphere grid. Stellar parameters ($\teff$, $\log g$) are indicated in the legend. See Section \ref{sec:atmos}.} 
\label{interror}
\end{figure}

The differences between the interpolated and the original $\md$ models are generally small and do not exceed ten percent, which is, in fact, remarkable, given that the dynamical range in the parameter space is very large: electron pressure varies by $\sim 5$ orders of magnitude and electron temperature by a factor of $5$ over the narrow depth range ($-5 \leq \log \tau_{5000} \leq 1$). The variations of T$(\tau)$  and P$_{\rm e}(\tau)$ with depth appear to be well captured by the chosen interpolation scheme, except the metal-poor model of a hot turnoff star, where electron pressure deviates by $\sim 40$ percent in the outer atmospheric layers, $\log \tau_{5000} \lesssim -3.5$ (Figure \ref{grid2}, panel (h)). While this difference is small compared to the differences between the 1D and $\md$ models, this may have an influence on the line profile and abundance determinations from the \mgi\ lines. 
To test this effect, we have performed a series of LTE calculations of \mgi\ line profiles using directly averaged $\md$ models from Figure \ref{grid2} and the $\md$ models derived by interpolation in the resampled grid. The abundance errors were evaluated by comparing the line equivalent widths (EW) and adjusting the Mg abundance in the calculations with the interpolated $\md$ models to fit the line EWs of the directly averaged $\md$ model. The results are shown in Figure \ref{interror}. The abundance errors caused by the interpolation in the $\md$ grid are within $0.04\,\mathrm{dex}$ for giants and within $0.02\,\mathrm{dex}$ for dwarfs. Interestingly, the larger P$_{\rm e}(\tau)$ error in the model of a metal-poor dwarf does not cause significant errors in the Mg abundance. This is because the line formation in very metal-poor atmospheres takes place at deeper layers (see section \ref{sec:bmk}). The errors are taken into account in the abundance analysis of the program stars in Section \ref{sec:results}.
%
%
% ---------------- atomic data table -------------------
%
\begin{table*}
\caption{Parameters of the \mgi\ lines used in the analysis. $\sigma_{\log gf}$ is the uncertainty of the oscillator strength, given in $\%$ if the estimates were taken from the NIST database. References to the transition probabilities and damping constants are given in columns 10 and 11, respectively. The parameters $\alpha$ and $\sigma$ are the dimensionless velocity parameter and the broadening cross-section in the atomic units of cross-section, a$_{\rm 0}^2$ \citep[as defined in][]{barklem2000}, respectively.}
\label{tab:at_data}
\renewcommand{\tabcolsep}{3pt}
\begin{center}
\begin{tabular}{l | cc cc | cc cc ll}
\hline
~~~$\lambda$ (air) & Lower & Upper &  $\Elow$ & $\Eup$ &  $\log gf$ & $\sigma_{\log gf}$ & $\sigma$ & $\alpha$ (ABO) & Reference ($\log gf$) & Reference (ABO) \\
~~\AA\  &  level & level & [eV] & [eV]  &  & &  &  &  \\ 
\noalign{\smallskip}\hline\noalign{\smallskip}
optical & & & & & &  & \\
4571.096  & \Mg{3s^2}{1}{S}{}{0}     & \Mg{3p}{3}{P}{\circ}{1} & 0.00  & 2.71  & $-$5.397  & 0.05 &   -- & --     &  \citet{pehl}$^{a}$  & \citet{unsold1955}     \\
4702.995  & \Mg{3p}{1}{P}{\circ}{1} &  \Mg{5d}{1}{D}{}{2}     & 4.35  &  6.98  & $-$0.456  & 0.05 &   2806 & 0.269     &  \citet{pehl}$^{a}$  & \citet{barklem2000}    \\
5172.684  & \Mg{3p}{3}{P}{\circ}{1} &  \Mg{4s}{3}{S}{}{1}     & 2.71  & 5.11  & $-$0.363  & 0.04  &  729 & 0.238  &  \citet{pehl}      & \citet{barklem2000}    \\
5183.604  & \Mg{3p}{3}{P}{\circ}{2} &  \Mg{4s}{3}{S}{}{1}     & 2.72  & 5.11  & $-$0.168  & 0.04  &  729 & 0.238  &  \citet{pehl}      & \citet{barklem2000}    \\
5528.405  & \Mg{3p}{1}{P}{\circ}{1} &  \Mg{4d}{1}{D}{}{2}     & 4.35  & 6.59  & $-$0.547  & 0.02 & 1461 & 0.312  &  \citet{pehl}      & \citet{barklem2000}    \\
5711.088  & \Mg{3p}{1}{P}{\circ}{1} &  \Mg{5s}{1}{S}{}{0}     & 4.35  & 6.52  & $-$1.742  & 0.05 & 1860 & 0.100  &  \citet{pehl}$^{a}$  & Barklem priv. comm     \\
8806.756   & \Mg{3p}{1}{P}{\circ}{1} &  \Mg{3d}{1}{D}{}{2}     & 4.35  & 5.75  & $-$0.144 & 0.03 &  530 & 0.277 &  \citet{pehl}       & \citet{barklem2000}    \\
Y-band & & & & & & & \\
10811.053   & \Mg{3d}{3}{D}{}{3}      & \Mg{5f}{3}{F}{\circ}{4} & 5.95  & 7.09  &  ~~0.052 & 0.04 & 2984 & 0.334 &   \citet{pehl}        & \citet{barklem1998}    \\
10811.076   & \Mg{3d}{3}{D}{}{2}      & \Mg{5f}{3}{F}{\circ}{3} & 5.95  & 7.09  & $-$0.137 & 10\% & 2961  & 0.336    &   \citet{butler}$^{b,c}$ & \citet{barklem2000}     \\
10811.097   & \Mg{3d}{3}{D}{}{3}      & \Mg{5f}{3}{F}{\circ}{3} & 5.95  & 7.09  & $-$1.038 & 18\% & 2961  & 0.336    &   \citet{butler}$^{b}$ & \citet{barklem2000}     \\
10811.122   & \Mg{3d}{3}{D}{}{2}      & \Mg{5f}{3}{F}{\circ}{2} & 5.95  & 7.09  & $-$1.036 & 18\% & 2960  & 0.336    &   \citet{butler}$^{b}$ & \citet{barklem2000}     \\
10811.143   & \Mg{3d}{3}{D}{}{3}      & \Mg{5f}{3}{F}{\circ}{2} & 5.95  & 7.09  & $-$2.587 & 50\% & 2960   & 0.336    &   \citet{butler}$^{b}$ & \citet{barklem2000}     \\
10811.158   & \Mg{3d}{3}{D}{}{1}      & \Mg{5f}{3}{F}{\circ}{2} & 5.95  & 7.09  & $-$0.321 & 0.04 &  2960  & 0.336    &   \citet{pehl}         & \citet{barklem2000}     \\
10965.386   & \Mg{4p}{3}{P}{\circ}{2} & \Mg{5d}{3}{D}{}{1}      & 5.93  & 7.06  & $-$2.184 & 0.05 &  --  & --    &   \citet{pehl}$^{a}$  & \citet{unsold1955} \\
10965.414   & \Mg{4p}{3}{P}{\circ}{2} & \Mg{5d}{3}{D}{}{2}      & 5.93  & 7.06  & $-$1.008 & 0.05 &  --  & --    &   \citet{pehl}$^{a}$  & \citet{unsold1955} \\
10965.450   & \Mg{4p}{3}{P}{\circ}{2} & \Mg{5d}{3}{D}{}{3}      & 5.93  & 7.06  & $-$0.260 & 0.05 & 3328 & 0.238 &   \citet{pehl}$^{a}$  & \citet{barklem1997}$^d$ \\
J-band & & & & & & & \\                                                                                                  
11828.185 & \Mg{3p}{1}{P}{\circ}{1}  & \Mg{4s}{1}{S}{}{0}      & 4.35  & 5.39  & $-$0.350  & 0.03  &  862 & 0.225  & \citet{pehl}  & \citet{barklem2000}  \\
12083.278 & \Mg{3d}{1}{D}{}{2}       & \Mg{4f}{3}{F}{\circ}{3} & 5.75  & 6.78  & $-$1.347  & 0.05   & -- & -- &  \citet{pehl}   & \citet{unsold1955}  \\
12083.346 & \Mg{3d}{1}{D}{}{2}       & \Mg{4f}{3}{F}{\circ}{2} & 5.75  & 6.78  & $-$1.500  & --   & -- & --  &   \citet{bergemann2015}   & \citet{unsold1955}  \\
12083.662 & \Mg{3d}{1}{D}{}{2}       & \Mg{4f}{1}{F}{\circ}{3} & 5.75  & 6.78  & ~~0.377  & 0.04  & 1466 & 0.329  &  \citet{pehl}  & \citet{barklem2000}  \\
H-band  & & & & & & & \\                                                                                                 
15024.992  & \Mg{4s}{3}{S}{}{1}      & \Mg{4p}{3}{P}{\circ}{2} & 5.11  & 5.93  &  ~~0.334  & 0.03  &  952 & 0.255  & \citet{pehl}  & \citet{anstee1995}$^e$  \\
15748.988 & \Mg{4p}{3}{P}{\circ}{1} & \Mg{4d}{3}{D}{}{2}      & 5.93  & 6.72  &  ~~0.129  & 0.05  & -- & --  &   \citet{pehl}$^{a}$  & \citet{unsold1955} \\
15765.645 & \Mg{4p}{3}{P}{\circ}{1} & \Mg{4d}{3}{D}{}{3}      & 5.93  & 6.72  &  $-1$.524  & 0.05  & -- & --  &   \citet{pehl}$^{a}$  & \citet{unsold1955} \\
15765.747 & \Mg{4p}{3}{P}{\circ}{1} & \Mg{4d}{3}{D}{}{3}      & 5.93  & 6.72  &  $-0$.348  & 0.05 & -- & --  &   \citet{pehl}$^{a}$  & \citet{unsold1955} \\
15765.842 & \Mg{4p}{3}{P}{\circ}{1} & \Mg{4d}{3}{D}{}{3}      & 5.93  & 6.72  &  ~~0.400   & 0.05  & 1636 & 0.280  &  \citet{pehl}$^{a}$  & \citet{barklem1997}$^e$ \\
15886.183 & \Mg{3d}{3}{D}{}{2}      & \Mg{5p}{3}{P}{\circ}{1} & 5.95  & 6.73  &  $-$1.465 & 0.05 & 1960 & 0.239  & \citet{pehl}$^{a}$    & \citet{barklem1997}$^d$ \\
15886.261 & \Mg{3d}{3}{D}{}{2}      & \Mg{5p}{3}{P}{\circ}{1} & 5.95  & 6.73  &  $-$1.942 & 0.05 & -- & --  &  \citet{pehl}$^{a}$   & \citet{unsold1955} \\
\noalign{\smallskip}\hline\noalign{\smallskip}
\end{tabular}
%\medskip
Note: $^a$ theoretical transition probabilities from \citet{pehl}; $^b$ theoretical f-values from \citet{butler}; $^c$ the transition probability computed from the multiplet value using a pure LS-couping; $^d$ calculated according to the prescription given in the reference; $^e$ interpolated from the tables given in the reference.
\end{center}
%\end{minipage}
\end{table*}

\subsection{NLTE statistical equilibrium and spectrum synthesis}
\label{sec:statisticalequilibrium}
The model atmospheres described above were used in combination with NLTE statistical equilibrium calculations to compute atomic number densities for Mg as a function of depth. The Mg atomic model has two ionization stages and was compiled by \citet{zhao1998}. The model was subsequently updated by \citet{mashonkina2013a}, who included the quantum mechanical collision rates with \hi,  and further tested on the spectra of the Sun and Arcturus in \citet{bergemann2015}. In this work, we have replaced the energy levels with $l \geq 3$ (instead of $l \geq 5$ in \citealt{zhao1998}) by the estimates computed using the polarisation formula by \citet{chang1983} that was recommended for the upper Mg states by \citet{carlsson1992}. The atom is constructed with $86$ energy levels ($85$ \mgi\ and closed by \mgii\ ground state), $65$ bound-free channels with full frequency dependence of the cross-sections as provided by the Opacity project \citep{butler1993}, and $453$ bound-bound transition channels. For the electron-impact excitation, we use the data by \citet{mauas1988}, if available, and \citet{zhao1998} for the remaining transitions. Ionization by electronic collisions was calculated from the \citet{seaton1962}'s formula with the effective Gaunt factor $\bar{g}$ set equal to $0.1$ for \mgi\ and to $0.2$ for \mgii. The quantum mechanical calculations of \citet{barklem2012} were used to compute the rate coefficients for \hi\ impact excitations and charge transfer processes. These data are available for the transitions between 7 \mgi\ states with the lower level excitation potential $E_{\rm low} \leq 5.93$ eV and the \mgii\ ground state; collisional coupling of the other energy states is provided only by electrons. The NLTE departure coefficients were computed using the DETAIL code \citep{butler1985}. The code was updated with the new linelists for molecular opacity calculations, including TiO, which are important for modelling RGB and red supergiant stars. The background opacity lists include about 10 million lines. Validation of the code with different types of 1D hydrostatic and mean 3D models was presented in \citet{bergemann2012b}.

The NLTE departure coefficients were used in the SIU spectrum synthesis code \citep{reetz1999, bergemann2012a} to correct the line opacities and line source functions for NLTE effects. Line profile fitting was done by adjusting the abundance and macro-turbulence for each \mgi\ line individually. The main updates in the SIU code include the improved molecular linelists and implementation of the damping constants from \citet{barklem2000} in the form of $\alpha$ and $\sigma$ coefficients. The line width is thus computed using the correct temperature exponent\footnote{Note that in the previous versions of SIU, the damping was implemented using the Uns\"{o}ld formalism, however, with the width computed using the ABO theory data for a fixed value of temperature.}.
\subsection{Atomic data}
The atomic data we adopted in the spectrum synthesis calculations are given in Table \ref{tab:at_data}. Wavelengths and oscillator strengths were taken from the \citet{pehl}, who provide experimental and theoretical transition probabilities derived by combining new branching fractions measured using the Fourier transform spectrometer at the Lund Observatory and theoretical level lifetimes. Although for some \mgi\ lines only theoretical $\log gf$ were available, we adopt their estimates, too. \citet[][their Figure 4]{pehl} show that their experimental results agree very well with their theoretical calculations, to better than a few percent. The uncertainties of the $\log gf$ estimates are typically very small, and for some \mgi\ lines in our list the errors are within $0.03\,\mathrm{dex}$. The uncertainties were carefully evaluated by combining the uncertainties of the branching fractions and the level lifetimes, for the former including the uncertainties of the measured intensities, instrumental effect (intensity of the calibration lamp), self-absorption correction, and the uncertainty of the normalisation factor. When experimental lifetimes were used, the uncertainties were estimated taking into account the statistical scattering errors and systematic effects \citep{jonsson1984}. For theoretical lifetimes, \citet{pehl} estimated the error by comparing with the experimental lifetimes in the literature.

The data from \citet{pehl} are generally consistent with the earlier estimates, e.g. \citet{froese2006}, with some exceptions. There are the \mgi\ intercombination line at 4571 \AA\ and the strong optical lines at $5172$ and $5183$ \AA. For the optical triplet lines, $5172$ \AA\ and $5183$ \AA, the oscillator strengths by \citet{pehl} are $\sim 0.1\,\mathrm{dex}$ higher. This difference is caused by the revision of the upper level lifetime, which \citet{pehl} determine to be $9.63$ ns for the \Mg{4s}{3}{S}{}{1} level, whereas the earlier experimental measurement by \citet{aldenius2007} gives $11.5 \pm 1.0$ ns. For the \mgi\ line at 4571 \AA, the oscillator strength determined by \citet{pehl} is $0.2\,\mathrm{dex}$ higher than the earlier estimate by \citet{tachiev} that was computed using the multiconfiguration Hartree-Fock method, $\log gf = -5.623$. The latter value is quite uncertain; according to the NIST database \citep{kramida2015} the error is $50 \%$. On the other hand, the theoretical estimate by \citet{tachiev} is closer to the experimental measurement by \citet{kwong1982}, $\log gf = -5.686$. Hence although we prefer, for consistency, the data from \citet{pehl} for all \mgi\ lines in our list, the results based on the intercombination line should be treated with caution.

To compute the broadening caused by elastic collisions with \hi\ atoms, we used the $\alpha$ and $\sigma$ coefficients from the quantum-mechanical calculations by \citet{barklem2000}, where available. For several Mg I lines (5711.088, 10811.076,  10965.45, 15024.992, 15765.842, 15886.183 \AA), we adopted $\alpha$ and $\sigma$ coefficients kindly provided by P. Barklem. The values of the assumed damping parameters are given in Table \ref{tab:at_data}.

Some \mgi\ lines are multi-component features. In particular, the $6$ lines around 10811 \AA\ originate in the transitions between the fine structure components of the \Mg{3d}{3}{D}{}{} and \Mg{5f}{3}{F}{\circ}{} levels. Fine structure splitting is also seen at $10965$, $12083$ \AA, and $15765$ \AA. One of the lines in the optical, the $8806$ \AA, is known to be affected by isotopic shift \citep{meissner1938}. The line is therefore represented by three isotopic components with the wavelengths $8806.757$ \AA~($^{24}$Mg), $8806.736$ \AA~($^{25}$Mg), and $8804.703$ \AA~($^{26}$Mg). All fine structure components were included in the spectrum synthesis using the wavelengths and oscillator strengths from \citet{pehl}, where available. Alternatively, we used the data from \citet{butler}.
%
%----------------------------------------------------------------------------------------------------
%
\subsection{Full 3D NLTE calculations}\label{sec:nlte3d}
Before proceeding with the abundance analysis, we would like to point out one important aspect of our study. By using mean 3D models we account for hydrodynamic cooling associated with surface convective overshooting in the simulations. However, information about horizontal inhomogeneities is inevitably lost with the averaging, and therefore cannot be accounted for directly. To quantify the effect, we turned to detailed full 3D NLTE radiative transfer calculations.
\begin{figure}[htp] 
\begin{minipage}[h]{1.0\linewidth}
	\centering
	\includegraphics[width=0.7\textwidth, angle=-90]{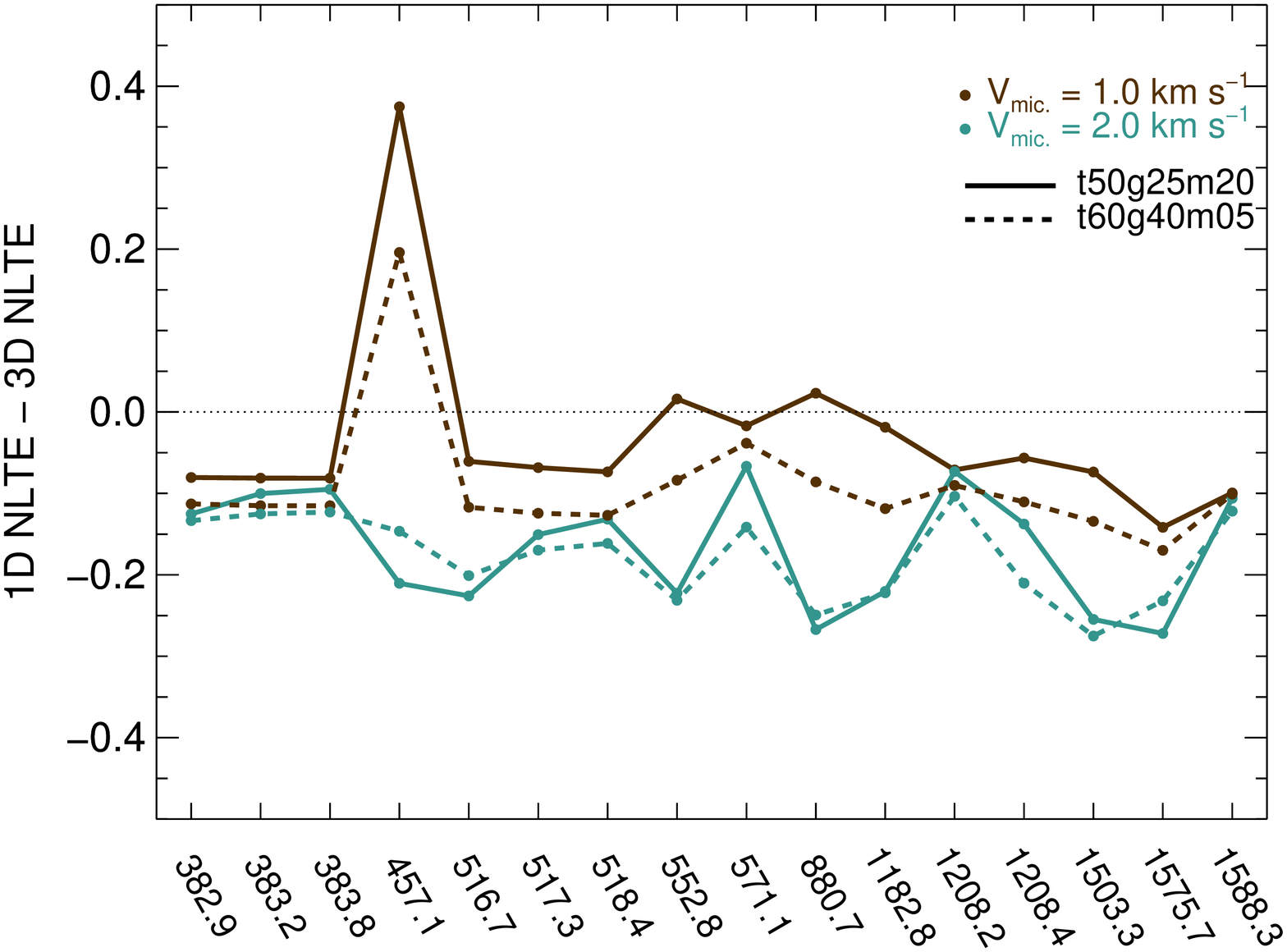}
	\vspace{0.5cm}
\end{minipage}
\begin{minipage}[h]{1.0\linewidth}
	\centering
	\includegraphics[width=0.7\textwidth, angle=-90]{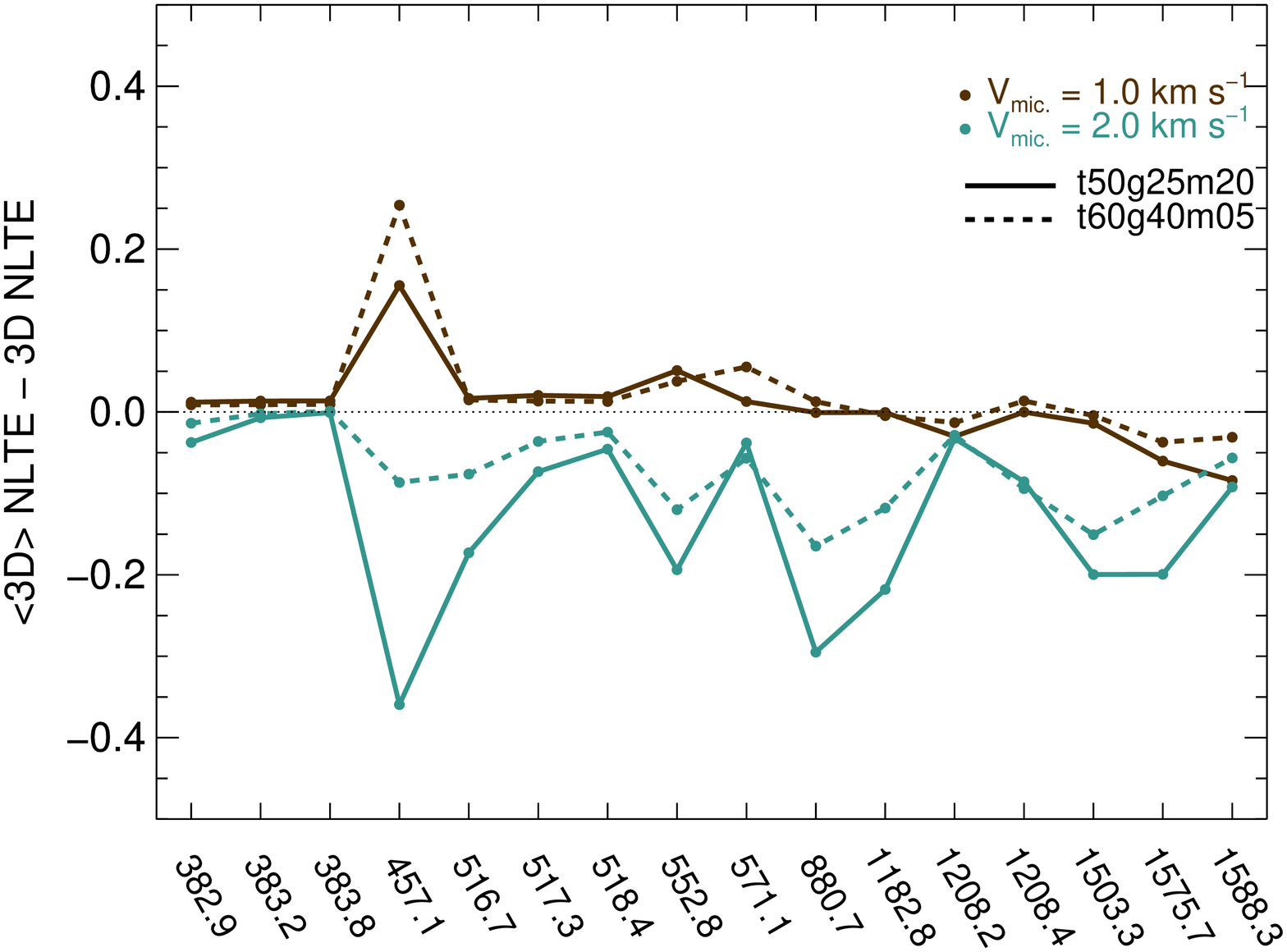}
	\vspace{0.5cm}
	\end{minipage}
\caption{Differences between Mg abundances inferred from equivalent widths using full 3D NLTE and 1D NLTE radiative transfer (top panel), and using full 3D NLTE and $\md$ NLTE radiative transfer (bottom panel). The calculations were done for two model atmospheres with the following parameters: a metal-poor giant $\teff  = 5000$ K, $\log g = 2.5$, and $\feh = -2$ (solid); and a moderately metal-poor dwarf $\teff  = 6000$ K, $\log g = 4.0$, and $\feh = -0.5$ (dashed). See Section \ref{sec:nlte3d}.}
\label{multi}
\end{figure}

The calculations were performed using \textsc{multi3d} \citep{leenaarts2009}, albeit with some updates to the equation-of-state and opacity package \citep{amarsi2016b}. The calculations were performed for two of the stars in the \textsc{stagger}-grid of 3D hydrodynamical model atmospheres that have stellar parameters representative of our large stellar sample in paper 2: a metal-poor giant $\teff  = 5000$ K, $\log g = 2.5$, and $\feh = -2$ and a metal-rich dwarf $\teff  = 6000$ K, $\log g = 4.0$, and $\feh = -0.5$.  Radiative transfer calculations were performed across 11 snapshots of each sequence. To reduce the computational cost, the horizontal resolution of each snapshot was reduced by selecting every third grid point, which reduces the number of grid points in a given layer from $240\times240$~to $80\times80$. Furthermore, the optically thick layers were trimmed such that the vertical logarithmic optical depth at $5000$ \AA\ satisfied $\log\tau_{5000}\lesssim3$, and interpolated onto a new depth scale such that there were $100$~grid points in a given column. The mean intensity was calculated by numerical integration over 26 rays across the unit sphere.

To minimise the computational cost of the 3D NLTE radiative transfer calculations, the complexity of the Mg model atom was reduced. All energy levels above the \mgii\ ground state and all lines above $30\mu$m~were discarded. Sharp resonances in the radiative bound-free cross-sections were smoothed, after which the cross-sections were interpolated onto new wavelength grids that were up to ten times more sparse than the original grids provided by the Opacity project. These reductions have a negligible impact on the results presented in this section. 
\begin{figure*}[ht!]
\centering
\hbox{
\includegraphics[width=0.32\textwidth, angle=0]{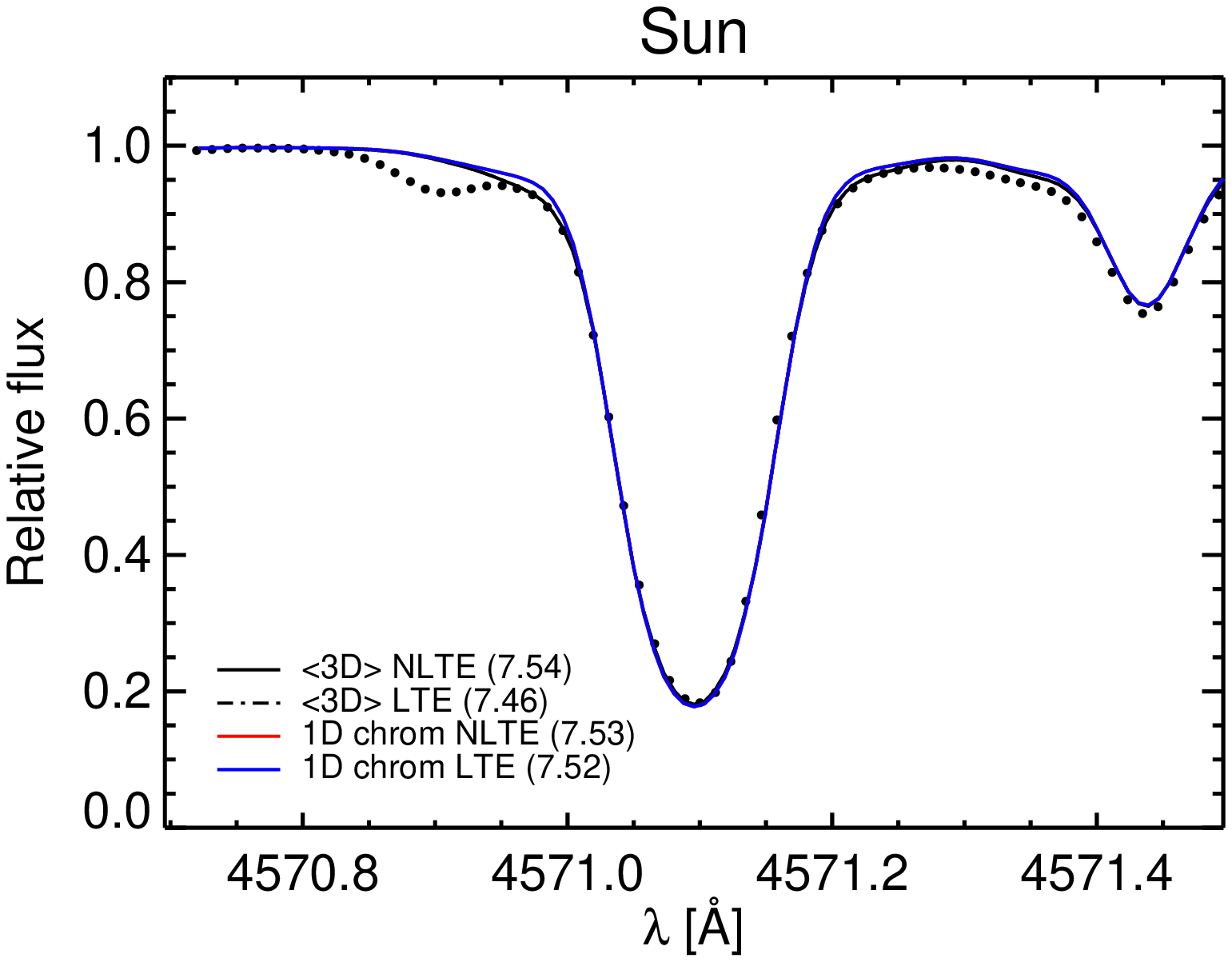}
\includegraphics[width=0.32\textwidth, angle=0]{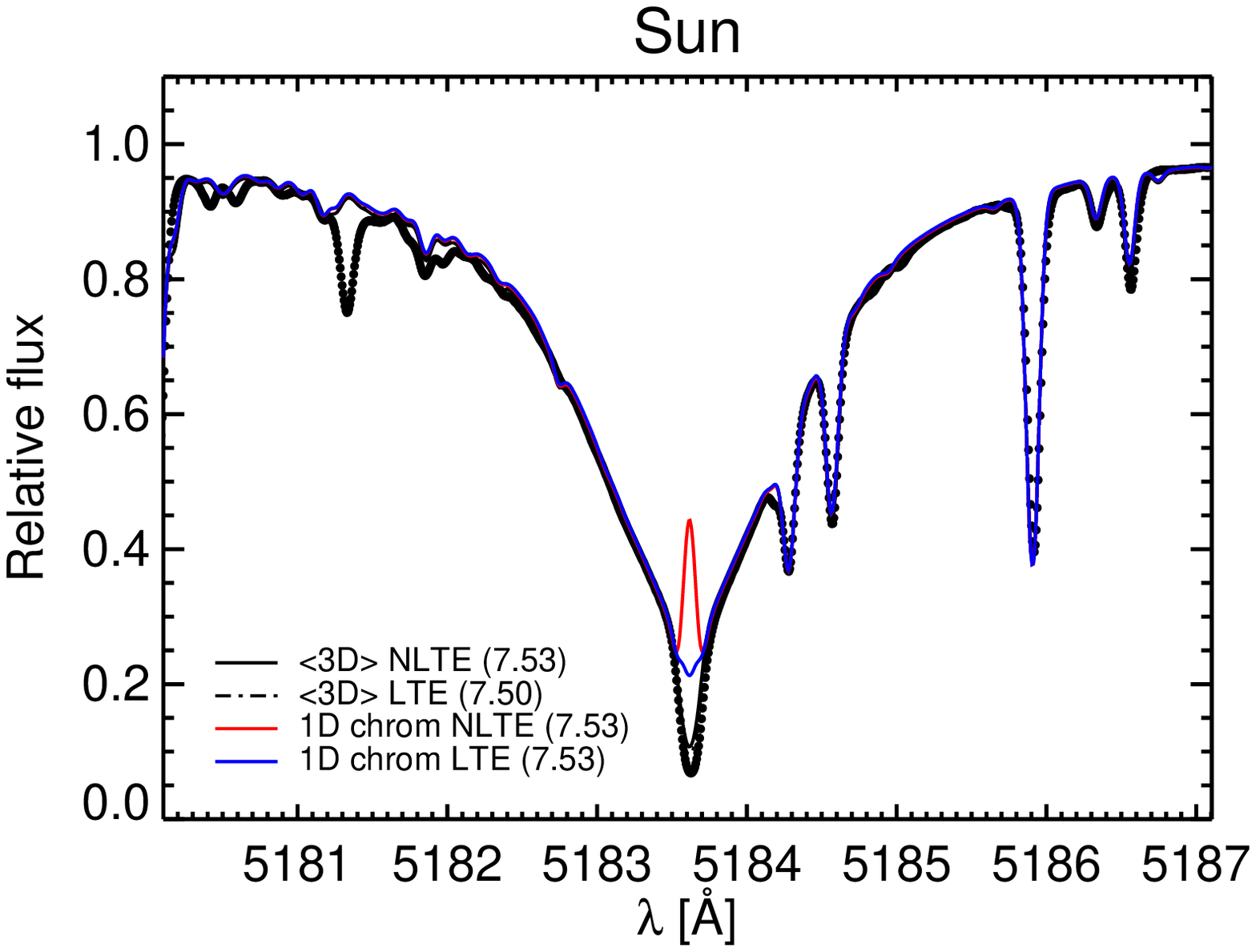}
\includegraphics[width=0.32\textwidth, angle=0]{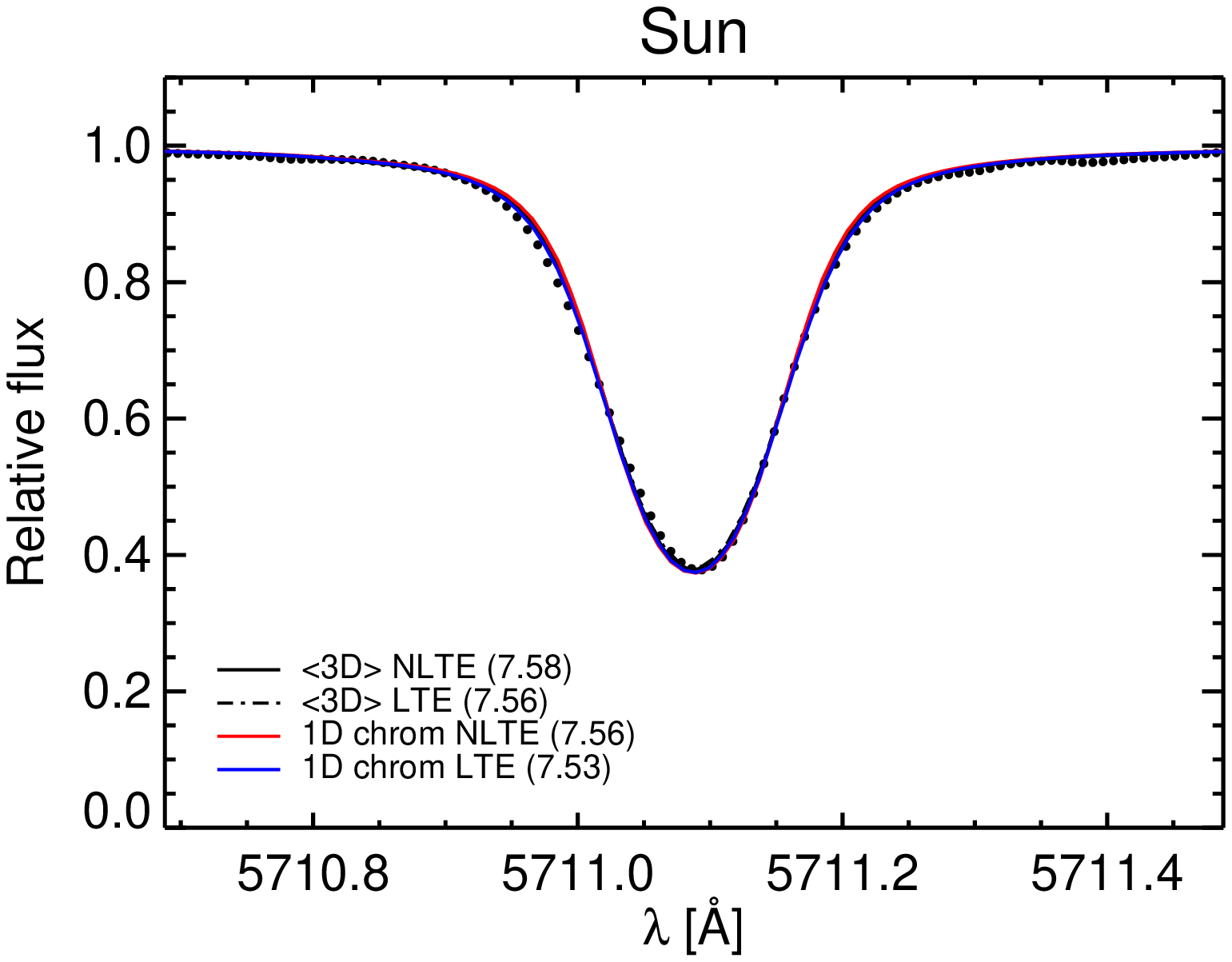}
}
\hbox{
\includegraphics[width=0.32\textwidth, angle=0]{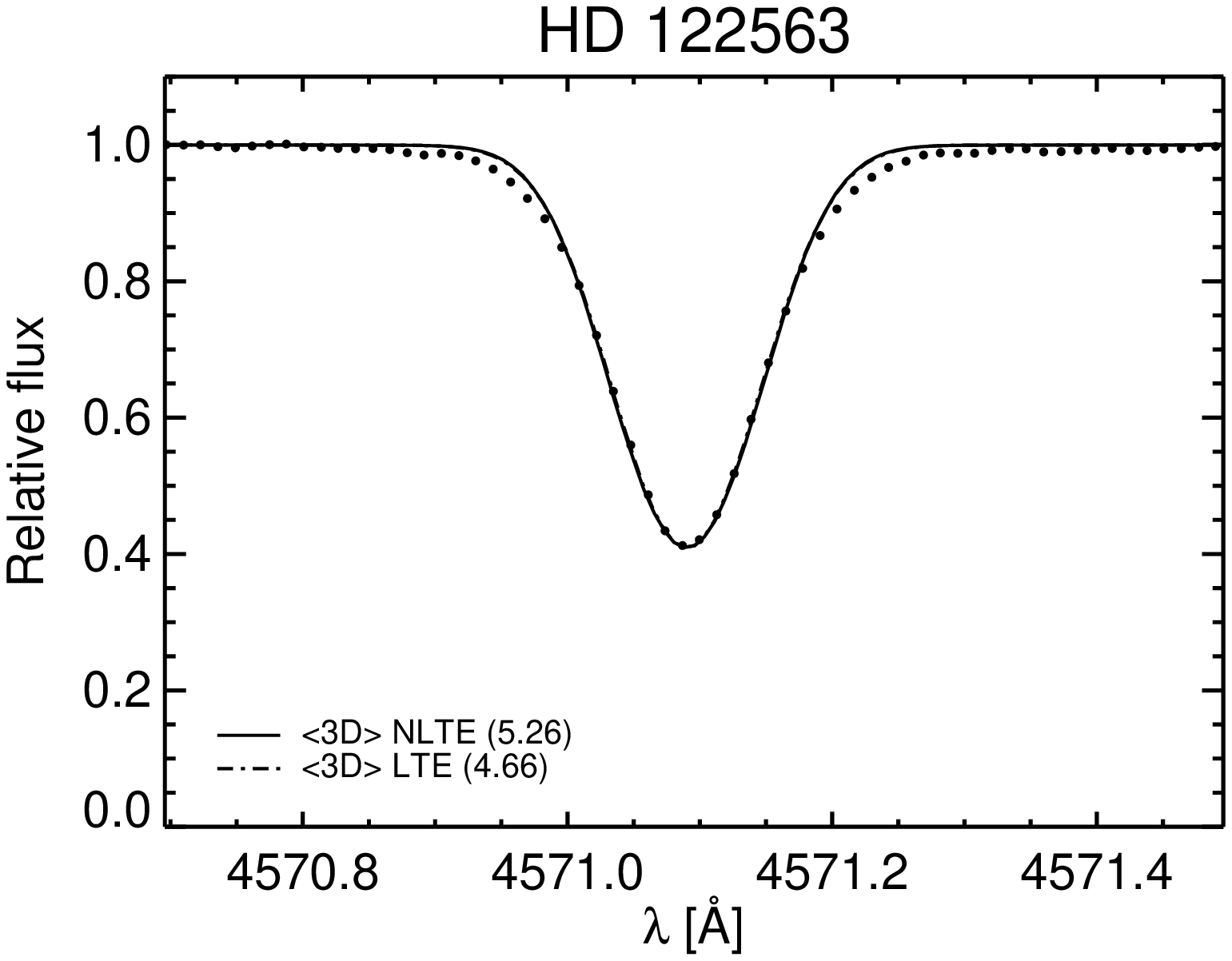}
\includegraphics[width=0.32\textwidth, angle=0]{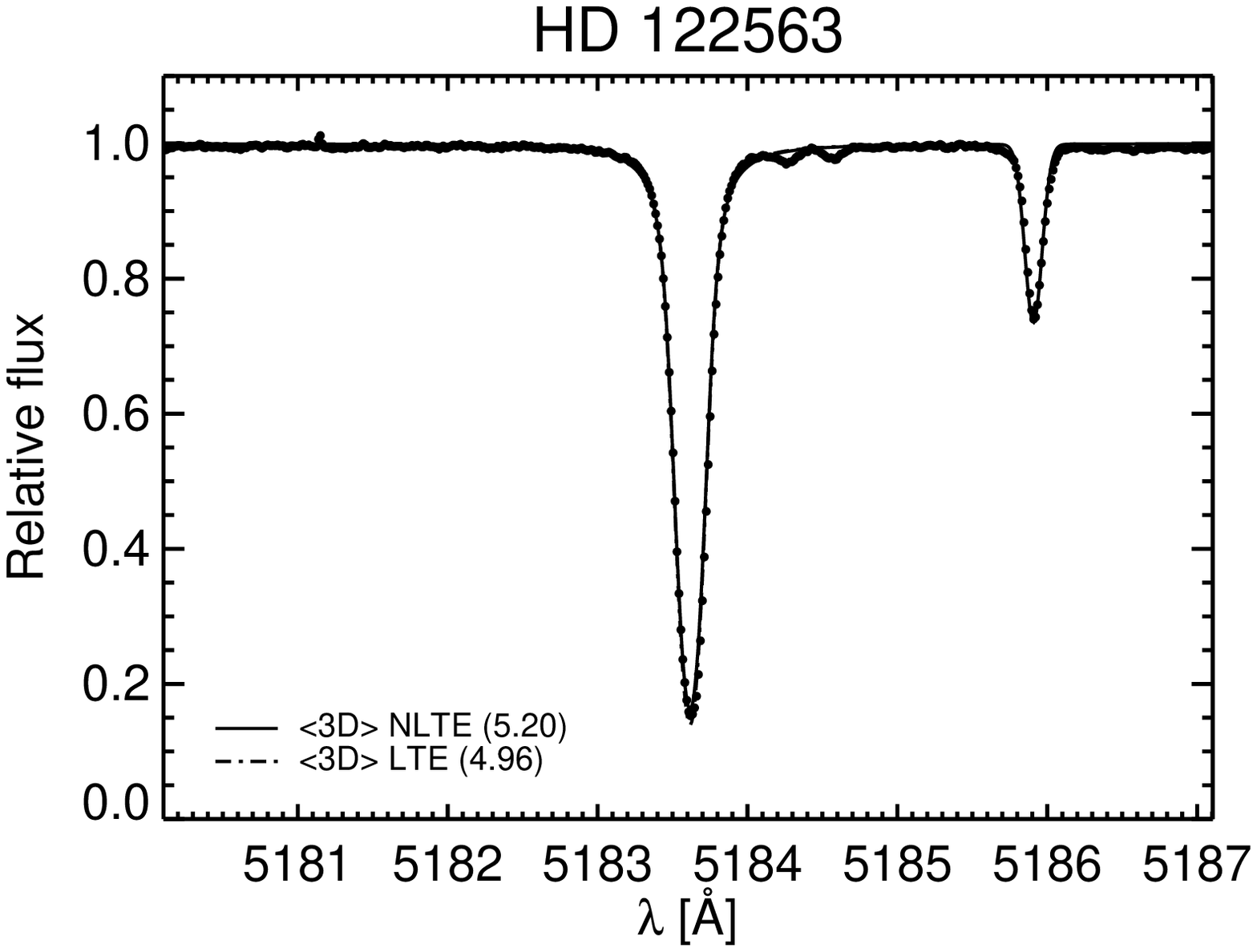}
\includegraphics[width=0.32\textwidth, angle=0]{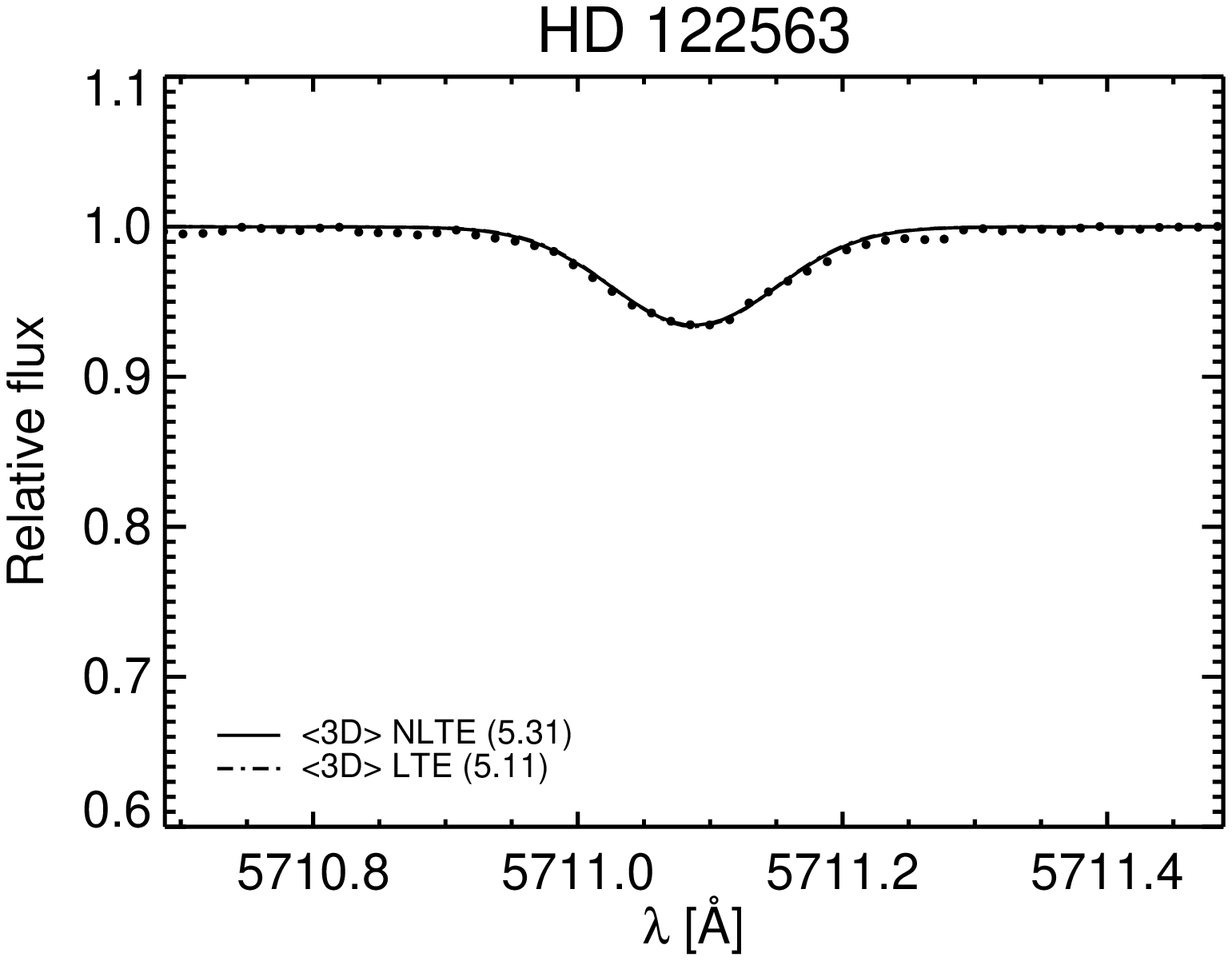}
}
\caption{Selected \mgi\ line profiles in the spectrum of the Sun (top panel) and the metal-poor red giant HD 122563 (bottom panel). The observed data are shown with filled black circles. The Mg abundance derived using $\md$ LTE, $\md$ NLTE, or the MACKKL model (for the Sun only) is also indicated in brackets.}
\label{fig:profiles}
\end{figure*}

In Figure \ref{multi} we compare the line strengths obtained from these 3D model atmospheres with those obtained from the 1D and $\md$~model atmospheres, by plotting the abundance errors obtained by matching the equivalent width of the 1D/$\md$~model line to the 3D model line:
$\rm A (\rm Mg )_{\rm{1D NLTE}} - \rm A (\rm Mg)_{\rm{3D NLTE}}$~and
$\rm A (\rm Mg )_{\rm{\md NLTE}} - \rm A (\rm Mg)_{\rm{3D NLTE}}$.
The J-band IR line at 12083 \AA\ shows the smallest differences between mean 3D NLTE and full 3D NLTE calculations. For the 5172 \AA, 5183 \AA, and 5711 \AA, the abundance errors are small, and typically around $\pm0.1\,\mathrm{dex}$. We note however that the $4571$ \AA, $5528$  \AA\ and $8806$ \AA\ show more significant abundance errors that are sensitive to the approximation of micro-scale velocity fields.
%
%
%------------------------------------------ Results --------------------------------
%
%
\section{Results} \label{sec:results}
In what follows, we describe our results of LTE and NLTE line formation, profile fits and abundance determinations of Mg in the program stars. We compare LTE with NLTE calculations using 1D hydrostatic and $\md$ model atmospheres, and discuss the advantages and shortcomings of each approach in different wavelength regimes. We also comment on the solar Mg abundance, and compare the results for all program stars with recent estimates. 

Throughout the text, we use the standard astronomical notation to express the abundance of the element:
\begin{equation}
\label{eq1}
\rm A (\rm Mg) = \log \dfrac{N_{\rm Mg}}{N_{\rm H}} + 12 
\end{equation}
%
%
%----------------------------------------------------------------------------------------------------
%
\subsection{NLTE abundance corrections} \label{sec:abundancecorrections}
NLTE abundance correction is a useful measure of the NLTE effect on a spectral line\footnote{We note, however, that this measure ignores the NLTE effect in the profile of the spectral line, which may not be alike in LTE and in NLTE, even when their equivalent widths are very similar.} and is commonly used in the literature. The parameter is defined as
\begin{equation}
\label{eq2}
\Delta \rm A (\rm Mg) = \rm A (\rm Mg )_{\rm{NLTE}} - \rm A (\rm Mg)_{\rm{LTE}}
\end{equation} 
and is derived through matching the equivalent width of the LTE model line to the NLTE line using the same model atmosphere.

The NLTE effects on \mgi\ lines as a function of stellar parameters have been discussed before in the literature \citep{merle2011, mashonkina2013a, bergemann2015, osorio2016}. Here, we focus on the NLTE abundance corrections in our program stars, and explore the sensitivity of NLTE corrections to model atmosphere structure. 

Table \ref{tab:nltecor} compares our NLTE corrections for selected models with the NLTE corrections determined by \citet{osorio2016}. The agreement is good. The main difference between their model atom and ours is the recipe for the rates of transitions caused by collisions with electrons. We assumed the collision induced rates from \citet{zhao1998} computed using the \citet{vanreg1962} and \citet{allen1973} formulae (and with the collision strength set equal to unity), while Osorio et al. adopted the  $\Upsilon_{ij}/g_{ij}$ approach separating the transitions involving electron exchange from those without electron exchange. Despite these differences, their and our results appear to be very consistent, although their NLTE corrections for the 4571 and 5528 \AA~lines are slightly larger. This could be, in effect, related to the fact that we are using different statistical equilibrium codes. \citet{osorio2016} employed MULTI 2.3 \citep{carlsson1986,carlsson1992}. This code was also used in \citet{bergemann2012b}, and was shown to give slightly stronger NLTE effects than DETAIL, which was traced back to the differences in background opacity in \citet{bergemann2012b}. DETAIL treats line opacity consistently at all frequencies of the bound-bound and bound-free transitions, while MULTI includes line opacity in the calculation of bound-free rates, but not in the calculation of the bound-bound radiative rates. For the other \mgi\ transitions, their NLTE corrections agree with ours to better than $0.03\,\mathrm{dex}$ across the full range of stellar parameters that we consider as acceptable given the differences in the model atom, atmospheres, and statistical equilibrium codes.
%
% -------------------------------- FINAL PARAM ---------------------------------
%
\begin{table}
\centering
\caption{Comparison of the NLTE abundance corrections from this work with those of \citet{osorio2016}. The data were computed using 1D hydrostatic model atmospheres, MAFAGS-OS in this work and MARCS in \citet{osorio2016}.}
\label{tab:nltecor}
\renewcommand{\tabcolsep}{2pt}
\begin{tabular}{ccc c cc cc cc}
\hline\noalign{\smallskip}
 $T_{\rm eff}$ & $\log g$ & [Fe$/$H] & $\Vmic$ &      \multicolumn{2}{c}{4571 \AA}   &      \multicolumn{2}{c}{5528 \AA}     &   \multicolumn{2}{c}{5711 \AA} \\
  K & dex & dex & kms$^{-1}$ &  &  &   &  &  &  \\
\cline{5-10}
 &          &        &        &  Ber  &  Oso   &  Ber  &  Oso   &  Ber  &  Oso  \\
\hline\noalign{\smallskip}
  6000 & 4.0 & $-$1.0 &  1  &  0.04  &  0.03  & $-$0.03   & $-$0.03  &  ~~0.03 & 0.03  \\
  6000 & 4.0 & $-$2.0 &  1  &  0.07  &  0.05  &  ~~0.04   & $-$0.01  &  ~~0.05 & 0.03  \\
  6000 & 4.0 & $-$3.0 &  1  &  0.12  &  0.12  &  ~~0.09   &  ~~0.10  &  ~~0.09 & 0.10  \\
  5000 & 2.0 & $-$1.0 &  2  &  0.08  &  0.20  & $-$0.06   & $-$0.07  & $-$0.03 & 0.02  \\
  5000 & 2.0 & $-$2.0 &  2  &  0.13  &  0.13  & $-$0.02   & $-$0.13  &  ~~0.04 & 0.02  \\
  5000 & 2.0 & $-$3.0 &  2  &  0.24  &  0.23  &  ~~0.13   &  ~~0.06  &  ~~0.09 & 0.07  \\
\noalign{\smallskip}\hline
\end{tabular}
\end{table}

We have also explored the sensitivity of the abundance corrections to the numerical uncertainties caused by the interpolation in the model atmosphere grid. As shown in Section \ref{sec:atmos}, interpolation in very low-metallicity models may cause non-negligible errors in the model structure. We can test the corresponding errors in $\Delta \rm A (\rm Mg)$ by assuming that the uncertainty of the $T(\tau)$ relationship can be approximated by a systematic shift in $\teff$. This is, of course, not a fully correct assumption, because also the shape of the $T(\tau)$ matters. However, this gives a rough idea of the sensitivity of the NLTE abundance corrections to the thermodynamic parameters of stellar atmospheres. Our test calculations, assuming a shift of $\teff$ of $100\,\mathrm{K}$ around the mean value, show that in NLTE the mean uncertainty is of the order $\pm 0.002\,\mathrm{dex}$ for 1D models and about $0.03\,\mathrm{dex}$ for $\md$ models (Table \ref{tab:nltesen}). The error is so small, because the NLTE abundance corrections describe the relative difference between LTE and NLTE spectral line profiles for a given type of  model atmospheres, and these quantities remain nearly invariant with small linear changes in the model structure. The sensitivity is larger in $\md$, because of the differences in the thermodynamic structure of 1D hydrostatic and $\md$ models.
%
% -------------------------------- FINAL PARAM ---------------------------------
%
\begin{table}
\centering
\caption{The sensitivity of NLTE abundance corrections defined in Equation 2 in 1D (columns 5,6,7) and in $\md$ (columns 8,9,10) to the change in $\teff$.}
\label{tab:nltesen}
\renewcommand{\tabcolsep}{1pt}
\begin{tabular}{cccc ccc | ccc }
\hline\noalign{\smallskip}
 $T_{\rm eff}$ & $\log g$ & [Fe$/$H] & $\Vmic$ &  \multicolumn{6}{c} {$+$ 100 K}  \\
              &          &        &  &   \multicolumn{3}{c} {1D} & \multicolumn{3}{c} {$\md$} \\
\cline{5-10}
 \multicolumn{3}{c} {wavelength  (\AA) $\rightarrow$}  & & ~~5183  &  ~~5528  &  ~~5711  &  ~~5183 &  ~~5528 & ~~5711 \\
  4500 & 1.5 & $-$2.0 & 2 &   0.000 & $-$0.002 & $-$0.001  &  $-$0.044 &  $-$0.018 &   $-$0.027\\
  4500 & 1.5 & $-$1.0 & 2 &   0.001 & $-$0.008 &  ~~0.006  &  $-$0.049 &  $-$0.032 &   $-$0.035\\
  4500 & 1.5 & $-$0.5 & 2 &   0.000 & $-$0.001 & $-$0.005  &  $-$0.045 &  $-$0.037 &   $-$0.031\\
\cline{5-10}
        &      &          &          &   \multicolumn{6}{c} {$-$ 100 K}  \\
        &      &          &          &   \multicolumn{3}{c} {1D} & \multicolumn{3}{c} {$\md$} \\
\cline{5-10}
  4500 & 1.5 & $-$2.0 & 2 &   0.000 &  ~~0.003 &  ~~0.000 & ~~0.002 & ~~0.021 & ~~0.017   \\
  4500 & 1.5 & $-$1.0 & 2 &$-$0.001 &  ~~0.008 & $-$0.005 & ~~0.020 & ~~0.001 & ~~0.003   \\
  4500 & 1.5 & $-$0.5 & 2 &   0.000 &  ~~0.001 &  ~~0.005 & ~~0.022 & ~~0.010 & ~~0.006   \\
\noalign{\smallskip}\hline
\end{tabular}
\end{table}

\subsection{NLTE effects in \mgi\ lines}\label{sec:lineform}
Figure \ref{fig:profiles} shows selected \mgi\ lines in the spectrum of the Sun (top panel) and in the UVES spectrum of HD 122563 (bottom panel) overlaid with the best-fit theoretical models. The derived Mg abundance is also shown in the inset. For the Sun, we show the $\md$ LTE (dashed-dotted curve) and $\md$ NLTE (solid curve) profiles, as well as the line profiles computed using the semi-empirical MACKKL solar model with a chromosphere. The fits obtained with 1D hydrostatic models are very similar and are not plotted. As seen from Figure \ref{fig:profiles}, both $\md$ LTE and $\md$ NLTE model successfully describe the observed \mgi\ line profiles in the solar spectrum,  even though the abundances measured through the fits are notably different. 

\begin{figure}[htp] 
\begin{minipage}[h]{1.0\linewidth}
	\centering
	\includegraphics[width=1\textwidth, angle=0]{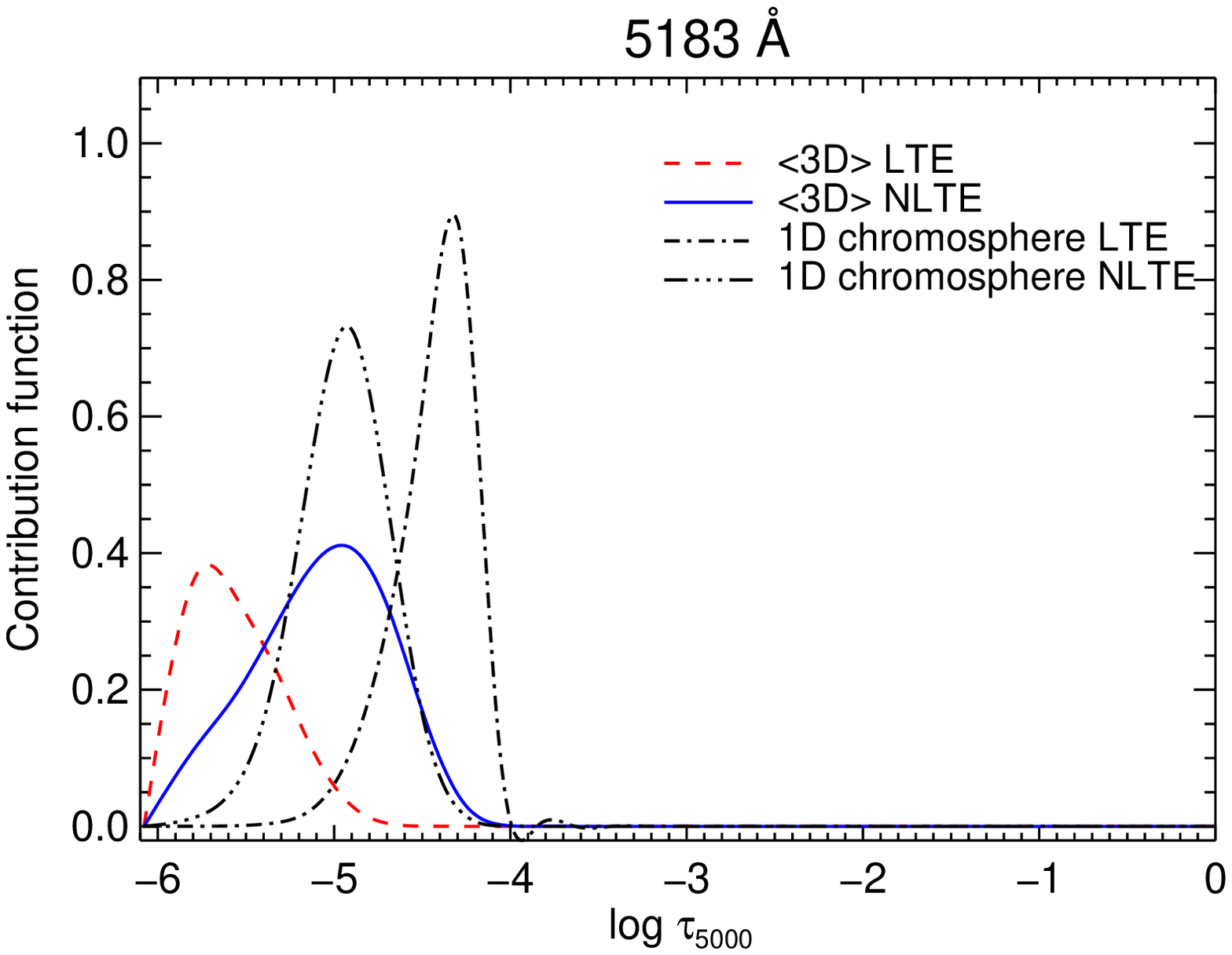}
	\vspace{0.1cm}
\end{minipage}
\begin{minipage}[h]{1.0\linewidth}
	\centering
	\includegraphics[width=1\textwidth, angle=0]{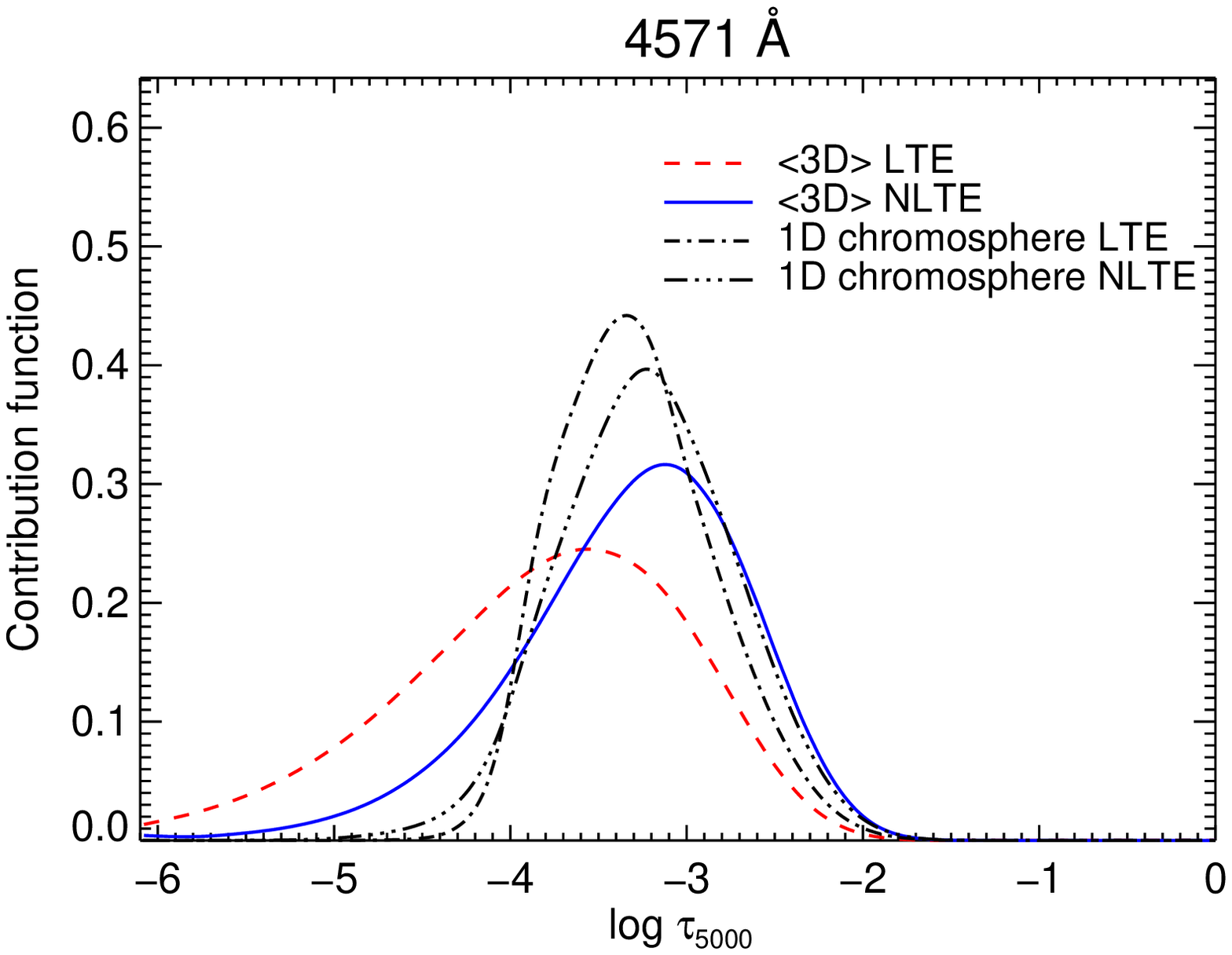}
	\vspace{0.1cm}
	\end{minipage}
\begin{minipage}[h]{1.0\linewidth}
	\centering
	\includegraphics[width=1\textwidth, angle=0]{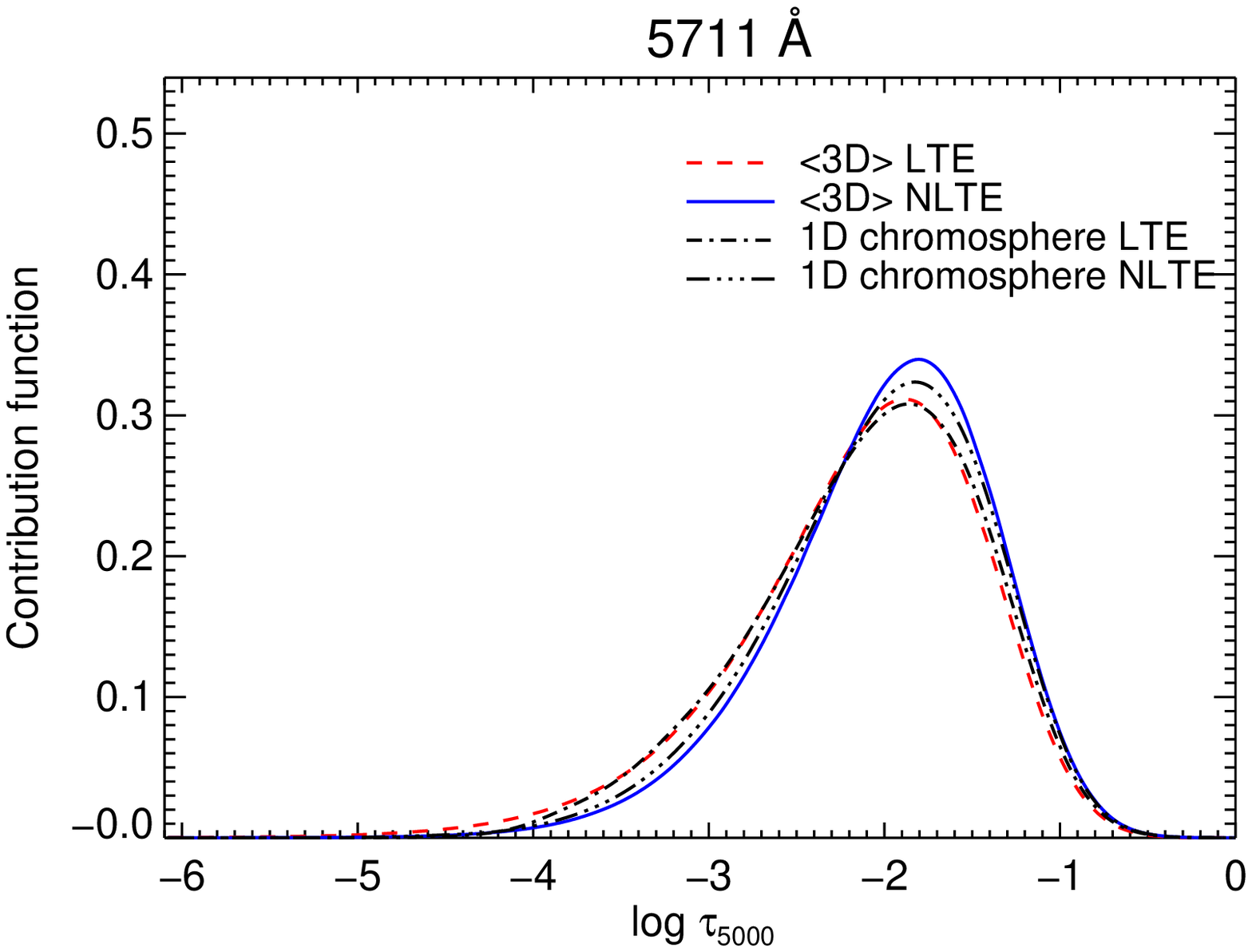}
	\vspace{0.1cm}
	\end{minipage}
\caption{LTE and NLTE contribution function to the flux depression for the selected \mgi\ lines in the $\md$ and MACKKL solar atmosphere models in arbitrary units.}
\label{suncf}
\end{figure}

The cores of the strong lines, such as the 5183 \AA, are slightly brighter in LTE, and are better fitted with NLTE models, at least when the hydrostatic or $\md$ models are used. We should stress, however, that the cores of these lines are formed in the outer layers, which are affected by the solar chromosphere. To illustrate this effect, Figure \ref{suncf} shows the line center contribution function (CF) for the 5183 \AA\ line, and, for comparison, the CFs for the weaker features at 4571 \AA\ and 5711 \AA\, computed using the MACKKL and the $\md$ solar model atmosphere. The contribution function is defined as in \citet[][eq. 15]{albrow1996}: 
\begin{equation}
CF_{\tau, \nu}  = \dfrac{(\ln10)\tau_0}{\kappa_0}\kappa_{\rm l, \nu} \int_{0}^1 \left(I_{\rm c} - S_{\rm l}\right) \exp^{-\tau/\mu} d\mu
\end{equation}
where $\kappa_{\rm l,\nu}$ is the line opacity at a given frequency, $S_{\rm l}$ the line source function, $I_{\rm c}$ the intensity in the continuum, $\mu = \cos \theta$ the angle between the ray and the direction to the observer, and the subscript $0$ in $\kappa_0$ and $\tau_0$ refers to the reference wavelength at $5000$ \AA. As seen in Figure \ref{suncf}, the core of the \mgi\ line at 5183 \AA\ forms at $\log\tau_{5000} \sim -5$ in 1D NLTE and at $\log\tau_{5000} \sim -6$ in LTE (Figure \ref{suncf}, top panel). This is the region where temperature rises to $\sim 6000\,\mathrm{K}$in the chromospheric solar model. Since in LTE the line source function is coupled to the local atmospheric structure, the LTE calculations with the MACKKL model lead to an unphysical emission in the line core (Figure \ref{fig:profiles}). Taking NLTE into account does not help much to resolve the discrepancy and the line core remains too bright compared to the observations. The intercombination line at 4571 \AA\  forms deeper, but even in NLTE it samples the optical depth range, where the transition to the chromosphere occurs in the MACKKL model (at $\log\tau_{5000} \sim -3.2$, see \citealt[][Figure 1]{bergemann2011}). The core of the high-excitation 5711 \AA\ line forms at $\log\tau_{5000} \sim -2$ (Figure \ref{suncf}, bottom panel). This \mgi\ line has a purely photospheric origin and can be safely used in the abundance determinations using model atmospheres, which do not include chromospheres.

To understand the effect of NLTE on Mg abundance determinations using 1D and $\md$ model atmospheres, it is useful to explore the properties of the line formation. This is basically an interplay between the effect of NLTE on the line source function and on the line opacity that can be conveniently described using the concept of level departure coefficients, defined as
\begin{equation}
b_i = n_i^{\rm NLTE}/n_i^{\rm LTE}
\end{equation}
where $n_i^{\rm NLTE}$ and $n^{\rm LTE}$ are the number densities of an atomic energy level $i$ computed using NLTE and LTE, respectively. 

The line opacity $\kappa_{\rm l}$ depends on the number density of the lower level of the transition, hence $\kappa_{\rm l}^{\rm NLTE} /\kappa_{\rm l}^{\rm LTE} \sim b_i$. The line source function $S_{\nu}$ has a thermal component, determined by the local atmospheric structure, and a non-local component representing the scattering of radiation. In NLTE, the ratio of the source function to the Planck function is proportional to the ratio of departure coefficients for the lower and upper levels of the transition, $S_{\nu} /B_{\nu} \sim b_j/b_i$. Hence, both the source function and the opacity can be very sensitive to different radiative and collisional processes populating or depopulating the atomic energy levels. 

\begin{figure}
\centering
\includegraphics[width=0.5\textwidth, angle=0]{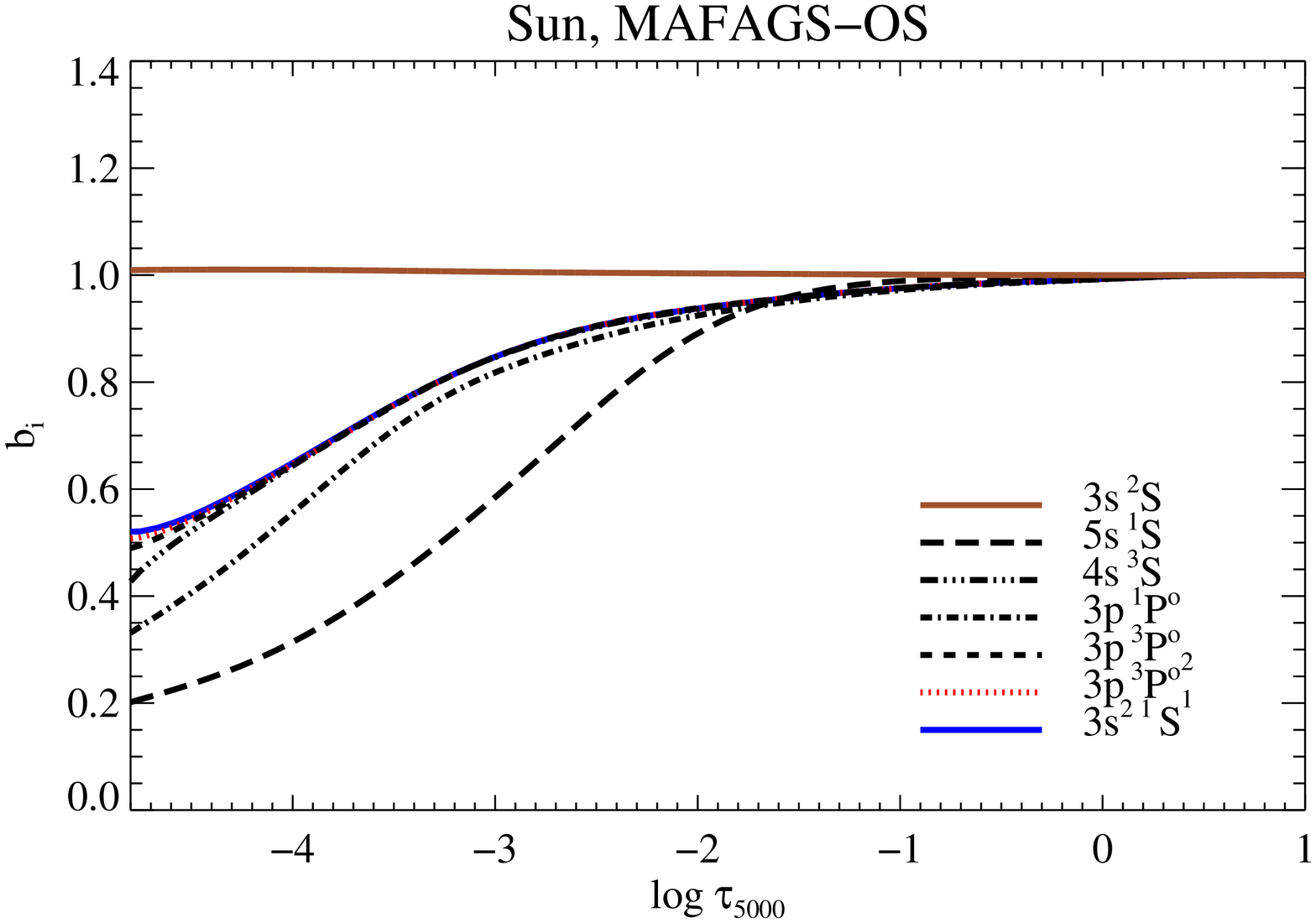}
\includegraphics[width=0.5\textwidth, angle=0]{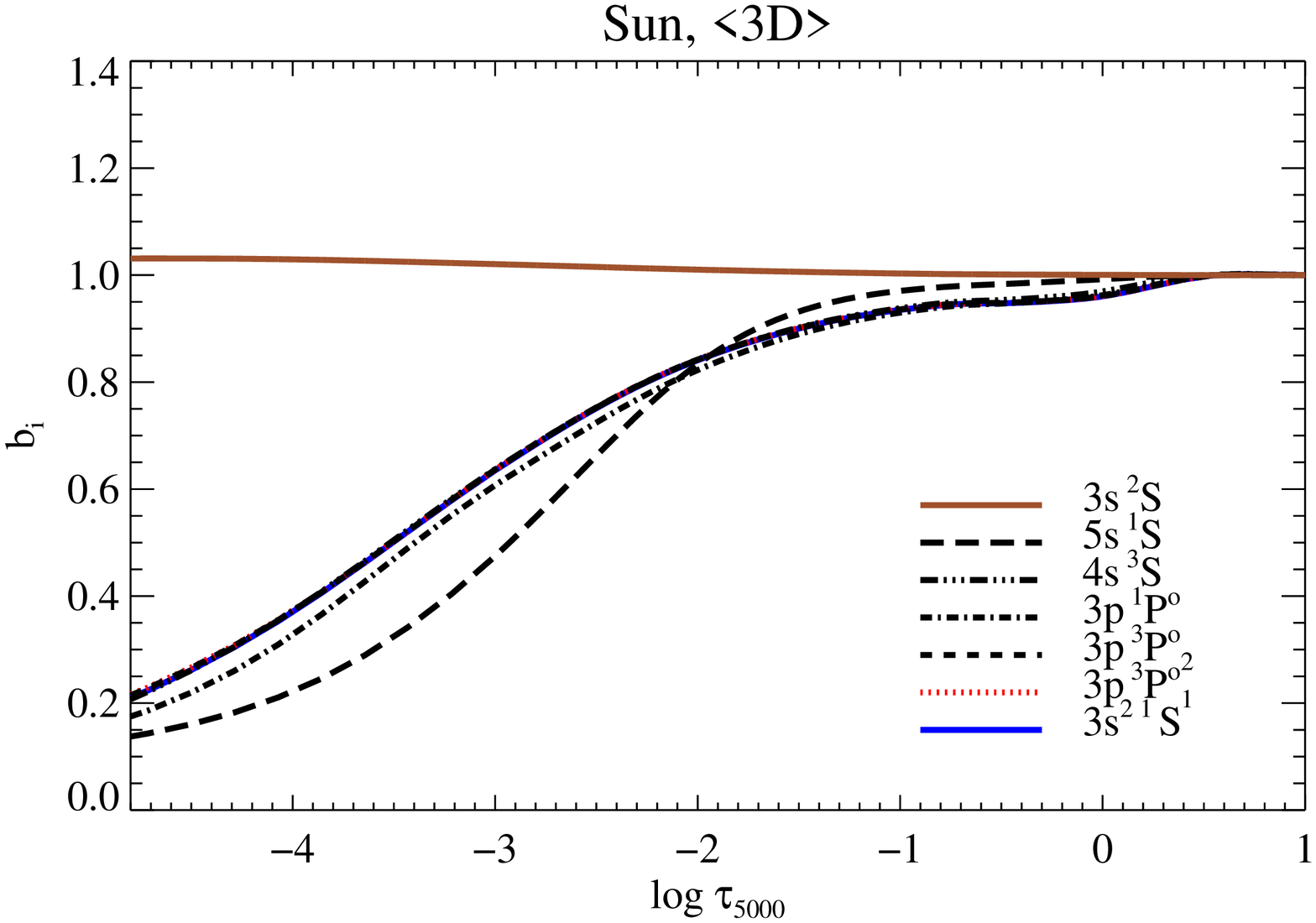}
\caption{Departure coefficients $b_{i}$ of the selected \mgi\ levels for the MAFAGS-OS (top panel) and $\md$ (bottom panel) solar model atmospheres.}
\label{fig:dep}
\end{figure}

Figure \ref{fig:dep} shows the departure coefficients for the selected \mgi\ levels, which are involved in the transitions of interest: \Mg{3s^2}{1}{S}{}{0} - \Mg{3p}{3}{P}{\circ}{1} (4571 \AA), \Mg{3p}{3}{P}{\circ}{2} - \Mg{4s}{3}{S}{}{1} (5183 \AA), and \Mg{3p}{1}{P}{\circ}{1} - \Mg{5s}{1}{S}{}{0} (5711 \AA). The \mgi\ levels are underpopulated compared to LTE, $b_i < 1$, at all depths above $\log\tau_{5000} \sim -1$, even more so in the $\md$ solar model atmosphere. This is almost entirely caused by the overionization, a typical NLTE phenomenon that is driven by the disbalance between the mean intensity of the radiation field and the Planck function. This disbalance is slightly stronger in the $\md$ model atmosphere, because of its slightly steeper T($\tau$) relationship. As a consequence, the NLTE line opacity in 1D hydrostatic and in the $\md$ models is reduced compared to LTE. It is also interesting that the levels, except for the \Mg{5s}{1}{S}{}{0} state (the level has a very high excitation energy, $6.52$ eV), are thermalised with respect to each other that implies thermal source functions.

Overionization has a large effect on the formation of the low-excitation lines even in the solar atmosphere. This is the reason, for example, why the $\md$ NLTE abundance derived from the 4571 \AA\ line is significantly (by $+0.08\,\mathrm{dex}$) higher than the $\md$ LTE estimate. Overionization also affects the levels of the \Mg{3p}{3}{P}{o}{} term (excitation energy $2.7$ eV), which are the lower levels of the optical triplet lines. The $\md$ NLTE abundance determined from the 5172 and 5183 \AA\ is thus $\sim 0.03\,\mathrm{dex}$ higher than the $\md$ LTE abundance. 

For the higher-excitation \mgi\ lines, such as  the 5711 \AA\ line, the NLTE effect on the line source function is also important. Analysis of the departure coefficients of the \Mg{3p}{1}{P}{\circ}{1} and \Mg{5s}{1}{S}{}{0} levels shows that at the depths, where the line wings form, $-1.5 \lesssim \log\tau_{5000} \lesssim 0$, the source function is superthermal ($b_j > b_i$, hence $S_{\nu} > B_{\nu}$, Figure \ref{fig:dep}). However, $S_{\nu}$ drops below $B_{\nu}$ in the outer regions of the solar atmosphere ($b_j < b_i$, $S_{\nu} < B_{\nu}$). As a result, the NLTE overionization, which leads to line brightening, is compensated by the line darkening caused by the sub-thermal source function. Hence, the LTE and NLTE line equivalent widths, as well as the abundances, are similar.

As it also follows from the departure plots (Figure \ref{fig:dep}), the NLTE effects on the \mgi\ levels in the MAFAGS-OS model are less pronounced compared to the calculations with the $\md$ model atmosphere. This implies that the NLTE abundance correction computed using the $\md$ model would be larger than the NLTE abundance correction derived using the hydrostatic model (see Equation \ref{eq2}), which was also pointed out by \citet{osorio2015}. On the other hand, 1D LTE and 1D NLTE Mg abundances for some lines (5528, 5711 \AA) are very similar to the $\md$ NLTE abundances. This is the consequence of the effect, which is conceptually similar to the "NLTE-masking" phenomenon \citep{rutten1982}. The thermal structure of the 1D hydrostatic models makes up for NLTE effects.

The NLTE effects in the \mgi\ lines in the other program stars are qualitatively similar, allthough the differences between the LTE and NLTE equivalent widths are typically larger that is caused by stronger overionization at lower metallicities and higher effective temperatures. NLTE effects grow with increasing $\teff$, and decreasing $\log g$ and $\feh$, similar to Fe \citep[e.g.][]{bergemann2012b}.

To illustrate the sensitivity of spectral lines to stellar parameters, Figure \ref{fig:profiles} (bottom panel) also shows the observed and theoretical line profiles for the metal-poor red giant HD 122563. The line profiles computed $\md$ LTE and $\md$ NLTE are nearly identical, despite large differences in the resulting abundance. The spectral lines do not change uniformly with stellar parameters. As the most striking example, the intercombination line at 4571 \AA\ has the EW of $110$ m\AA\ in the spectrum of the Sun and $87$ m\AA\ in the spectrum of HD 122563. In contrast, the EW of the 5711 \AA\ line, $\sim 104$ m\AA\, in the solar spectrum, drops to 10 m\AA\ at $\feh = -2.5$, also the EW of the optical triplet line at 5183 \AA\ drops by a factor of $10$ (from $1670$ to $270$ m\AA) from the solar parameters to that of HD 122563. 
\subsection{The solar Mg abundance}
\begin{figure*}
\centering
\hbox{
\includegraphics[width=0.5\textwidth, angle=0]{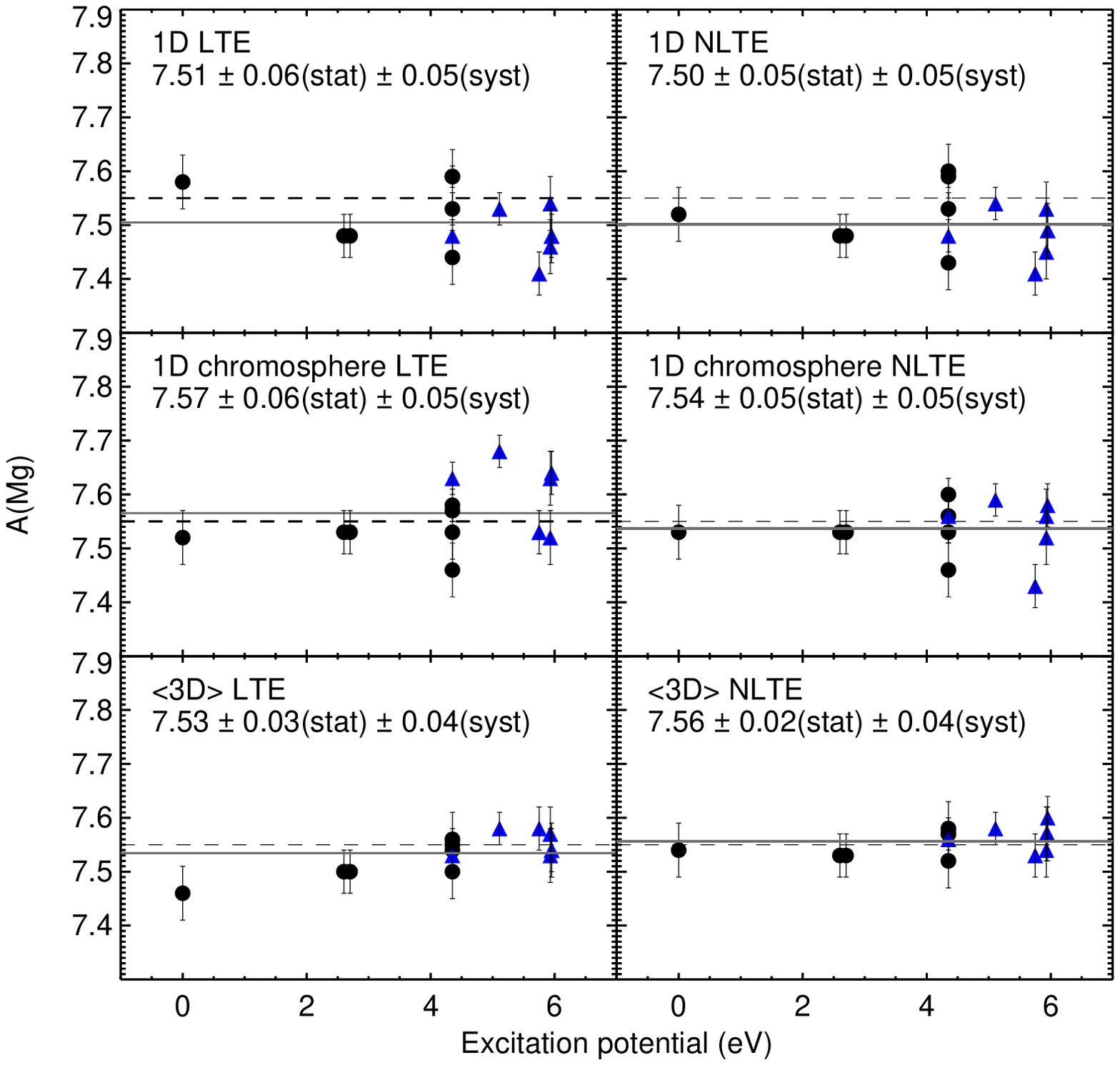}
\includegraphics[width=0.5\textwidth, angle=0]{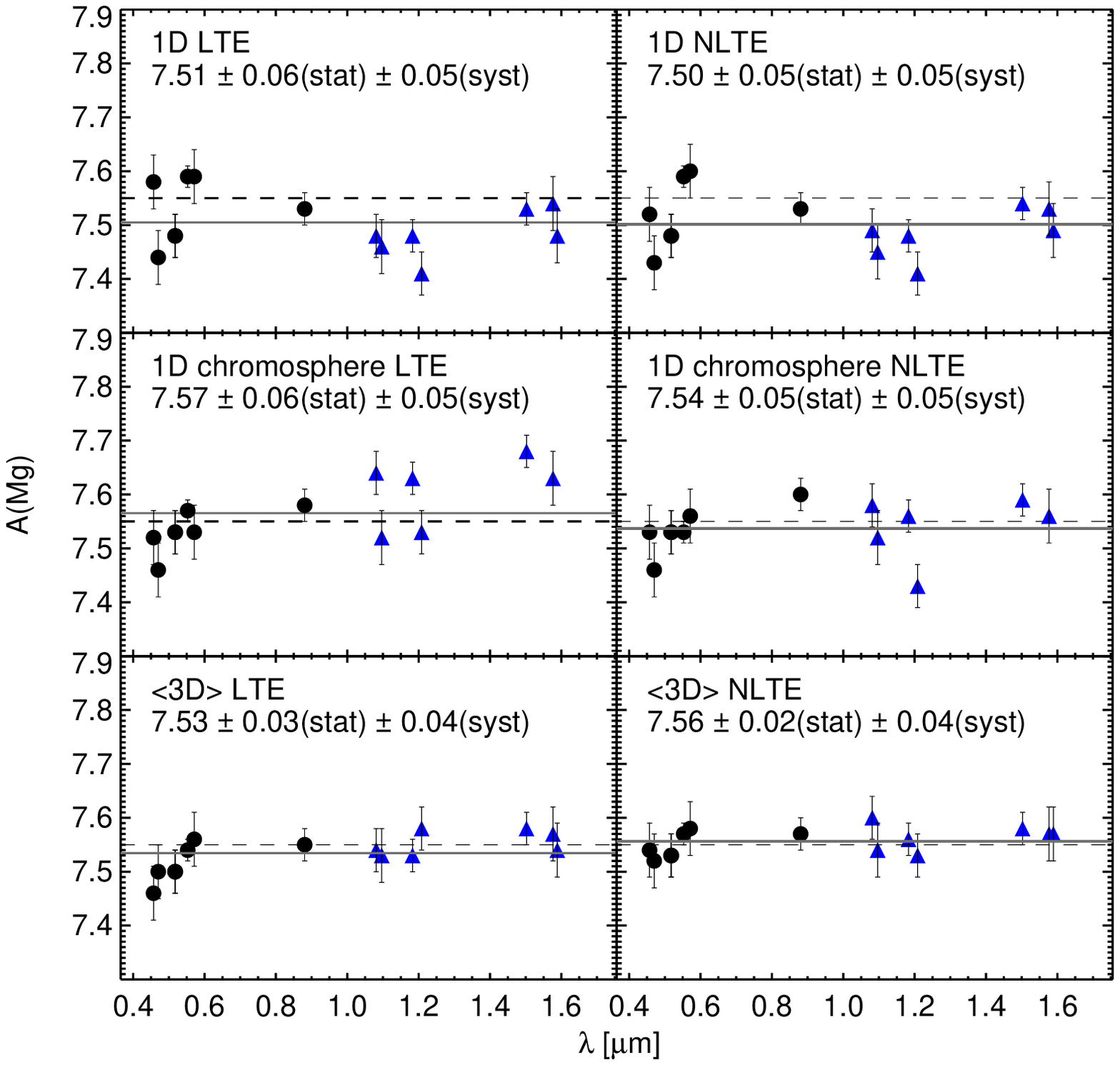}
}
\caption{Abundances determined using the \mgi\ lines in the solar spectrum as a function of the excitation potential of the lower level of a transition (left panel) and as a function of wavelength (right panel). Optical \mgi\ lines are shown with black circles, and blue triangles represent the infrared \mgi\ lines. The solid line is the mean of all abundance measurements, also indicated in the top left corner in each panel. The dashed line represents the meteoritic Mg abundance from \citet{lodders2009}. The uncertainties of the individual \mgi\ lines correspond to the errors of the oscillator strengths (Table \ref{tab:at_data}).}
\label{fig:solar}
\end{figure*}

To determine the solar Mg abundance, we used all lines from Table \ref{tab:at_data}, except for the IR line at 15748 \AA. This feature is 
extremely contaminated by blends, which contribute about $50 \%$ of the total equivalent width in the solar spectrum. However, this line is useful in the abundance diagnostic of metal-poor stars, and can be measured in the APOGEE spectra of HD 84937 and HD 122563. The results of the solar abundance calculations for individual \mgi\ lines are shown in Figure \ref{fig:solar}.

When 1D hydrostatic solar model is used (Figure \ref{fig:solar}, top panels), the optical and infrared lines show a significant scatter around the mean value:  $\rm A (\rm Mg)_{\rm  1D~LTE} = 7.51 \pm 0.06{(\rm stat)} \pm 0.05{(\rm syst)}\,\mathrm{dex}$, $\rm A (\rm Mg)_{\rm  1D~NLTE} = 7.50 \pm 0.05{(\rm stat)} \pm 0.05{(syst)}\,\mathrm{dex}$, where the statistical uncertainty is represented by one standard deviation of all measurements (Table \ref{table3}). The systematic error is slightly larger and is dominated by the uncertainties of the oscillator strengths. The results of 1D LTE and 1D NLTE calculations showed no obvious trend with the equivalent width or lower level excitation potential, but strong optical and infra-red lines tend to underestimate Mg abundance. This issue has already been noticed by \citet{bergemann2015}, and is related to the fact that these features are very sensitive to the atmospheric structure. \citet{mashonkina2013a} found that the $4702$ line in the solar spectrum gives a systematically low abundance, but they attributed the problem to the erroneous $\log gf$ and damping values. It seems, however, that adopting the transition probabilities from \citet{pehl} and the damping constants from \citet{barklem2000}, we can achieve an adequate description of their line formation in the solar atmosphere using NLTE calculations with $\md$ models that bring the strong lines in good concordance with the other weaker \mgi\ features. 

The calculations with the MACKKL semi-empirical solar model (Figure \ref{fig:solar}, middle panels) show that in LTE, the 8806 \AA\ line and the IR Mg lines are very sensitive to the chromosphere, over-estimating the Mg abundance compared to the mean. The LTE solar abundance determined with the MACKKL model is $\rm A (\rm Mg ) = 7.57 \pm 0.06{(\rm stat)} \pm 0.05{(\rm syst)}\,\mathrm{dex}$. The NLTE result is slighly lower, $\rm A (\rm Mg ) = 7.54 \pm 0.05{(\rm stat)} \pm 0.05{(\rm syst)}\,\mathrm{dex}$. On the other hand, these calculations also suggest that in NLTE the optical \mgi\ lines, with the exception of the 4702 \AA\ feature, are weakly sensitive to the chromospheric temperature rise. The NLTE MACKKL abundance determined from the $5$ optical lines (4571, 5172, 5183, 5528, 5711 \AA) is only 0.012 dex lower than the $\md$ NLTE abundance based on these lines.

LTE calculations with the $\md$ model atmosphere (Figure \ref{fig:solar}, bottom panels) reveal a clear tendency for the low excitation lines to under-estimate the Mg abundance compared to the high-excitation features. However, the line-to-line scatter is reduced compared to 1D LTE or 1D NLTE calculations. The $\md$ LTE solar abundance is $\rm A (\rm Mg ) = 7.53 \pm 0.03{(\rm stat)} \pm 0.04{(\rm syst)}\,\mathrm{dex}$ (Table \ref{table3}).
%
%abundances
\begin{table*}
\caption{Measurements of Mg abundances for the program stars. The wavelengths of \mgi\ lines are given in \AA. The last columns give the mean abundance, the standard deviation of individual measurements, and the systematic abundance error that includes the uncertainty due to the stellar parameters, the atomic data, and model atmosphere interpolation. [Mg$/$Fe] estimates were derived using the solar estimates of $\rm A({\rm Mg})$ given in the table and metallicity estimates from \citet{bergemann2012b} computed using 1D LTE, 1D NLTE, $\md$ LTE, and $\md$ NLTE. See Section \ref{sec:bmk}.}
\label{table3}
\renewcommand{\tabcolsep}{2pt}
\begin{center}
\begin{tabular}{lcccc ccc | cccc cccc ccc c}
\hline
 &  4571 & 4702 & 5172 & 5183 & 5528 & 5711 & 8806 & 10811 & 10965 & 11828 & 12083 & 15024 & 15765 & 15748 &  15886 & A(Mg) & $\sigma_{\rm stat}$ & $\sigma_{\rm syst}$ & [Mg$/$Fe] \\
\hline
 Sun & &  &  &  &  &  &  &  &  &  &  &  &  &  &  &  & & & \\
 1D LTE & 7.58 & 7.44 & 7.48 & 7.48 & 7.59 & 7.59 & 7.53 & 7.48 & 7.46 & 7.48 & 7.41 & 7.53 & 7.54 &   -- & 7.48 & 7.51 & 0.06 & 0.05 & 0.00 \\
1D NLTE & 7.52 & 7.43 & 7.48 & 7.48 & 7.59 & 7.60 & 7.53 & 7.49 & 7.45 & 7.48 & 7.41 & 7.54 & 7.53 &   -- & 7.49 & 7.50 & 0.05 & 0.05 & 0.00 \\
 $\md$ LTE & 7.46 & 7.50 & 7.50 & 7.50 & 7.54 & 7.56 & 7.55 & 7.54 & 7.53 & 7.53 & 7.58 & 7.58 & 7.57 &   -- & 7.54 & 7.53 & 0.03 & 0.04 & 0.00 \\
$\md$ NLTE & 7.54 & 7.52 & 7.53 & 7.53 & 7.57 & 7.58 & 7.57 & 7.60 & 7.54 & 7.56 & 7.53 & 7.58 & 7.57 &   -- & 7.57 & 7.56 & 0.02 & 0.04 & 0.00 \\
 Procyon & &  &  &  &  &  &  &  &  &  &  &  &  &  &  &  & & & \\
 1D LTE & 7.20 & 7.57 & 7.46 & 7.47 & 7.51 & 7.49 & 7.65 &   -- &   -- &   -- &   -- &   -- &   -- &   -- &   -- & 7.48 & 0.14 & 0.13 &  0.00 \\
1D NLTE & 7.23 & 7.55 & 7.48 & 7.49 & 7.48 & 7.51 & 7.60 &   -- &   -- &   -- &   -- &   -- &   -- &   -- &   -- & 7.48 & 0.12 & 0.13 &  0.01 \\
 $\md$ LTE & 7.20 & 7.63 & 7.50 & 7.58 & 7.65 & 7.51 & 7.69 &   -- &   -- &   -- &   -- &   -- &   -- &   -- &   -- & 7.54 & 0.16 & 0.12 &  0.04 \\
$\md$ NLTE & 7.55 & 7.62 & 7.57 & 7.60 & 7.66 & 7.60 & 7.70 &   -- &   -- &   -- &   -- &   -- &   -- &   -- &   -- & 7.61 & 0.05 & 0.12 &  0.08 \\
 HD 84937 & &  &  &  &  &  &  &  &  &  &  &  &  &  &  &  & & & \\
 1D LTE & 5.61 & 5.78 & 5.77 & 5.83 & 5.76 & 5.84 & 5.77 &   -- &   -- &   -- &   -- &   -- & 5.80 & 5.77 &   -- & 5.77 & 0.07 & 0.11 &  0.29 \\
1D NLTE & 5.68 & 5.85 & 5.77 & 5.83 & 5.85 & 5.85 & 5.79 &   -- &   -- &   -- &   -- &   -- & 5.90 & 5.89 &   -- & 5.82 & 0.07 & 0.11 &  0.35 \\
 $\md$ LTE & 5.48 & 5.72 & 5.56 & 5.58 & 5.75 & 5.76 & 5.66 &   -- &   -- &   -- &   -- &   -- & 5.75 & 5.78 &   -- & 5.67 & 0.11 & 0.10 &  0.17 \\
$\md$ NLTE & 5.68 & 5.86 & 5.85 & 5.86 & 5.90 & 5.85 & 5.89 &   -- &   -- &   -- &   -- &   -- & 5.84 & 5.88 &   -- & 5.85 & 0.07 & 0.10 &  0.32 \\
HD 122563 & &  &  &  &  &  &  &  &  &  &  &  &  &  &  &  & & & \\
 1D LTE & 5.31 & 5.33 &   -- & 5.36 & 5.43 & 5.40 & 5.71 &   -- &   -- &   -- &   -- &   -- & 5.35 & 5.32 &   -- & 5.40 & 0.13 & 0.16 &  0.46 \\
1D NLTE & 5.38 & 5.34 &   -- & 5.39 & 5.44 & 5.38 & 5.43 &   -- &   -- &   -- &   -- &   -- & 5.33 & 5.31 &   -- & 5.38 & 0.05 & 0.16 &  0.45 \\
 $\md$ LTE & 4.66 & 5.06 & 4.97 & 4.96 & 5.16 & 5.11 & 5.21 &   -- &   -- &   -- &   -- &   -- & 5.13 & 5.10 &   -- & 5.04 & 0.16 & 0.16 &  0.08 \\
$\md$ NLTE & 5.26 & 5.48 & 5.22 & 5.20 & 5.26 & 5.31 & 5.29 &   -- &   -- &   -- &   -- &   -- & 5.33 & 5.31 &   -- & 5.30 & 0.08 & 0.16 &  0.31 \\
 HD 140283 & &  &  &  &  &  &  &  &  &  &  &  &  &  &  &  & & & \\
 1D LTE & 5.18 & 5.43 & 5.38 & 5.41 & 5.49 & 5.45 & 5.48 & 5.52 &   -- &   -- &   -- &   -- &   -- &   -- &   -- & 5.42 & 0.11 & 0.11 &  0.31 \\
1D NLTE & 5.27 & 5.51 & 5.43 & 5.45 & 5.61 & 5.53 & 5.49 & 5.58 &   -- &   -- &   -- &   -- &   -- &   -- &   -- & 5.48 & 0.11 & 0.11 &  0.38 \\
 $\md$ LTE & 5.00 & 5.49 & 5.27 & 5.30 & 5.52 & 5.50 & 5.41 & 5.60 &   -- &   -- &   -- &   -- &   -- &   -- &   -- & 5.39 & 0.19 & 0.11 &  0.26 \\
$\md$ NLTE & 5.33 & 5.56 & 5.46 & 5.48 & 5.58 & 5.53 & 5.48 & 5.63 &   -- &   -- &   -- &   -- &   -- &   -- &   -- & 5.51 & 0.09 & 0.11 &  0.35 \\
 G 64-37 & &  &  &  &  &  &  &  &  &  &  &  &  &  &  &  & & & \\
 1D LTE &   -- &   -- & 4.78 & 4.79 & 4.86 &   -- & 4.78 &   -- &   -- &   -- &   -- &   -- &   -- &   -- &   -- & 4.80 & 0.04 & 0.09 &  0.29 \\
1D NLTE &   -- &   -- & 4.84 & 4.85 & 4.90 &   -- & 4.88 &   -- &   -- &   -- &   -- &   -- &   -- &   -- &   -- & 4.87 & 0.03 & 0.09 &  0.37 \\
 $\md$ LTE &   -- &   -- & 4.63 & 4.60 & 4.80 &   -- & 4.77 &   -- &   -- &   -- &   -- &   -- &   -- &   -- &   -- & 4.70 & 0.10 & 0.08 &  0.17 \\
$\md$ NLTE &   -- &   -- & 4.88 & 4.87 & 4.93 &   -- & 4.94 &   -- &   -- &   -- &   -- &   -- &   -- &   -- &   -- & 4.91 & 0.04 & 0.08 &  0.35 \\
\hline
\end{tabular}
\end{center}
\end{table*}

%
%errors
% -------------------------------- ERRORS ---------------------------------
%
\begin{table*}
\centering
\caption{Sensitivities of the Mg abundances based on individual \mgi\ lines to the stellar parameters. The changes of $\teff$, $\log g$, $\feh$ and $\Vmic$ reflect the typical uncertainty of stellar parameters for the program stars.}
\label{tab:abusen}
\renewcommand{\tabcolsep}{5pt}
\begin{tabular}{l cc | cc | cc | cc}
\hline\noalign{\smallskip}
 Star  &  \multicolumn{2}{c} {$\teff$} & \multicolumn{2}{c} {$\log g$} & \multicolumn{2}{c} {[Fe$/$H]} & \multicolumn{2}{c} {$\Vmic$}   \\
\cline{2-9}
       &  $+100$ & $-100$   & $+0.1$ & $-0.1$ &  $+0.1$ & $-0.1$  & $+0.3$   & $-0.3$   \\
       &       K &        K &  dex   &   dex  &    dex  &     dex & kms$^{-1}$ & kms$^{-1}$  \\
\cline{2-9}
\hline\noalign{\smallskip}
Procyon   &    &    &    &    &    &    &     \\                       
 4571  & ~~0.11  & $-$0.10  &  ~~0.00  &  ~~0.00  & $-$0.10  & ~~0.10 & $-$0.06 & ~~0.06 \\
 5711  & ~~0.09  & $-$0.09  & $-$0.04  &  ~~0.04  & $-$0.10  & ~~0.10 & $-$0.09 & ~~0.09 \\
 5183  & ~~0.10  & $-$0.10  & $-$0.07  &  ~~0.07  & $-$0.10  & ~~0.10 & $-$0.02 & ~~0.02 \\
 5528  & ~~0.07  & $-$0.07  & $-$0.02  &  ~~0.02  & $-$0.10  & ~~0.10 & $-$0.08 & ~~0.08 \\
 5711  & ~~0.05  & $-$0.05  &  ~~0.00  &  ~~0.00  & $-$0.10  & ~~0.10 & $-$0.05 & ~~0.05 \\
 8806  & ~~0.07  & $-$0.07  & $-$0.02  &  ~~0.02  & $-$0.10  & ~~0.10 & $-$0.06 & ~~0.06 \\
\cline{2-9}
\hline\noalign{\smallskip}
HD 84937    &    &    &    &    &    &    &     \\
 4571  & ~~0.10  & $-$0.10  &  ~~0.00  &  ~~0.00  & $-$0.08  & ~~0.11 &  ~~0.00 & ~~0.00  \\
 4702  & ~~0.05  & $-$0.04  & $-$0.01  &  ~~0.01  & $-$0.10  & ~~0.10 & $-$0.02 & ~~0.03  \\
 5183  & ~~0.11  & $-$0.11  & $-$0.05  &  ~~0.05  & $-$0.10  & ~~0.10 & $-$0.03 & ~~0.03  \\
 5528  & ~~0.04  & $-$0.04  & $-$0.01  &  ~~0.01  & $-$0.10  & ~~0.10 & $-$0.02 & ~~0.02  \\
 5711  & ~~0.04  & $-$0.04  &  ~~0.01  & $-$0.01  & $-$0.10  & ~~0.10 &  ~~0.00 & ~~0.00  \\
 8806  & ~~0.02  & $-$0.03  &  ~~0.00  &  ~~0.00  & $-$0.10  & ~~0.10 & $-$0.03 & ~~0.03  \\
15748  & ~~0.04  & $-$0.04  & $-$0.01  &  ~~0.01  & $-$0.10  & ~~0.10 & $-$0.02 & ~~0.02  \\
15765  & ~~0.07  & $-$0.07  & $-$0.02  &  ~~0.02  & $-$0.10  & ~~0.10 & $-$0.01 & ~~0.01  \\
\cline{2-9}
\hline\noalign{\smallskip}
HD 140283  &    &    &    &    &    &    &     \\
 4571  & ~~0.12  & $-$0.12  &  ~~0.00  &  ~~0.00  & $-$0.09  & ~~0.10 &  ~~0.00 & ~~0.00  \\
 4702  & ~~0.04  & $-$0.05  & $-$0.01  &  ~~0.01  & $-$0.10  & ~~0.10 & $-$0.02 & ~~0.02  \\
 5183  & ~~0.13  & $-$0.13  & $-$0.05  &  ~~0.05  & $-$0.10  & ~~0.10 & $-$0.03 & ~~0.03  \\
 5528  & ~~0.05  & $-$0.05  &  ~~0.00  &  ~~0.00  & $-$0.09  & ~~0.09 & $-$0.02 & ~~0.02  \\
 5711  & ~~0.04  & $-$0.03  &  ~~0.01  & $-$0.01  & $-$0.10  & ~~0.10 &  ~~0.00 & ~~0.00  \\
 8806  & ~~0.05  & $-$0.05  & $-$0.01  &  ~~0.01  & $-$0.10  & ~~0.10 & $-$0.01 & ~~0.01  \\
10811  & ~~0.02  & $-$0.02  &  ~~0.00  &  ~~0.00  & $-$0.10  & ~~0.10 &  ~~0.00 & ~~0.00  \\
\cline{2-9}
\hline\noalign{\smallskip}
HD 122563  &    &    &    &    &    &    &     \\
 4571  & ~~0.18  & $-$0.21  & $-$0.02  &  ~~0.02  & $-$0.11  & ~~0.11 & $-$0.12 & ~~0.12  \\
 4702  & ~~0.07  & $-$0.07  & $-$0.01  &  ~~0.01  & $-$0.11  & ~~0.10 & $-$0.06 & ~~0.05  \\
 5183  & ~~0.17  & $-$0.20  & $-$0.05  &  ~~0.05  & $-$0.10  & ~~0.10 & $-$0.02 & ~~0.02  \\
 5528  & ~~0.08  & $-$0.08  & $-$0.01  &  ~~0.01  & $-$0.10  & ~~0.10 & $-$0.03 & ~~0.03  \\
 5711  & ~~0.07  & $-$0.08  & $-$0.01  &  ~~0.01  & $-$0.11  & ~~0.11 & $-$0.01 & ~~0.01  \\
 8806  & ~~0.10  & $-$0.10  &  ~~0.00  &  ~~0.00  & $-$0.10  & ~~0.10 & $-$0.08 & ~~0.08  \\
15748  & ~~0.04  & $-$0.04  &  ~~0.00  &  ~~0.00  & $-$0.10  & ~~0.10 & $-$0.02 & ~~0.02  \\
15765  & ~~0.06  & $-$0.06  & $-$0.01  &  ~~0.01  & $-$0.10  & ~~0.10 & $-$0.03 & ~~0.03  \\
\cline{2-9}
\hline\noalign{\smallskip}
G 64-37  &    &    &    &    &    &    &     \\
 5172  & ~~0.06  & $-$0.06  &  ~~0.00  &  ~~0.00  & $-$0.10  & ~~0.10 & $-$0.05 & ~~0.05  \\
 5183  & ~~0.05  & $-$0.05  &  ~~0.00  &  ~~0.00  & $-$0.10  & ~~0.10 & $-$0.04 & ~~0.04  \\
 5528  & ~~0.03  & $-$0.03  &  ~~0.00  &  ~~0.00  & $-$0.10  & ~~0.10 & $-$0.00 & ~~0.00  \\
 8806  & ~~0.03  & $-$0.03  &  ~~0.00  &  ~~0.00  & $-$0.10  & ~~0.10 & $-$0.00 & ~~0.00  \\
%
%\cline{4-9}
%
%
\noalign{\smallskip}\hline
\end{tabular}
\end{table*}

\begin{figure*}
\centering
\hbox{
\includegraphics[width=0.5\textwidth, angle=0]{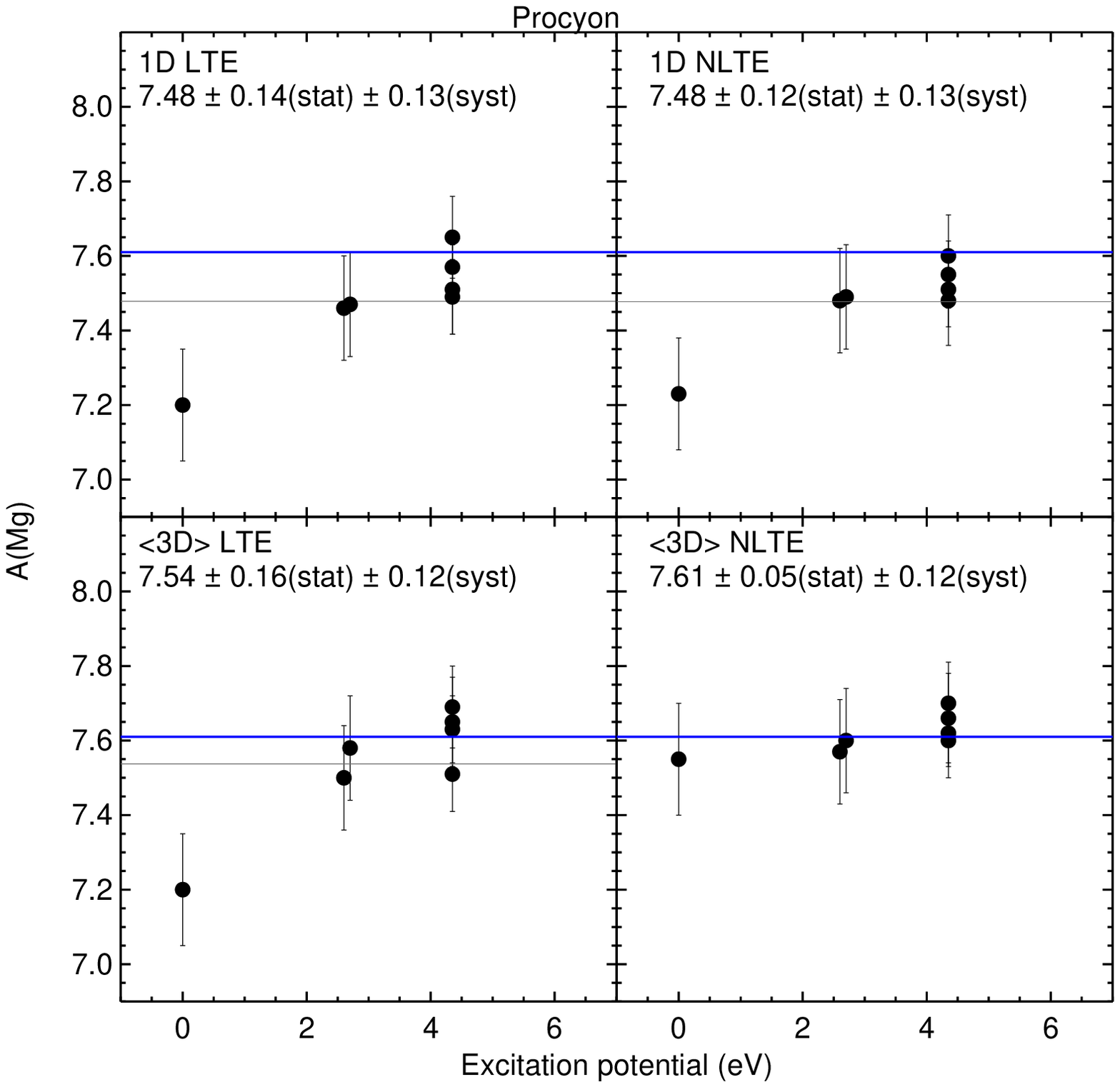}
\includegraphics[width=0.5\textwidth, angle=0]{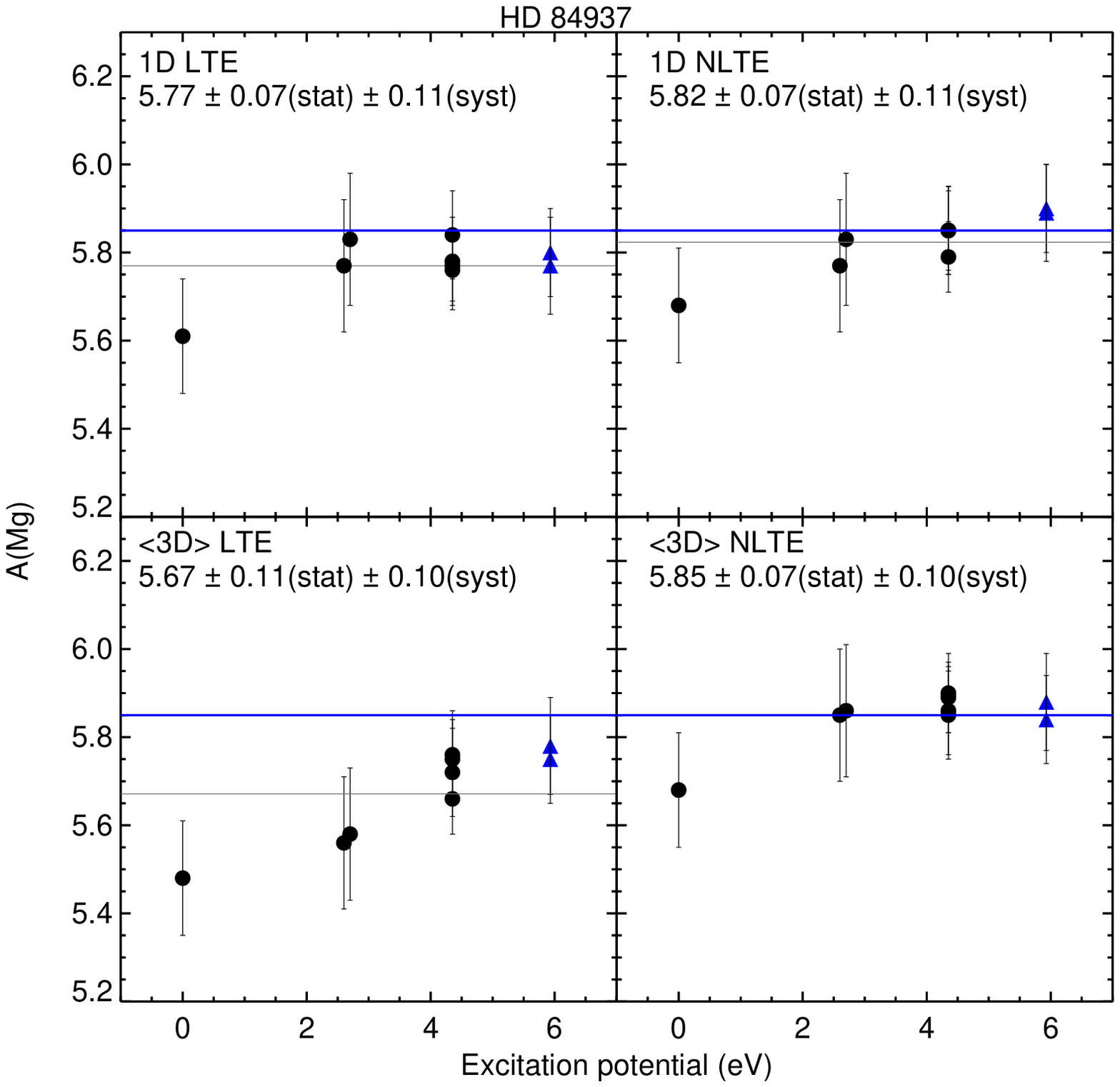}
}
\hbox{
\includegraphics[width=0.5\textwidth, angle=0]{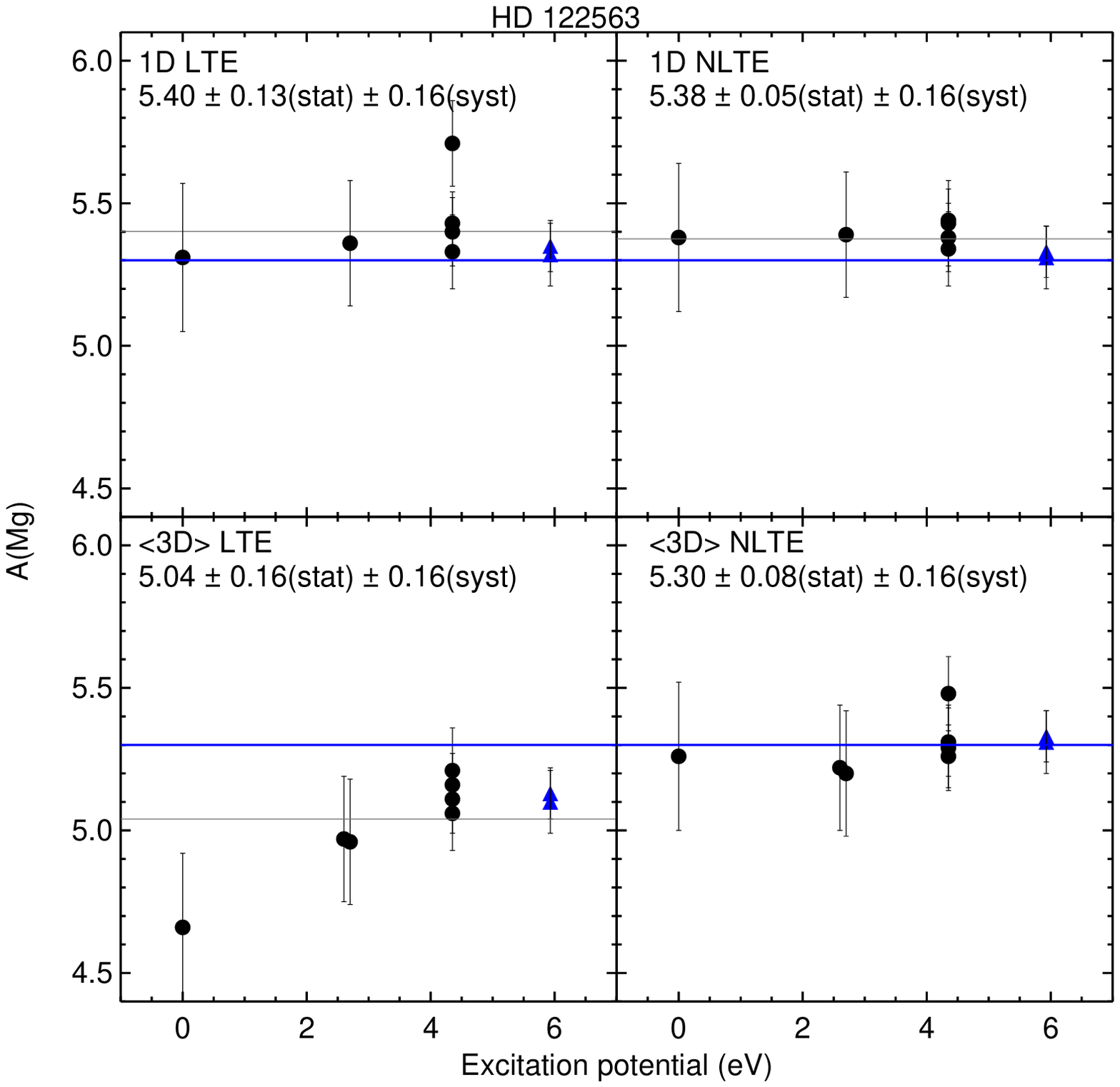}
\includegraphics[width=0.5\textwidth, angle=0]{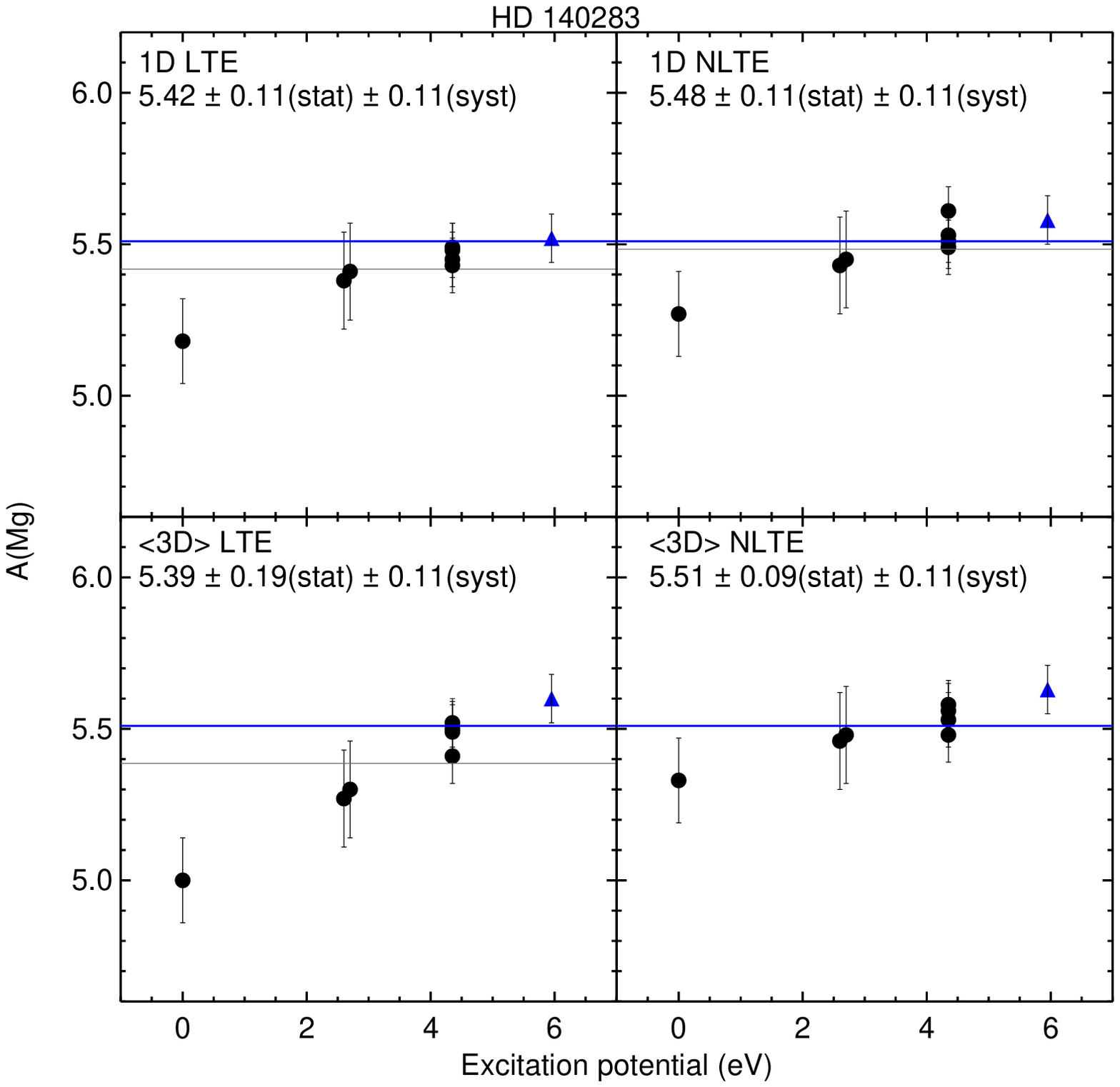}
}
\caption{Abundances determined using different \mgi\ lines as a function of the lower level excitation potential for Procyon (top left), HD 84937 (top right), HD 122563 (bottom left), and HD 140283 (bottom right). The results are shown for the case of 1D LTE, 1D NLTE,  $\md$ LTE, and $\md$ NLTE. The average Mg abundances and the standard deviations of the individual measurements are indicated in the legend. The average $\md$ NLTE abundance is shown with thick blue line. The uncertainties of individual \mgi\ lines reflect the systematic error, which includes the uncertainties due to stellar parameters, $\log gf$, and model atmosphere interpolation.}
\label{fig:stars}
\end{figure*}

The $\md$ NLTE solar Mg abundance (Figure \ref{fig:solar}, bottom panels) determined using all \mgi\ lines from Table \ref{tab:at_data} is $7.56$ $\pm~0.02{(\rm stat)} \pm 0.04{(\rm syst)}\,\mathrm{dex}$. For comparison, the Mg abundance measured in CI chondritic meteorites is $7.55 \pm 0.02$ \citep{lodders2009}\footnote{We note that this value is affected by the solar Si abundance, which was recently revised from $7.52 \pm 0.06$ \citep{shi2008} to $7.51 \pm 0.03$ \citep{amarsi2017}.}. The estimates of the solar Mg abundance in the literature show a large scatter around the meteoritic value. The $\md$ NLTE solar photospheric estimate by \citet{scott2015} is $\rm A (\rm Mg) = 7.59 \pm 0.04\,\mathrm{dex}$\footnote{We note, however, that \citet{scott2015} do not carry out full 3D NLTE calculations for Mg, and compute the Mg abundance by applying a 1D NLTE correction to the 3D LTE result.}. \citet{osorio2015} obtain much higher $\md$ NLTE solar abundance, $7.66 \pm 0.07\,\mathrm{dex}$, but $7.57 \pm 0.08\,\mathrm{dex}$ in 1D LTE, where the uncertainty is the line-to-line dispersion. \citet{mashonkina2013a} studied five \mgi\ lines in the optical solar spectrum using the MAFAGS-OS model and NLTE line formation. Taking the mean of the values in their Table 2, the solar abundance is $\rm A (\rm Mg) = 7.47\,\mathrm{dex}$ in LTE and $\rm A (\rm Mg) = 7.49\,\mathrm{dex}$ in NLTE. The lower estimates derived from several  lines were attributed to the problems with the atomic data. The 1D LTE solar abundance derived by \citet{jofre2015} is also somewhat higher, $7.65 \pm 0.08\,\mathrm{dex}$. It seems that typically larger uncertainties arise when multiple estimates obtained with different codes are combined, as e.g. in the latter study, or when the line selection includes less reliable diagnostic features. \citet{osorio2015}, for example, employed a variety of different Mg lines, mainly to explore the effect of departures from 1D LTE.

The internal consistency among all solar \mgi\ lines in $\md$ NLTE, as well as the remarkable agreement between our $\md$ NLTE photospheric abundance and meteorites, suggests that the effect of atmospheric inhomogeneities on the formation of \mgi\ lines in the solar photosphere is minor and significant systematic errors seem to be excluded. 
%
%
% --------------------------------
\subsection{Mg abundances in the benchmark stars}\label{sec:bmk}
The measured Mg abundances and their errors are listed in Tables \ref{table3} and \ref{tab:abusen}, and are plotted in Figure \ref{fig:stars}. We also provide the standard deviation given by the line-to-line scatter, $\sigma_{\rm stat}$, and the systematic error of abundance measurements, $\sigma_{\rm syst}$, computed by adding the individual errors in quadrature:
\begin{multline}
\label{eq:errors}
\sigma_{\rm syst} = ( \sigma_{\teff}^2 + \sigma_{\rm \log g}^2 + \sigma_{\rm [Fe/H]}^2+ \sigma_{\rm \Vmic}^2 + \\
  \sigma_{\log gf}^2  + \sigma_{\rm model}^2)^{1/2}
\end{multline}
where $\sigma_{\rm model}$ is the abundance error caused by the interpolation in the resampled $\md$ grid (Section \ref{sec:compmodels}), and other components reflect the uncertainty due to the stellar parameters and due to the oscillator strengths. The systematic error $\sigma_{\rm syst}$ was computed by averaging over all used \mgi\ lines. For 1D hydrostatic models, we assume $\sigma_{\rm model}$ of $0.05\,\mathrm{dex}$. The uncertainty due to $\teff$, $\log g$, $\feh$, and $\Vmic$ was estimated using the sensitivities of abundance measurements to the variation of stellar parameters (Table \ref{tab:abusen}).
\begin{figure*}
\centering
\hbox{
\includegraphics[width=0.5\textwidth, angle=0]{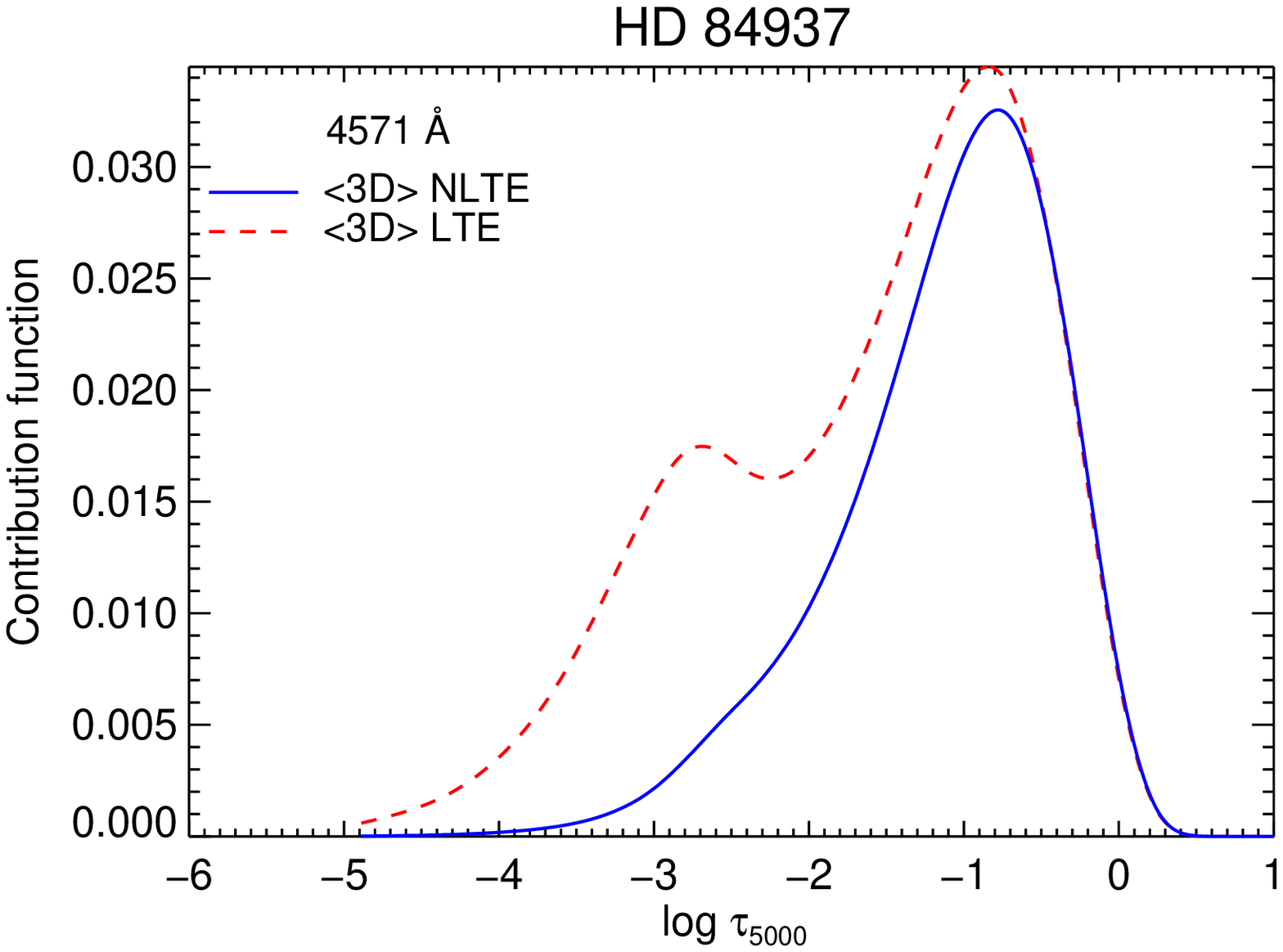}
\includegraphics[width=0.5\textwidth, angle=0]{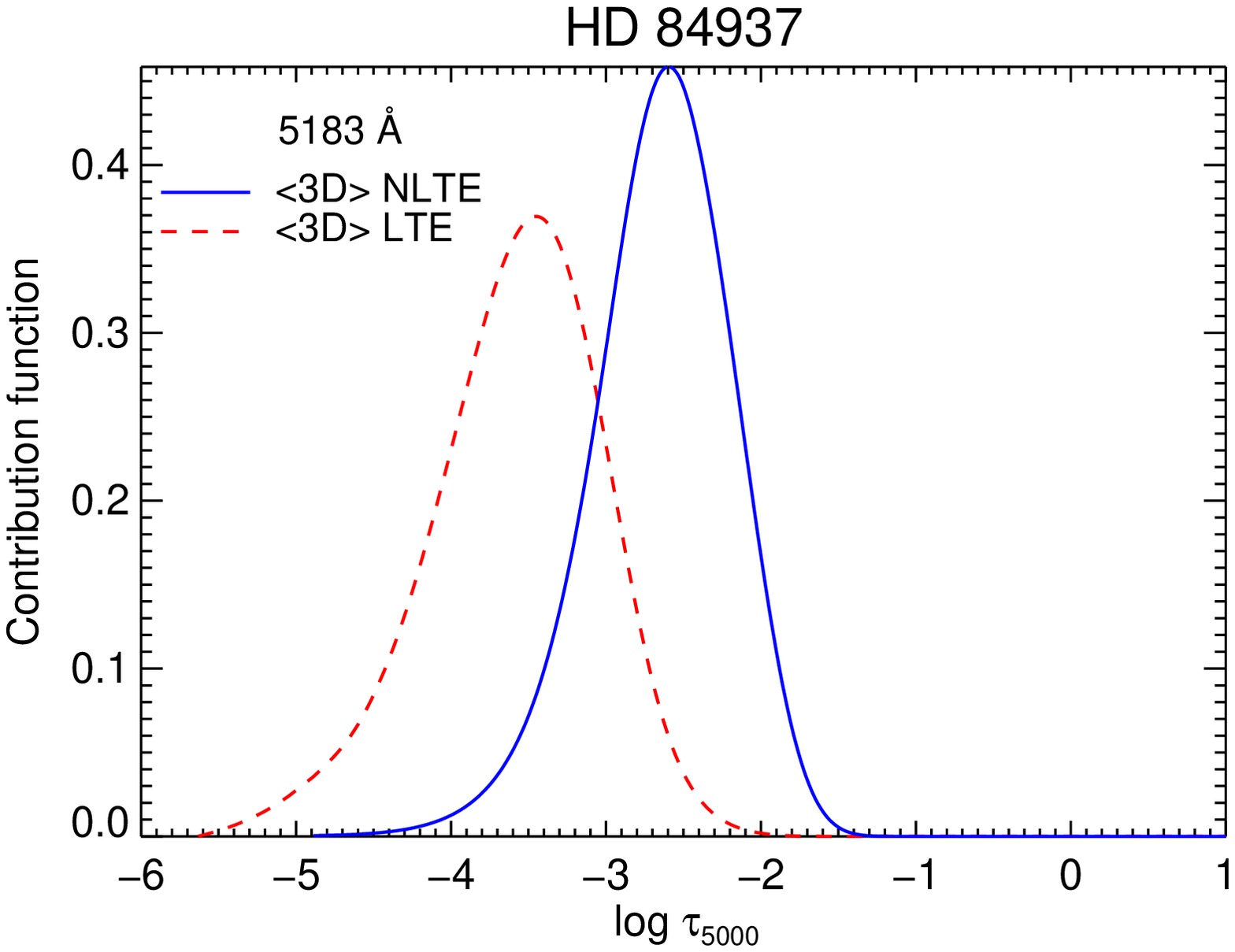}}
\hbox{
\includegraphics[width=0.5\textwidth, angle=0]{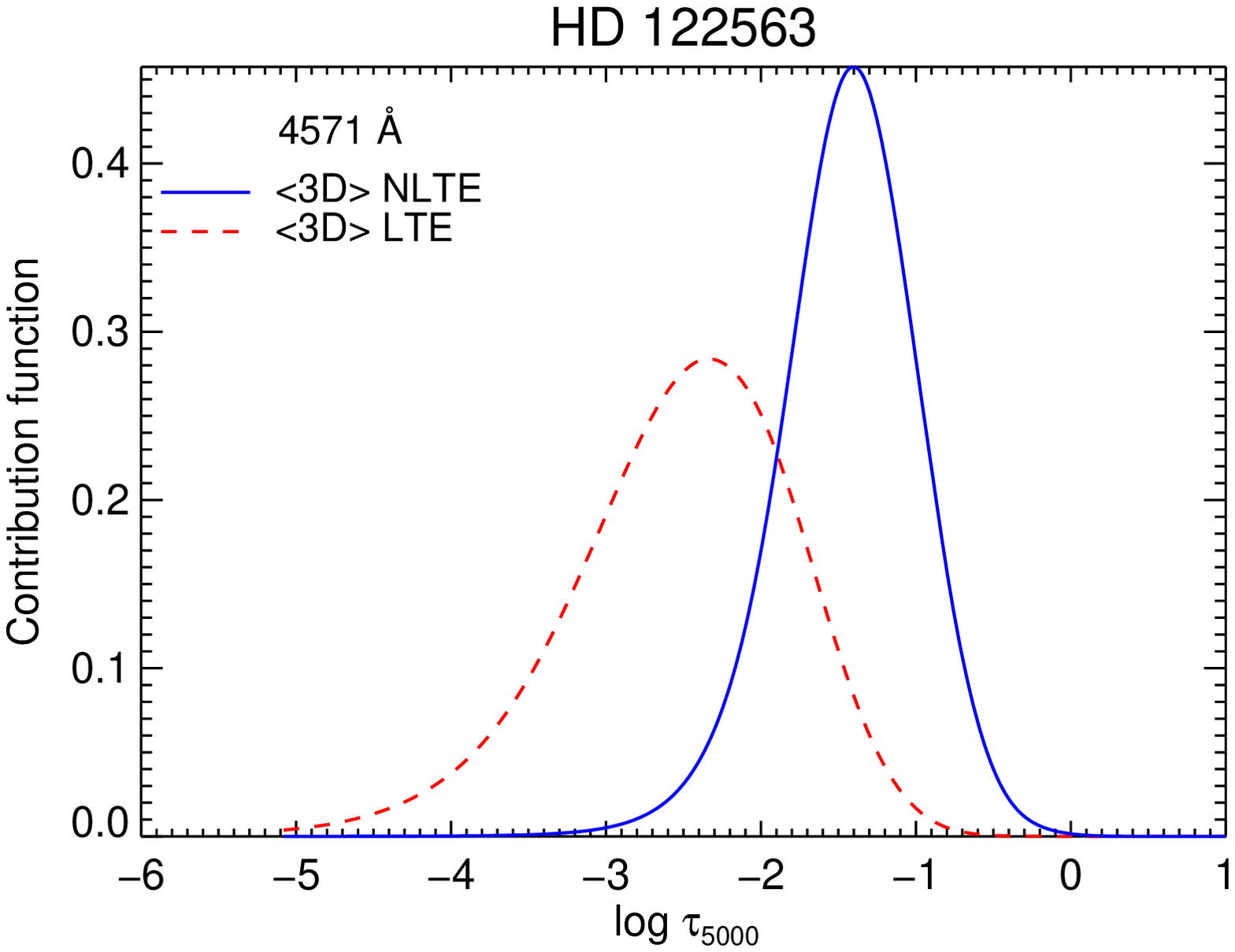}
\includegraphics[width=0.5\textwidth, angle=0]{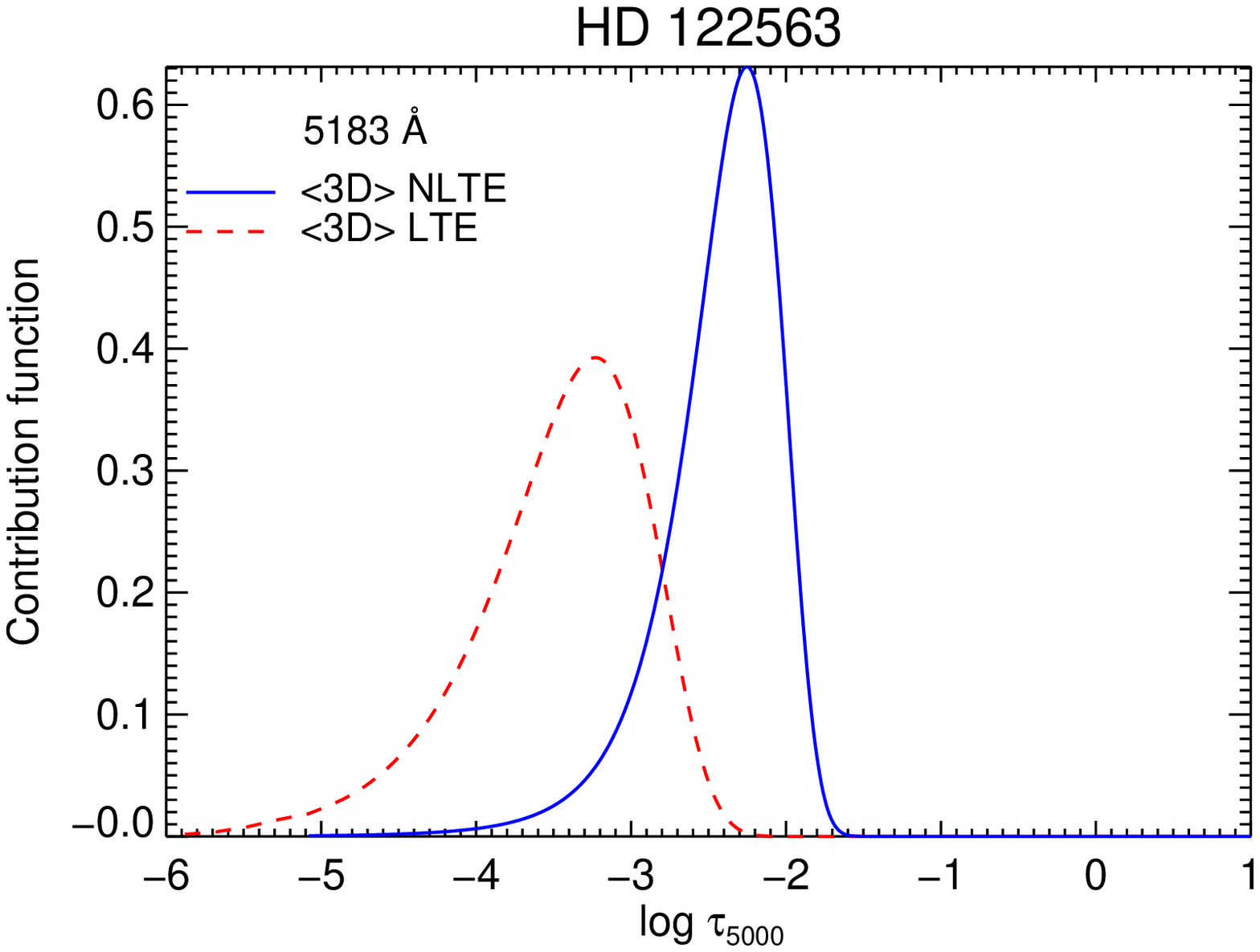}}
\caption{Contribution function in the cores of the 4571 \AA\ (left panels) and 
5183 \AA\ (right panels) lines for the metal-poor stars HD 84937 (top row) and 
HD 122563 (bottom row). The contribution function is defined according to 
\citet{albrow1996}.}
\label{fig:cont}
\end{figure*}
\subsection{Procyon}
Procyon (top left panel, Figure \ref{fig:stars}) is a known spectroscopic binary. Similar to the Sun, $\md$ LTE, and to a lesser degree 1D LTE and 1D NLTE, results in a prominent excitation imbalance, with Mg abundance determined from the intercombination line being too low. Taking NLTE and $\md$ effects into account brings the intercombination line and the Mg triplet lines in a much better agreement with the high-excitation features and reduces the line-to-line scatter from $0.14$ dex (1D LTE) to $0.05\,\mathrm{dex}$ ($\md$ NLTE). The improved excitation balance is the consequence of the shift of the deph of line formation for the low-excitation \mgi\ lines. In particular, in the $\md$ model the 4571 \AA\ intercombination line forms at $\log\tau_{5000}= -2.5$ in LTE, which corresponds to the local temperature of $5263$ K. But the depth of line formation is shifted to $\log\tau_{5000} = -1.8$ in $\md$ NLTE, the local temperature of the $\md$ model at this optical depth point being $5471$ K. It is clear from Table \ref{tab:abusen} that the 4571 feature reacts sensitively to the changes in temperature. For the $\sim +250$ K difference, the abundance would change by $+0.25$ dex, which is exactly the difference between the $\md$ NLTE and $\md$ LTE abundances. This $0.25$ dex abundance correction for the 4571 \AA\ line solves the excitation balance problem for Procyon.

1D LTE and 1D NLTE abundances are nearly identical, but $\sim 0.1\,\mathrm{dex}$ lower than the $\md$ NLTE Mg abundance, $7.61 \pm 0.05{(\rm stat)} \pm 0.12{(\rm syst)}\,\mathrm{dex}$.

Our $\md$ NLTE abundance for Procyon is in a reasonable agreement with the $\md$ NLTE measurement by \citet{osorio2015}. Their analysis gives $\rm A (\rm Mg) = 7.56 \pm 0.06\,\mathrm{dex}$ that is only $0.05\,\mathrm{dex}$ lower than our result. Also their 1D LTE ($7.47 \pm 0.04\,\mathrm{dex}$) and 1D NLTE ($7.45 \pm 0.04\,\mathrm{dex}$) values agree well with our measurements, $7.48 \pm 0.14{(\rm stat)} \pm 0.13{(\rm syst)}$ and $7.48 \pm 0.12{(\rm stat)} \pm 0.13{(\rm syst)}\,\mathrm{dex}$, respectively. Our uncertainity estimates are larger, because we inlcude different sources of errors (Equation \ref{eq:errors}), while \citet{osorio2015} quote the line-to-line dispersion as abundance error.

\citet{jofre2014} derived [Mg$/$H] $ = -0.037 \pm 0.07$ (1D LTE) and [Mg$/$H] $ = -0.035 \pm 0.07$ (1D NLTE), which translates to $\rm A (\rm Mg) = 7.613 \pm 0.07\,\mathrm{dex}$ and  $\rm A (\rm Mg) = 7.615 \pm 0.07\,\mathrm{dex}$. Both estimates are somewhat too high, compared to the 1D results by \citet{osorio2015} and our study. One may relate the discrepancy to the choice of the lines and to the way the analysis is done, by combining the measurements obtained using different abundance pipelines. One should point out, however, that the 1D LTE and NLTE Mg abundances derived by \citet{jofre2014} are consistent with our $\md$ NLTE Mg abundance.
\subsection{Metal-poor dwarfs HD 84937 and G 64-37}
Mg lines in the spectrum of the metal-poor turnoff star HD 84937 are reasonably well described by all models, but 
the resulting abundances are quite different (top right panel, Figure \ref{fig:stars}). $\md$ NLTE abundance is $0.08\,\mathrm{dex}$ 
higher than the 1D LTE estimate, and $0.18\,\mathrm{dex}$ higher than the $\md$ LTE 
estimate. In all cases, but $\md$ LTE, the abundances determined using the 
optical and the infrared lines are consistent within the uncertainties, the 4571 
\AA\ line being the only problematic exception. Whether this is a consequence of 
its weakness in this regime of stellar parameters or sensitivity to atmospheric 
inhomogeneities, is not entirely clear. Variation of microturbulence has 
virtually no effect on its strength (Table \ref{tab:abusen}), but the line is 
somewhat more sensitive to the variation of temperature than the 
higher-excitation lines. Figure \ref{fig:cont} shows that the contribution 
function of the 4571 \AA\ line core is very broad and extends over two 
orders of magnitude in optical depth. Even in NLTE, the line core is sensitive 
to conditions at $-2 \lesssim \log \tau_{\rm{5000}} \lesssim 0$. The LTE contribution function shows 
a characteristic secondary bump at $\log \tau_{\rm{5000}} \sim -3$ corresponding 
to the depth where temperature drops in the $\md$ model because of adiabatic 
cooling at the surface. Interestingly, the recent study by \citet{spite2017} 
suggested that HD 84937 has a solar-like chromosphere, which has an effect on 
the formation of molecular lines in 3D. We cannot exclude the possibility that 
chromosphere also has an effect on \mgi\ lines; on the other hand, the optical 
triplet lines (e.g. the 5183 \AA\ line) form even higher out in the $\md$ model 
(Figure \ref{fig:cont}, top right panel) and these lines do not stand out compared 
to the mean $\md$ NLTE estimate, being fully consistent with the high-excitation 
lines.

The difference between the 1D NLTE and $\md$ NLTE results for HD 84937 is 
$\sim 0.03\,\mathrm{dex}$, which is too small to be considered as significant. This is 
also true for the very metal-poor dwarf G 64-37, for which we estimate $\rm A 
(\rm{Mg}) = 4.87 \pm 0.03{(\rm stat)} \pm 0.09{(\rm syst)}\,\mathrm{dex}$ in 1D NLTE and $\rm A (\rm{Mg}) = 4.91 \pm 0.04{(\rm stat)} \pm 0.08{(\rm syst)}\,\mathrm{dex}$ in $\md$ NLTE. Thus, within the rather generous margin for uncertainty, of the order $\sim 0.1\,\mathrm{dex}$, the optical triplet lines (5172, 5183 \AA) and the lines arising from $4.35$ eV levels (5528, 5711, 8806 \AA) appear to be reliable diagnostic of Mg abundances in main-sequence stars across a wide metallicity range.

There are several accurate studies of Mg abundance in HD 84937, as this is a Gaia benchmark star and is commonly used to assess the accuracy of large-scale surveys, e.g. the Gaia-ESO stellar spectroscopic survey \citep{gilmore2012,randich2013}. Here we have chosen to compare our results with those studies, which employ the same 1D or $\md$ atmospheres and/or NLTE. We stress, however, that our work is the first that makes use of the new transition probabilities from \citet{pehl} that arguably leads to systematic differences with the earlier work. 
Among the studies that include NLTE line formation, it is interesting to compare with \citet{gehren2006}, who estimated $\mgfe = 0.24\,\mathrm{dex}$ in LTE and $\mgfe = 0.32$ in NLTE, well in agreement with our results, $\mgfe = 0.29 \pm 0.12\,\mathrm{dex}$ in LTE and $\mgfe = 0.35 \pm 0.12$ in NLTE, respectively. 
 \citet{mashonkina2013a} determined somewhat lower values for HD 84937. Using their individual measurements (Table 4, BBSGF case for NLTE), one may obtain the mean abundance of $\rm A (\rm Mg) = 5.65$ in LTE and $\rm A (\rm Mg) = 5.66$ in NLTE. The offset from our measurement could be explained by the differences in the stellar parameters and atomic data between our and their study. 
 The analysis by \citet{jofre2015} using the same $\feh$ and $\Vmic$ as in our study, yielded [Mg$/$H] $ = -1.76 \pm 0.11\,\mathrm{dex}$ in LTE and [Mg$/$H] $ = -1.77 \pm 0.11$ in NLTE. Our estimates on the [Mg$/$H] scale are $-1.74 \pm 0.07\,\mathrm{dex}$ (1D LTE) and $-1.68 \pm 0.07$ (1D NLTE). Our 1D LTE estimates are in excellent agreement with \citet{jofre2015}, but the offset in 1D NLTE could be explained by the fact that \citet{jofre2015} applied NLTE abundance corrections to the LTE measurements, while we perform NLTE spectrum synthesis.
Finally, by comparing the results with \citet{osorio2015}, we find good agreement with their 1D LTE, 1D NLTE, and $\md$ NLTE estimates, although, our results are $\sim 0.05\,\mathrm{dex}$ higher. Since \citet{osorio2015} used the same $\md$ model atmosphere and stellar parameters, the small difference likely stems from the different selection of lines, atomic data, and abundance analysis techniques. 

G 64-37 was recently analysed by \citet{ishi2012}. The latter study used 1D LTE to determine the surface gravity, metallicity, and $\mgfe = 0.15 \pm 0.07\,\mathrm{dex}$, which is lower than our 1D LTE result, $\mgfe = 0.29 \pm 0.09\,\mathrm{dex}$. However,  we note that the stellar parameters adopted in this work are considerably different from our estimates. In particular, the surface gravity is $\log g = 4.60\,\mathrm{dex}$, $0.3\,\mathrm{dex}$ higher than to the value based on the NLTE ionization equilibrium of Fe \citep{hansen2013}. Also, their adopted $\teff$ and microturbulence values, $6621\,\mathrm{K}$ and $2.50$ kms$^{-1}$ respectively, are notably different from our estimates of $6494\,\mathrm{K}$ and $1.4$ kms$^{-1}$. 
\subsection{Metal-poor giant HD 122563}
The most internally consistent results for the metal-poor giant HD 122563 
(bottom left panel, Figure \ref{fig:stars}) are obtained using 1D NLTE, which 
results in a remarkable scatter of only $0.05\,\mathrm{dex}$, contrasting with the $\md$ 
LTE scatter of $0.16\,\mathrm{dex}$. $\md$ NLTE abundance is $5.30 \pm 0.08{(\rm stat)} \pm 0.16{(\rm syst)}\,\mathrm{dex}$, $0.26$ dex higher than $\md$ LTE and $0.1\,\mathrm{dex}$ lower than 1D NLTE. The individual \mgi\ 
lines yield consistent results within the respective uncertainties. In $\md$ 
NLTE, the 4702 \AA\ line stands out slightly above the other features. This line 
forms in the transition between the \Mg{3p}{1}{P}{\circ}{1} ($4.35$ eV) and 
\Mg{5d}{1}{D}{}{2} ($6.98$ eV) levels, and represents the transition with the 
largest energy difference between the lower and the upper level. There remains a 
possibility that our representation of the model atom at these energies, 
separated by only $0.6$ eV from the $1$st ionization threshold of \mgii, is not 
complete. In particular, the lack of inelastic hydrogen collisions for the 
levels above $\sim 5.9$ eV is unfortunately still a bottleneck in the 
calculations. On the other hand, incompleteness of the model 
atom would be readily seen in the failure of the NLTE calculations to provide 
consistent abundances from the other spectral lines, too, in particular from the 
infra-red \mgi\ lines that connect even higher excitation states. This is not 
the case, neither in 1D NLTE nor in $\md$ NLTE. Nevertheless, presently we 
cannot rule out the possibility that very high excitation levels, $\gtrsim 6$ 
eV, require a more accurate knowledge of collisions, and do not recommend using 
the lines connecting these levels in abundance determinations. The 4571 \AA\ and 
optical triplet lines have very well-defined contribution functions in $\md$ 
NLTE (Figure \ref{fig:cont}, bottom panels), and yield abundances consistent
the other diagnostic features in 1D LTE, 1D NLTE, and $\md$ NLTE.

Table \ref{table3} shows that, in accord with our full 3D NLTE calculations 
(section \ref{sec:nlte3d}), the line at 5711 \AA\ is least affected by $\md$ 
NLTE, but it is also very weak at low metallicity (Figure \ref{fig:profiles}) and 
requires very high-resolution and high S$/$N observations. In the spectrum of HD 
122563, the line has an equivalent width of only $10$ m\AA, and is barely 
distinguishable at the resolving power of the UVES instrument at the VLT 
($R \sim 47\,000$). The other useful diagnostic features, the 5528 \AA\ and 5183 \AA\ 
lines, give slightly lower abundances compared to the mean $\md$ NLTE estimate, 
although still consistent within the error bars.  However, we should also note 
that the optical triplet lines are very sensitive to temperature (Table 
\ref{tab:abusen}) that makes them an unreliable abundance diagnostic in 
metal-poor red giants; this sensitivity also implies that the effect of 
inhomogeneities would be more important. It thus appears that the analysis of 
Mg abundances in metal-poor evolved stars should be done using the 5711 and 
5528 \AA\ features.

HD 122563 is a classical metal-poor halo giant and several estimates of its Mg abundance are available in the literature. Among other studies, we may compare our results with \citet{jofre2015}, whose estimates, [Mg$/$H] $ = -2.354\,\mathrm{dex}$ (1D LTE) and [Mg$/$H] $ = -2.359\,\mathrm{dex}$ (1D NLTE), are significantly lower than our results, but this offset may reflect signficant differences in the assumed metallicity, $\feh = -2.66$, and microturbulence, $\Vmic = 1.92$ kms$^{-1}$, in \citep{jofre2015}. As shown in Table \ref{tab:abusen}, most \mgi\ lines in metal-poor red giants are very sensitive to $\Vmic$ and $0.3$ kms$^{-1}$ difference may affect the abundance at the level of $\sim 0.1\,\mathrm{dex}$. The $\md$ NLTE abundance determined by \citet{osorio2015}, $\rm A(\rm Mg) = 5.27 \pm 0.08\,\mathrm{dex}$, is in excellent agreement with our result, $5.30 \pm 0.08{(\rm stat)} \pm 0.16{(\rm syst)}\,\mathrm{dex}$. We note that the error of our estimate includes various sources of uncertainty, including the uncertainties due to stellar parameters, atomic data, and model interpolation. The error quoted by \citet{osorio2015} reflects the line-to-line dispersion only.
\subsection{Metal-poor giant HD 140283}

The results for the metal-poor sub-giant HD 140283 are shown in bottom right panel in Figure \ref{fig:stars}. All models yield a clear residual trend of abundance with excitation potential, although $\md$ NLTE performs better 
compared to $\md$ LTE in this respect. We have considered the possibility that 
the excitation disbalance is caused by the uncertainties of stellar parameters. 
\citet[][]{heiter2015} suggest that the $\teff$ 
estimate for HD 140283 is poorly constrained. Their photometric and 
spectroscopic estimates, which are fully consistent with our adopted value, are 
nonetheless $\sim$ 200 K higher than the $\teff$ value based on the  
interferometric angular diameter, $5500$ K. However, if we were to adopt their  
lower $\teff$, the excitation imbalance would be even more extreme (Table 
\ref{tab:abusen}), because the low-excitation lines strengthen with decreasing 
temperature thus requiring lower abundance to fit the observed profiles. 
Inspection of the spectral line sensitivity to surface gravity and metallicity 
shows that none of these parameters could reconcile the abundances derived from 
the low- and high-excitation lines. The failure to establish excitation balance 
 may imply that the assumption of quasi-static (1D or $\md$) atmospheric  
structure is no longer adequate in this regime of stellar parameters and full 3D 
NLTE analysis is necessary in order to determine accurate abundances from the 
low-excitation lines. Nevertheless, the difference between the average 1D NLTE 
and $\md$ NLTE Mg abundances is, in fact, minor: $5.48 \pm 0.11{(\rm stat)} \pm 0.11{(\rm syst)}$ dex and $5.51 \pm 0.09{(\rm stat)} \pm 0.11{(\rm syst)}$ dex, respectively. Avoiding the intercombination line, all other \mgi\ features can be used in the NLTE Mg abundance analysis of metal-poor sub-giants with 1D or $\md$ models.

HD 140283 was analysed using 1D LTE and 1D NLTE by \citet{jofre2015}, and using $\md$ NLTE by \citet{osorio2015}. According to the latter study, Mg abundance in $\md$ NLTE is $5.45 \pm 0.09\,\mathrm{dex}$. Our $\md$ NLTE estimate is $5.51 \pm 0.09{(\rm stat)} \pm 0.11{(\rm syst)}\,\mathrm{dex}$, $\sim 0.06\,\mathrm{dex}$ higher than the measurement by \citet{osorio2015}, but compatible with the latter within the rather generous margin of uncertainty. The 1D LTE estimate by \citet{jofre2015} based is [Mg$/$H] $ = -2.326 \pm 0.046\,\mathrm{dex}$ (1D LTE) and [Mg$/$H] $ = -2.336 \pm 0.046\,\mathrm{dex}$ (1D NLTE), or $\rm A(\rm Mg) = 5.324$ (1D LTE) and  $\rm A(\rm Mg) = 5.314$ (1D NLTE). Our values are $\sim 0.1\,\mathrm{dex}$ higher, although still consistent within the respective uncertainties. It should be noted that the estimate by \citet{jofre2015} is based on the 5711 \AA\ line only, and there is significant abundance dispersion between different nodes, e.g. the iSpec method provided $\rm A(\rm Mg) = 5.38\,\mathrm{dex}$, while using the Porto pipeline they obtain $\rm A(\rm Mg) = 5.16\,\mathrm{dex}$ \citep[][see Section 3.3, also online material]{jofre2015}. Our estimate is based on 8 \mgi\ lines, which makes us believe that our estimate is more reliable.
\section{Conclusions}\label{sec:conclusions}
We perform chemical abundance analysis of $6$ standard stars using the most recent experimental and theoretical transition probabilities from \citet{pehl}. The abundances of Mg are determined using $15$ \mgi\ lines in the optical and infrared with four different techniques: LTE and NLTE with 1D hydrostatic model atmospheres, as well as with the averages of 3D hydrodynamical model atmospheres.

We may summarise our results as follows:
\begin{itemize}
\item $\md$ NLTE solar Mg abundance determined using 14 \mgi\ lines in the optical and IR is $7.56 \pm 0.02 {(\rm stat)} \pm 0.04{(\rm syst)}\,\mathrm{dex}$. The systematic error is dominated by the uncertainty of the atomic data, while the line-to-line dispersion is small. The abundance is in agreement with the meteoritic estimate $7.55 \pm 0.02\,\mathrm{dex}$ \citep{lodders2009}, although we note that the latter may undergo a slight downward revision following the new analysis of the solar Si abundance by \citet{amarsi2017}. 1D LTE and 1D NLTE estimates of the solar Mg abundance are $7.51 \pm 0.06{(\rm stat)} \pm 0.05{(\rm syst)}\,\mathrm{dex}$ and $7.50 \pm 0.05{(\rm stat)} \pm 0.05{(\rm syst)}\,\mathrm{dex}$, respectively.
\item The optical \mgi\ lines in the solar spectrum, with the exception of the 4702 and 8806 \AA\ features, are barely sensitive to the chromosphere. Using the optical lines only, the NLTE abundance derived using the semi-empirical solar model with chromosphere \citep{maltby1986} is consistent with the $\md$ NLTE Mg abundance, with the mean abundance difference of $\rm A (\rm{Mg, chromosphere~NLTE}) - \rm A (\rm{Mg, \md~NLTE}) = 0.012$ dex.
\item Mg abundances determined from the infrared stellar spectra are as accurate as the optical diagnostics, if 1D hydrostatic or $\md$ model atmospheres are used. On the other hand, test calculations using the \citet{maltby1986} solar model show that chromosphere has a stronger effect on the infra-red \mgi\ lines, if the LTE assumption is used, with some lines deviating by more than $0.1$ dex.

\item $\md$ NLTE offers an improvement to the accuracy of Mg abundance determinations in cool stars, despite persistent problems with the intercombination line at 4571 \AA, which is very sensitive to temperature and likely requires full three-dimensional treatment with account of NLTE. For the high-excitation lines, $\md$ NLTE abundance scatter in all program stars is reduced compared to 1D LTE. 
\item The difference between 1D LTE and 1D NLTE abundances is not large, of the order $0.06\,\mathrm{dex}$ for the most metal-poor stars in the sample, such as G 64-37 with $\feh = -3\,\mathrm{dex}$. Line-to-line abundance discrepancies are evidently related to the poor representation of the atmospheric structure in 1D hydrostatic calculations.
\item Low-excitation \mgi\ lines, 4571, 5172, and 5183 \AA\, are sensitive to the atmospheric structure. For these lines, LTE calculations with $\md$ models lead to significant systematic errors in abundance. As a consequence, $\md$ LTE calculations leave a strong residual trend of abundance with excitation potential of the lower level of the transition. This supports our earlier results on Fe. In \citet{bergemann2012b}, we showed that $\md$ LTE approach fails to provide satisfactory solutions for stellar parameters based on the excitation-ionization equilibrium of \fei\ and \feii\ lines.
\item The case of the metal-poor subgiant HD 140283 is not conclusive: a clear correlation between Mg abundances and the line excitation potential also remains in 1D and $\md$ NLTE calculations. While one could suspect erroneous $\teff$, our value $5777\,\mathrm{K}$is in a good agreement with other photometric and spectroscopic estimates~\citep{heiter2015}. The $\teff$ estimate based on the interferometric measurement of the star's angular diameter is significantly lower, $5500\,\mathrm{K}$\citep{heiter2015}, and it is not supported by our calculations. 
\item 1D NLTE and $\md$ NLTE abundances are consistent within the errors of individual measurements. For the metal-poor dwarfs and sub-giant, the difference between $\md$ NLTE and 1D NLTE abundances amounts to $0.04\,\mathrm{dex}$, which is small compared to standard sources of error in stellar spectroscopy, such as, for example, the uncertainties of stellar parameters and atomic data. For the metal-poor giant, HD 122563, 1D NLTE approach over-estimates abundance by $\sim 0.1\,\mathrm{dex}$ compared to $\md$ NLTE. This error should be taken into account in abundance analyses of metal-poor red giants with hydrostatic model atmospheres.
\end{itemize}
 
Full 3D NLTE calculations suggest that the high-excitation line 5711 \AA\ is the most robust diagnostic of Mg abundance in cool FGK stars. It is least sensitive to the deviations from 1D and LTE. For this transition, the difference between $\md$ NLTE and full 3D NLTE abundance determinations is within $0.05\,\mathrm{dex}$ across the full metallicity range $-2.5 \leq \feh \leq 0$. Furthermore, the line can be safely modelled in 1D NLTE in dwarfs, and only a small negative correction of $\sim 0.1\,\mathrm{dex}$ should be applied to the 1D NLTE abundance based in this line in 
red giants. However, this line is detectable only in very high-quality (high resolution and signal-to-noise) spectra of metal-poor stars. Alternatively, one may use the optical triplet (5172, 5183 \AA) and 5528 \AA\ features, which are strong enough at low metallicity and, according to the $\md$ NLTE results, are in agreement with the abundances based on  5711 \AA\ line. However, full 3D NLTE calculations indicate that these lines are more sensitive to atmospheric inhomogeneities and should be used with caution in the spectra of red giants.

\acknowledgements
We thank T. Gehren and F. Grupp for providing the observed optical spectra and 
stellar atmosphere models used in this work. We thank Y. Takeda for providing 
the reduced infrared spectra for several stars. RC acknowledges partial support 
from a DECRA grant from the Australian Research Council (project DE120102940). 
GRR acknowledges support from the project grant "The New Milky Way" from the 
Knut and Alice Wallenberg Foundation. MB acknowledges support by the 
Collaborative Research Centre SFB 881 (Heidelberg University) of the Deutsche 
Forschungsgemeinschaft (DFG, German Research Foundation). This work was 
supported by a research grant (VKR023406) from VILLUM FONDEN. This research was 
undertaken with the assistance of resources from the National Computational 
Infrastructure (NCI), which is supported by the Australian Government. Funding 
for the Stellar Astrophysics Centre is provided by the Danish National Research 
Foundation. AMA acknowledges funds from the Alexander von Humboldt Foundation in the framework of the Sofja Kovalevskaja Award endowed by the Federal Ministry of Education and Research. We thank the anonymous referee for many useful suggestions.

\end{document}